\documentclass[useAMS,usenatbib]{mn2e}
\usepackage{myaasmacros}
\usepackage{amssymb}
\usepackage{graphicx}
\usepackage{xcolor}

\def\rhodm{\rho_\mathrm{dm}}
\def\rhotot{\rho_\mathrm{tot}}

\def\Sigmatot{\Sigma_\mathrm{tot,L}}
\def\Sigmas{\Sigma_\mathrm{s,L}}
\def\rhodmext{\rho_\mathrm{dm,ext}}
\def\rhos{\rho_\mathrm{s}}

\def\vztwo{\overline{v_z^2}}
\def\vthetatwo{\overline{v_\theta^2}}
\def\vRtwo{\overline{v_R^2}}
\def\vztwoi{\overline{v_{z,i}^2}}
\def\rhoeff{\rho_\mathrm{dm}^\mathrm{eff}}
\def\beq{\begin{equation}}
\def\eeq{\end{equation}}
\def\msun{\,M$_\sun$}

\def\urij{{\bf r}_{ij}}
\def\uri{{\bf r}_{i}}
\def\urj{{\bf r}_{j}}
\def\ur{{\bf r}}
\def\dthr{\mathrm{d}^3{ r}}
\def\dx{x_{ij}}
\def\dy{y_{ij}}
\def\dz{z_{ij}}

\title{Limits on the local dark matter density}

\author[S. Garbari et al.]{Silvia Garbari$^{1}$\thanks{e-mail:
\texttt{silvia@physik.uzh.ch}}, Justin I. Read$^{2,3}$ and George Lake$^1$\\
  $^1$Institute of Theoretical Physics, University of Z\"urich, Winterthurerstrasse 190, 8057 Z\"urich, Switzerland\\
  $^2$Department of Physics and Astronomy, University of Leicester, University Road, Leicester , LE1 7RH\\
  $^3$Institute for Astronomy, Department of Physics, ETH Z\"urich, Wolfgang-Pauli-Strasse 27, CH-8093 Z\"urich, Switzerland}

\begin{document}

\pagerange{\pageref{firstpage}--\pageref{lastpage}} 
\pubyear{2011}

\maketitle

\label{firstpage}

\begin{abstract}
We revisit systematics in determining the local dark matter density $\rhodm$ from the vertical motion of stars in the Solar Neighbourhood. Using a simulation of a Milky Way like galaxy, we determine the data quality required to detect $\rhodm$ at its expected local value. We introduce a new method for recovering $\rhodm$ that uses moments of the Jeans equations, combined with a Monte Carlo Markov Chain technique to marginalise over the unknown parameters. Given sufficiently good data, we show that our method can recover the correct local dark matter density even in the face of disc inhomogeneities, non-isothermal tracers and a non-separable 
distribution function. We illustrate the power of our technique by applying it to {\it Hipparcos} data. We first make the assumption that the A and F star tracer populations are isothermal. This recovers $
\rhodm=0.003^{+0.009}_{-0.007}$\msun\,pc$^{-3}$  ($\rhodm=0.11^{+ 0.34}_{-0.27}$GeV\,cm$^{-3}$, with 90 per cent confidence), consistent with previous determinations. However, the vertical dispersion profile of these tracers is poorly known. If we assume instead a non-isothermal profile similar to the recently measured blue disc stars from SDSS DR-7, we obtain a fit with a very similar $\chi^2$ value, but with $\rhodm=0.033^{+0.008}_{-0.009}$\msun\,pc$^{-3}$ ($\rhodm=1.25^{+0.30}_{-0.34}$GeV\,cm$^{-3}$ with 90 per cent confidence). This highlights that it is vital to measure the vertical dispersion profile of the tracers  to recover an unbiased estimate of $\rhodm$.
\end{abstract}



\section{Introduction}\label{intro}
There are two approaches to determine the local dark matter density: extrapolating its value from the Milky Way's rotation curve \citep[$\rhodmext$; e.g.][]{sofue_unified_2008,weber_determination_2010}; and using the kinematics of stars in the Solar Neighbourhood \citep[$\rhodm$; e.g.][]{oort_force_1932,oort_note_1960}. The first requires an assumption about the global and local shape of the dark matter halo.  Simple extrapolations that assume spherical symmetry, find $\rhodmext \simeq 0.01$\msun pc$^{-3}$ \citep{sofue_unified_2008}. However, uncertainties about the halo shape lead to errors of at least a factor of two \citep{weber_determination_2010}. Even larger uncertainties arise if the Milky Way has a dark matter disc \citep{lake_must_1989,read_thin_2008} as predicted by recent cosmological simulations. The second approach relies on fewer assumptions, and this is our focus in this paper. However, both approaches are complementary and, together, provide a powerful probe of Galactic structure. If $\rhodm < \rhodmext$, this suggests a prolate dark matter halo for the Milky Way; while $\rhodm > \rhodmext$ could imply either an oblate halo or a dark matter disc \citep{lake_must_1989,read_thin_2008,read_dark_2009}. 

The local dark matter density is needed for direct dark matter search experiments.  In the simplest case where
the dark matter is a Weakly Interacting Massive Particle, or WIMP \citep{jungman_supersymmetric_1996,baudis_dark_2006}, these experiments produce results that are degenerate between the WIMP interaction cross section and the local matter density \citep{gaitskell_direct_2004,aprile_xenon_2005,collaboration_search_2008}. Thus, extracting WIMP properties requires knowledge of $\rhodm$ \citep[e.g.][]{gaitskell_direct_2004}. 

To date, most limits on WIMP properties have assumed the `Standard Halo Model' (hereafter SHM) density: $\rhodm(R_\sun)=0.3$\,GeV\,cm$^{-3}$ ($\simeq 0.008$\msun\,pc$^{-3}$; \cite{jungman_supersymmetric_1996})\footnote{1\,GeV\,cm$^{-3}\simeq 0.0263158$\msun\,pc$^{-3}$.  The SHM is an isothermal sphere model for the Milky Way's dark matter halo with a value of the dark matter velocity dispersion assumed to be $\sigma_\mathrm{iso}\simeq 270$\,km\,s$^{-1}$.}. This is similar to the latest rotation curve extrapolated values that assume a spherical Milky Way halo. However, if the Milky Way halo is oblate, or there is a dark matter disc, then this could be a significant underestimate \citep[e.g.][]{weber_determination_2010}.

\begin{figure}
\center
\includegraphics[width=0.5\textwidth]{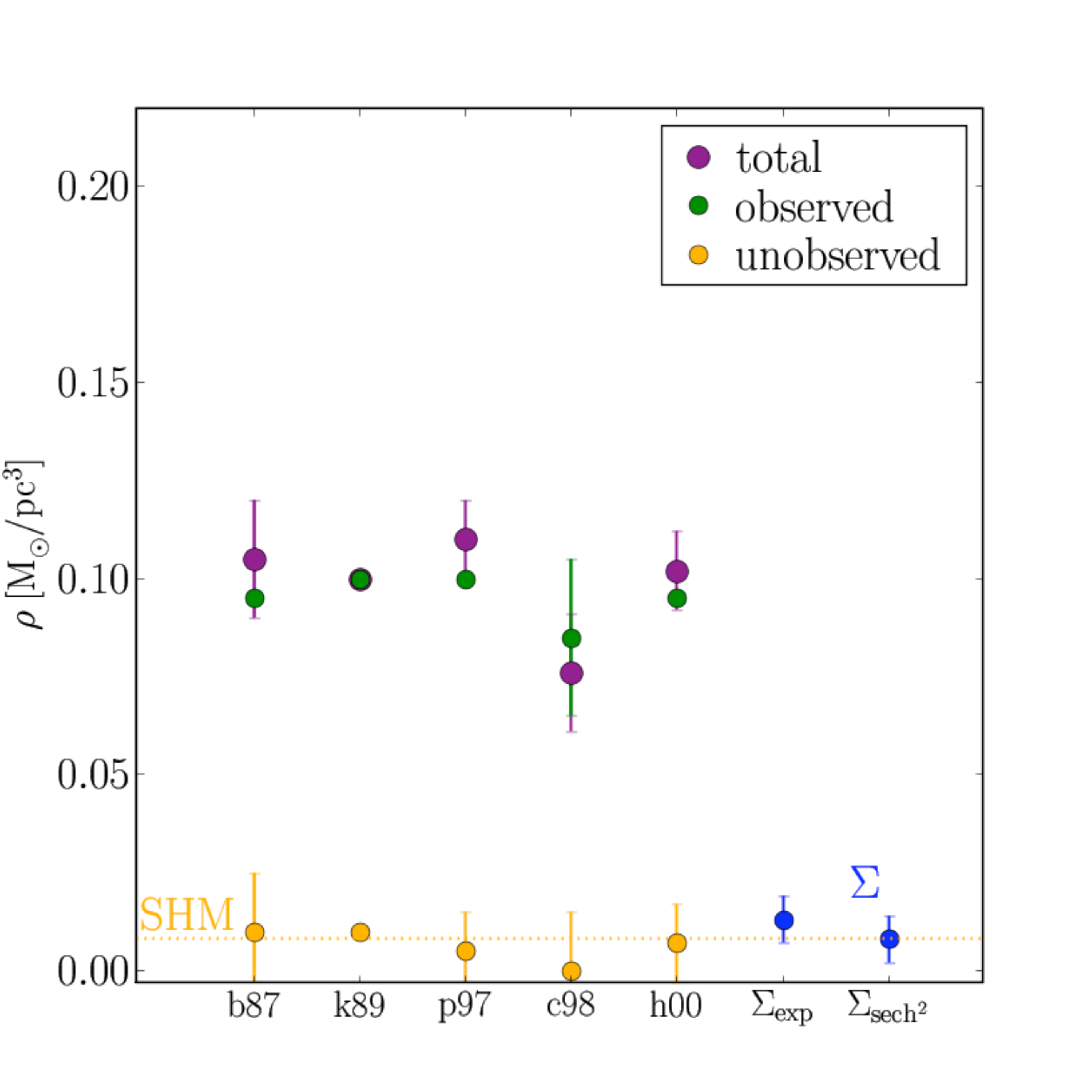}
\caption{A summary of recent determinations of total density $\rhotot$ (purple), dark matter density $\rhodm$ (yellow) and observed matter density (green) from the kinematics of Solar Neighbourhood stars in the literature. The yellow dotted line represents the dark matter density in the SHM. The blue points are the values of $\rhodm$ calculated from the local surface density (using an exponential and a $\mathrm{sech}^2$ profile for the disc; see footnote \ref{foot2}). Data are taken from:
b87: \protect\cite{bienayme_mass_1987}; k89: \protect\cite{kuijken_mass_1989-1}; p97: \protect\cite{pham_estimation_1997}; c98: \protect\cite{creze_distribution_1998}; h00: \protect\cite{holmberg_local_2000}.}
\label{fig:summary}
\end{figure}

Measuring the local matter and dark matter density from the kinematics of Solar Neighbourhood stars has a long history dating back to Oort \citep{oort_force_1932,oort_note_1960} who determined the total matter density $\rhotot(R_\sun)$ . Many studies since then have revisited the determination of both $\rhotot$ and $\rhodm$; we summarise recent results from the literature in Figure \ref{fig:summary}.

We can see from Figure \ref{fig:summary} that results have converged on no or very little disc dark matter\footnote{\label{foot2}We should be careful about what we mean by the terms  'local dark matter' and 'dark matter disc'.  In simulations,
the dark matter disk has a scaleheight of $\sim 1-2$\,kpc \citep{read_thin_2008}, but most importantly, it is just intermediate between the disc ($z_0\sim 250$\,pc) and the halo which has an effective scaleheight of $\sim R_\sun$. Here, we use `local dark matter' to mean dark matter within a local volume probed by the motions of stars in the solar neighborhood. Since this will only probe $\rhodm$  to  $|z|\sim 1$\,kpc, we can only separate a dark disk from a dark halo using another estimate of the dark matter halo's density. In the past, studies have talked about `disk dark matter' and meant dark matter with a scaleheight similar to the stellar disk.  Here, we would consider that to be just normalizing our stellar mass distribution rather than being a dark matter component.}.
In addition to the local volume density, several studies have measured the dynamical {\it surface density} of all gravitating matter -- $\Sigmatot$ -- rather than the volume density, typically probing up to heights of about $L \sim 1$\,kpc above the Galactic disc \citep[e.g.][]{kuijken_galactic_1991,holmberg_local_2004}. If we assume a constant dark matter density over this range, we can estimate the local volume density as $\rhodm = (\Sigmatot - \Sigmas)/L$. This gives\footnote{We derive the surface density of the visible matter at L as $\Sigma_{\mathrm{s},L}=\Sigma_{\mathrm{thin},L}+\Sigma_{\mathrm{thick},L}$, where $$\Sigma_{i,L}=2\int_0^{L}\rho_i(0)F(z) dz$$ with $i=$ thin, thick -- for the thin and the thick disc and $F(z)=\exp(-z/z_{0,i})$ or $F(z)=\mathrm{sech}^2(z/z_s)$ if we consider exponential or $\mathrm{sech}^2$ disc, respectively. The densities at the midplane $\rho_i(0)$ are taken from Table \ref{mmodel} and the exponential ($\mathrm{sech}^2$) disc scale heights $z_{0,i}$ ($z_{s,i}$) are calculated from the values in Table \ref{mw-tab}. The cited values of $\rhodm$ is obtained from a simple average of $\rhodm$ obtained using the dynamical $\Sigma_\mathrm{tot}$ from \cite{kuijken_galactic_1991} and \cite{holmberg_local_2004}.} $\rhodm = 0.013 \pm 0.006$\msun\,pc$^{-3}$ for an exponential and $\rhodm =  0.008 \pm 0.006$\msun\,pc$^{-3}$ for a $\mathrm{sech}^2$ disc profile, respectively.

The uncertainties on $\rhotot$ and $\rhodm$ quoted in Figure \ref{fig:summary} owe only to the sample size and observational errors.
With current/future surveys like GAIA \citep{jordan_gaia_2008,bailer-jones_what_2008}, RAVE \citep{steinmetz_rave:_2003,steinmetz_radial_2006,zwitter_radial_2008} and SEGUE \citep{yanny_segue:_2009} we expect a dramatic improvement in the number of precision astrometric, photometric and spectroscopic measurements. With this explosion in data, it is timely to revisit the systematic errors in determining $\rhodm$ from Solar Neighbourhood stars since these will  become the dominant source of error, if they are not already.  This is the goal of this paper.

Previous work in the literature has examined some of the possible systematics. \citet{statler_problems_1989} approximated the Galactic potential with a St\"ackel potential \citep{stackel_1895} and used the analytic third integral to treat cross terms in the Jeans equations. He applied this method to artificial data superficially resembling data available at the time, finding that systematic uncertainties were at least $30$ per cent, due mainly to sample size and uncertainties in the rotation curve. \cite{kuijken_mass_1989-2} reconsidered the determination of the volume density near the Sun with particular emphasis on possible systematic effects in the analyses of local F and K stars. They focused on the importance of modeling the velocity distribution of the stars near the plane (important for their method that assumes that the distribution function is separable; see Section \ref{theory}), and determining the density distribution as a function of height $z$ above the plane.

In this paper, we study systematic errors using high resolution N-body simulations. We first build an equilibrium N-body model approximating the Milky Way that satisfies all of the usual assumptions made in determining $\rhodm$ -- vertical isotropy in the velocity distribution, separability of the Galactic potential, constant local dark matter density and negligible radial gradient in the tilt of the velocity ellipsoid. We then evolve the disc over several dynamical times to form an inhomogeneous and complex disc structure that includes a strong bar and spiral waves similar to the Milky Way \citep{drimmel_3d_2001,dehnen_our_2002,binney_galactic_2008}.  This breaks many of the usual assumptions, providing a stringent test of different techniques.  We first use our simulation to test a standard method in the literature for recovering $\rhodm$. We then present and test a new method that: (i) relies only on a `minimal' set of assumptions; and (ii) that uses a Monte Carlo Markov Chain (MCMC) technique to marginalise over unknown parameters. The former makes the method -- given good enough data -- robust to model systematics. The latter allows us to cope with incomplete or noisy data and model degeneracies. Finally, we apply our new method to data from the literature to obtain a new measure of both $\rhotot$ and $\rhodm$. 

This paper is organised as follows. In Section \ref{theory}, we review the basic equations of the method and the assumptions used in past work. We present two methods that we test in detail: the `HF' method proposed by \cite{fuchs_kinematical_1993} and developed by \cite{holmberg_local_2000}; and a new more general method that assumes only equilibrium. In Sections \ref{testsim}, we describe the simulation that we use to test these two methods and we confront the different methods with our simulated Milky Way to assess the systematic uncertainties. In Section \ref{hf-data}, we apply our new method to data from the literature to determine more realistic errors on the local dark matter density. Finally in Section \ref{conclusion}, we present our conclusions.


\section{Determining the Local Matter Density}\label{theory}

Ideally, we should solve the Vlasov-Poisson equations to obtain the gravitational potential $\Phi$ from the distribution function of stars $f({\bf x},{\bf v})$:

\begin{equation}
\frac{\partial f}{\partial t} + \nabla_x f {\bf v} - \nabla_v f \nabla_x \Phi = 0 \label{eqn:vlasov}
\end{equation}
\begin{equation}
\nabla^2 \Phi = 4\pi G \rhotot \label{eqn:poisson}
\end{equation}
where $G$ is the gravitational constant; $\rhotot$ is the {\it total} matter density; and the density of tracers $\nu$ follows from the distribution function: $\nu = \int d^3 {\bf v} f({\bf x},{\bf v})$. If the system is in equilibrium, we may also assume that it is in a steady state such that $\frac{\partial f}{\partial t} = 0$. 

However, equations (\ref{eqn:vlasov}) and (\ref{eqn:poisson}) are difficult to solve in practice. The distribution function is six dimensional, requiring full phase space information. Worse still, we require its derivatives which amplifies any noise in the data (even a million stars will sample only ten points per phase space dimension). As a result, there have been two types of methods proposed in the literature: take velocity moments of the Vlasov equation and solve the resulting Jeans equations  \citep[e.g.][]{bahcall_distribution_1984, bahcall_k_1984, bahcall_self-consistent_1984}; or guess the form of the distribution function and ask if the data are consistent with this \citep{kuijken_mass_1989-1,kuijken_mass_1989-2,kuijken_mass_1989}. The first method has the advantage that we need not specify $f$, since we constrain it only through its moments. However, it throws away information about the shape of $f$. The latter method maximises the available information but comes at the price of potentially fatal systematic errors if an incorrect form for $f$ is assumed. Some mixed methods have also been proposed where the Jeans-Poisson system is solved, but the tracer density is closed by an integral over the measured (planar) distribution function \citep{fuchs_kinematical_1993,flynn_density_1994}. 

In this paper, we focus on the moment based methods that solve the Jeans-Poisson system of equations. This is because we want to make as few assumptions as possible to combat systematic errors. We do, however, also test the mixed method proposed by \cite{fuchs_kinematical_1993} and applied to {\it Hipparcos} data by \citet{creze_distribution_1998} and \cite{holmberg_local_2000}. This allows us to evaluate systematic errors introduced by assumptions about the form of $f$. 

In the following sections, we review methods for recovering $\rhos$ (the in-plane disc matter density) and $\rhodm$ from the simultaneous solution of the Jeans and Poisson equations. We present first a new method based on minimal assumptions -- our `MA' method. We then derive the method used in \cite{holmberg_local_2000} as a special case -- the `HF' method. We test both the MA and the HF methods on our Milky Way like simulation in Section \ref{testsim}. 

\subsection{The Minimal Assumption method (MA)}\label{fm-method}

The Jeans equations in cylindrical coordinates follow from velocity moments of the steady state Vlasov equation \citep[equation \ref{eqn:vlasov};][]{binney_galactic_2008}. Consider first just the $z$ Jeans equation:

\begin{equation}
\frac{1}{R}\frac{\partial}{\partial R}\left(R\nu_i \overline{v_{R,i} v_{z,i}}\right)
  + \frac{\partial}{\partial z}\left(\nu_i \vztwoi \right) + \nu_i
  \frac{\partial \Phi}{\partial z} = 0
\label{jeans}
\end{equation}
where $\nu_i$, $\vztwoi$ and $ \overline{v_{R,i} v_{z,i}}$ are the density and the velocity dispersion components of a {\it tracer} population $i$ moving in potential $\Phi$.

We now introduce our only assumptions:
\begin{enumerate}
\item The system is in equilibrium (steady state assumption).\label{eq-hyp}
\item The dark matter density is constant over the range of $|z|$ considered. \label{dm-hyp}
\item The `tilt' term: $\frac{1}{R}\frac{\partial}{\partial R}\left(R\nu_i \overline{v_{R,i} v_{z,i}}\right)$ is negligible compared to all other terms.  \label{tilt-hyp}
\end{enumerate}
The first assumption is necessary for any mass modelling method \citep[e.g.][]{sanchez_equilibrium_2011}. The second assumption requires that the disc scale height is much smaller than the dark matter halo scale length $z_d\ll r_h$, or for disc-like dark matter, that the scale height of dark disc is significantly larger than $z_d$. 

\cite{binney_galactic_2008} show that the `tilt' term is likely smaller than $(\overline{v_R^2}-\vztwo)(z/R)$
(see their discussion of the asymmetric drift in \S4.8.2a and \S4.9.3); so, assuming that $\overline{v_R^2}$ and $\vztwo$ both decline with $R$ as $\exp(-R/R_0)$ (applying also for our simulation, at least in the early stage, by construction), then the {\it tilt term} in equation \ref{jeans} is constrained by:
\beq
\left|\frac{1}{R} \frac{\partial(R\nu\overline{v_Rv_z})}{dR}\right| \simeq \frac{2\nu}{R_0}\overline{v_Rv_z}\lesssim \frac{2\nu z}{R_0}\frac{\overline{v_R^2}-\vztwo}{R_\sun}
\eeq
The second term in equation \ref{jeans} is of the order of $\nu\vztwo/z_0$ where $z_0\ll R_\sun$ and $z_0\ll R_0$ is the disc scale height. Hence the neglected term is smaller then the second term by at least a factor of $2zz_0/(R_0R_\sun)$. For these reasons we define these assumptions as a `minimal' set.

With the above assumptions, equation \ref{jeans} becomes a function only of $z$ and we can neglect the other two Jeans equations in $R$ and $\theta$. Our remaining Jeans equation becomes:
\beq
 \vztwoi \frac{\partial \nu_i}{\partial z} + 
 \nu_i \left(\frac{\partial \Phi}{\partial z} +\frac{\partial  \vztwoi}{\partial z} \right)= 0; 
 \label{jeans2}
\eeq
This is the Jeans equation for a one-dimensional slab. In principle, we should solve it for $R=\mathrm{constant}$. However, in practice we must average over some range $\Delta R$. We examine what is the maximum tolerable value of $\Delta R$ in Section \ref{box}.

For a given tracer population $i$, we can now write:
\beq
\frac{d\nu_i}{\nu_i}=-\frac{1}{\vztwoi}d(\vztwoi+\Phi)
\eeq
which can be solved straightforwardly:
\beq
\log\left(\frac{\nu_i}{\nu_i(0)}\right)=-\log\left(\frac{\vztwoi}{\vztwoi(0)}\right)-\int_0^z\frac{1}{\vztwoi}\frac{d\Phi}{dz}dz
\eeq
Thus, at each height above the disc $z_*$, the density of the tracer population $\nu_i(z_*)$ can be calculated:
\beq
\frac{\nu_i(z_*)}{\nu_i(0)}=\frac{\vztwoi(0)}{\vztwoi(z_*)}\exp\left(-\int_0^{z_*}\frac{1}{\vztwoi(z)}\frac{d\Phi}{dz}dz\right)
\label{new_nu}
\eeq
This general equation for $\nu_i(z)$ can be used to describe all the visible components of the disc. Given the density at the midplane $\nu_i(0)$ and the vertical velocity dispersion $\vztwoi(z)$ as a function of $z$ for each of the gas and stellar populations in the local disc, we can model the full disc density distribution as:
\beq
\rhos(z)=\sum_i m^*_i\nu_i(0)\frac{\vztwoi(0)}{\vztwoi(z)}\exp\left(-\int_0^{z}\frac{1}{\vztwoi}\frac{d\Phi}{dz}dz\right).
\label{new_rhodisc}
\eeq
where $m^*_i$ is the mass-to-light ratio for a given population $i$.
The Poisson equation then determines the potential $\Phi$ from the density. In cylindrical polar coordinates this is given by: 
\begin{eqnarray}
4\pi G \rho & = & \frac{\partial^2 \Phi}{\partial z^2} +
\frac{1}{R}\frac{\partial}{\partial R}\left(R\frac{\partial
    \Phi}{\partial R}\right) \nonumber \\
& = & \frac{\partial^2 \Phi}{\partial z^2} +
\frac{1}{R}\frac{\partial V_c^2(R)}{\partial R}
\end{eqnarray}
where $\rho$ is now the {\it total mass density} and $V_c(R)$ is the circular velocity at radius $R$. 

Splitting the matter density $\rho$ into disc contributions (gas+stars) that vary with $z$ ($\rhos(z)$\footnote{Note that we use throughout the notation $\rho_s = \rho_s(0)$ -- the in-plane baryonic mass density.}), and an effective dark matter contribution that includes the circular velocity term ($\rhoeff$), the Poisson equation becomes:
\begin{equation}
\frac{\partial^2\Phi}{\partial z^2}=4\pi G(\rhos(z)+\rhoeff)
\label{poisson}
\end{equation}
with:
\begin{equation}
\rhoeff=\rhodm(R)-(4\pi G R)^{-1}\frac{\partial}{\partial R}V^2_c(R)
\label{dmeff}
\end{equation}
where $\rhodm(R)$ is the halo mass density (following assumption \ref{dm-hyp}, this is assumed to be independent of $z$ in the volume considered); and $V_c(R)$ is the (total) circular velocity at a distance R (in the plane) from the centre of the Galaxy. For a flat rotation curve, the second term vanishes and $\rhoeff(R)=\rhodm(R)$. Note that there is an important difference between the vertical velocity dispersion of a tracer population, $\vztwoi(z)$ in equation \ref{new_nu}, and the same quantity as it appears in the mass model (equation \ref{new_rhodisc}). The former is something that we must measure for our chosen tracers, while the latter is simply a parameter that appears in our disc mass model. To put it another way, the tracers {\it must} satisfy equation \ref{new_nu}, but we could replace equation \ref{new_rhodisc} with some other mass model for the disc.

We may now solve equations \ref{new_rhodisc} and \ref{eqn:poisson} numerically for a given tracer population. We adopt the following procedure: first, we make initial trial guesses for $\rhos(0)$ (and any other unknowns in the star/gas disc), $\rhodm$, and the run of vertical velocity dispersion for the tracers $\vztwoi(z)$. Next, we solve equation \ref{new_rhodisc} to obtain $\Phi(z)$ and its first derivative $\frac{\partial\Phi}{\partial z}$, with $\Phi(0)=\left.\frac{\partial\Phi}{\partial z}\right|_{0}=0$. Then, we plug this result into equation \ref{new_nu} to obtain the vertical density fall-off the tracers $\nu_i(z)$. Finally, this is compared with the observed distribution to obtain a goodness of fit. In principle, each tracer population gives us an independent constraint on $\Phi(z)$.  A useful consistency check then follows since all tracers should yield the same potential, while combining different tracers gives smaller errors on the derived parameters. Note that the above procedure requires many input parameters that are typically poorly constrained, for example the normalisations and dispersions of each of the disc components and the vertical dispersion profile of the tracers. To efficiently explore this parameter space and marginalise over the uncertainties, we use a Monte Carlo Markov Chain (MCMC) method. This is described in Section \ref{sec:mcmcmethod}. 

Our Minimal Assumption (MA) method requires a measurement of $\vztwoi(z)$ for each tracer population considered. The HF method we derive next does not require $\vztwoi(z)$ -- using an additional assumption of separability instead. This has several advantages, but comes with a risk that this additional assumption will lead to systematic bias. We examine this in detail in Section \ref{testsim}.  

\subsection{The Holmberg and Flynn method (HF)}\label{hf_method}

The HF method \cite{fuchs_kinematical_1993, holmberg_local_2000} adds four additional assumptions: 
\begin{enumerate}
\item The potential is separable: $\Phi(R,z)=\Phi(R)+\Phi(z)$ 
\item The distribution function of tracers also separates.  At a fixed
cylindrical radius in the disc, it is a function only of the vertical energy: $f=f(E_z)$.
\item All disc components are isothermal.  
\item The rotation curve contribution to the Poisson equation -- $(4\pi G R)^{-1}\frac{\partial}{\partial R}V^2_c(R)$ -- is negligible. Thus $\rhodm = \rhoeff$ by construction. 
\end{enumerate}

The first two assumptions are critical for the method and also lie at the heart of the method proposed by \citet{kuijken_mass_1989}. Thus testing their validity applies to a wider range of past methods. Note that if these two assumptions are satisfied, then the `tilt' term in the Jeans equation is exactly zero, thus perfectly satisfying assumption \ref{tilt-hyp} of the MA method. However, the MA method makes the weaker assumption that the tilt term is {\it small} as compared to the other terms in the Jeans equations. Unlike the HF method, it requires no assumptions about the form of the potential or the distribution function. It is the latter that is the key difference between the two. If the motion is not separable, then the distribution function cannot be approximated by $f=f(E_z)$. As we will demonstrate in Section \ref{testsim}, this assumption leads to significant systematic errors even at $\sim 1.5$ disc scale heights above the plane. By contrast, assuming that the tilt term is simply small appears to be robust even up to several disc scale heights\footnote{Note that should the tilt term become large then in principle we could correct for it in the Jeans equation. This is perfectly possible in the MA method, but problematic for the HF method. In the HF method we would also have to correct for it in the distribution function. Such tilt corrections are, however, beyond the scope of this paper.}.

The HF method is a mixed method that uses the Jeans equations (as in the MA method), but assumes 
that each disc component is isothermal. This gives a Jeans equation as a function of $z$ similar to that in the MA method:
\begin{equation}
\vztwoi \frac{\partial\nu_i}{\partial z} + \nu_i \frac{\partial \Phi(z)}{\partial z} = 0\label{boltz}
\end{equation}
which is independent of $R$ and can then be straightforwardly solved to give:
\begin{equation}
\nu_i = \nu_{0,i} \exp\left(-\frac{\Phi(z)}{\vztwoi}\right)
\label{nuform}
\end{equation}
where $\nu_{0,i} = \nu_i(0)$. 

Thus, the density of the disc $\rhos$ can be written as a sum over isothermal components:
\begin{equation}
\rhos(z) = \sum_i m^*_i\nu_{i,0} \exp\left(-\frac{\Phi(z)}{\vztwoi}\right)
\label{rhodisc}
\end{equation}
where $m^*_i$ is the mass-to-light ratio for a given population $i$.
With the above decomposition, non-isothermality can still be modeled as a linear combinations of a larger number of isothermal distributions \citep{bahcall_distribution_1984}. However, this expansion is degenerate, and introduces many additional parameters that become expensive to explore \citep{kuijken_mass_1989}. 

Plugging equation \ref{rhodisc} into the Poisson equation \ref{poisson}, we can then calculate the gravitational potential, assuming a constant contribution for the dark matter density.

As in our MA method, the HF equations are closed by comparing the observed fall-off of the tracer population with the predicted one (given an initial guess of the disc model and dark matter density parameters). However, instead of using the solution to the moment equation \ref{new_nu} (or \ref{nuform}), they calculate the density fall-off of the tracers from the integral of the distribution function \citep{fuchs_kinematical_1993, flynn_density_1994}. Here they use the additional assumptions (reasonable close to the midplane) that the potential is separable: $\Phi(R,z)=\Phi(R)+\Phi(z)$ and that the distribution function of tracers is a function only of the vertical energy: $f=f(E_z)$. This has two key advantages. Firstly, it maximises the use of information in the data since it uses the {\it shape} of the distribution function, rather than just its lowest moments as in the MA method above. Secondly, one needs only measure $f$ at one height $z$ above the disc: $\vztwoi(z)$ is not required. We may understand this as follows. The density of the tracers is given by: 

\begin{equation}
\nu_i(z) = \int_{-\infty}^{\infty} dv_z f(E_z) = 2 \int_0^\infty
dv_z\left[f\left(\frac{1}{2}v_z^2+\Phi\right)\right]
\label{distfunc}
\end{equation}
And, since $f = f(E_z)$, we can rewrite equation \ref{distfunc} as an Abel integral:

\begin{equation}
\nu_i(z) = 2 \int_{\Phi(z)}^\infty d(\sqrt{2E_z})\sqrt{2E_z} \frac{f(\sqrt{2E_z})}{\sqrt{2\left(E_z-\Phi\right)}} .
\end{equation}

Then, substituting $|w_0| = \sqrt{2 E_z}$ and using $f(\sqrt{2E_z}) = f(w_0)$, we obtain: 
\begin{equation}
\nu_i(z)=2\int^{\infty}_{\sqrt{2\Phi}}\frac{f(w_0)w_0dw_0}{\sqrt{w_0^2-2\Phi}}
\label{den}
\end{equation}
where $w_0$ is the vertical velocity of stars in the midplane ($z=0$). Thus, we can measure $f(E_z)$ -- valid for all height about the disc $z$ -- from $f(|w_0|)$ measured only in the disc plane. 

Note that the above does not assume that the tracers are isothermal, though the mass model (equation \ref{rhodisc}) does. This will become inconsistent if the tracers comprise most of the mass of the disc. In practice, this is unlikely to be the case. However, the inconsistency can always be avoided by using the more general mass model derived in the MA method, while still closing the equations using equation \ref{den}. We test the effect of this inconsistency in Section \ref{testsim}. 

We stress that the assumption of $f=f(E_z)$ is likely to be valid close to the disc plane. Thus, the HF method as employed in \cite{holmberg_local_2000} -- where they probe only up to $\sim 1$ half mass scale height above the disc -- is unlikely to be biased. However, as we probe to heights greater than the disc scale height, systematics will creep in. Furthermore, probing to such heights -- as we shall show -- is necessary for breaking a degeneracy between $\rhodm$ and $\rhos$. We explore the effect of the $f=f(E_z)$ assumption in Section \ref{testsim}. 

\subsection{Determining $\rhodm$ and $\rho_s$ with an MCMC}\label{sec:mcmcmethod}
In summary, while the MA and HF methods differ in their underlying assumptions, the basic strategy for recovering the local matter density is the same: 
\begin{enumerate}
\item Build a mass model for the local mass distribution consisting of components $\nu_i$, defined by equation \ref{new_nu} or \ref{nuform}, for gas and stellar populations, and a constant contribution for dark matter $\rhodm$.
\item Use this mass model to integrate the Poisson (\ref{poisson}) and the Jeans equation (\ref{jeans2} or \ref{boltz}) simultaneously  to compute the local potential $\Phi$ (and its $z$-derivative).
\item Use the calculated potential $\Phi$ and the measured kinematics of the tracers to compute their density fall-off $\nu(z)$ (using equation \ref{new_nu} or \ref{den}). To predict the density fall-off of the tracers the HF method needs the measure of their vertical velocity distribution function in the mid-plane $f(w_0)$, while in the case of the MA method the vertical velocity dispersion as a function of $z$ - $\vztwoi(z)$ - is required.
\item Compare the predicted density profile(s) $\nu(z)$ with the observed one(s) $\nu_{\mathrm{obs}}(z)$ to reject or accept the model input parameters: $\rhodm$ and parameters governing each of the components $\nu_i$.
\end{enumerate}

In practical applications, the above implies many (degenerate) free parameters if the disc model has many non-isothermal components with parameters that are poorly known, while $\vztwoi(z)$ for the tracers may also be poorly constrained. A Monte Carlo Markov Chain (MCMC) provides an efficient way to rapidly explore this parameter space. It naturally deals with parameter degeneracies: all of the unknown parameters are `marginalised out' to leave the key parameters of interest in (the total matter density $\rhotot$ and the dark matter density $\rhodm$). In this way, the MCMC addresses some of the issues raised by \cite{kuijken_mass_1989-1,kuijken_mass_1989-2,kuijken_mass_1989} about degeneracies between parameters in very complex models making such models unworkable. 

\begin{figure*}
\center
\includegraphics[height=.4\textwidth]{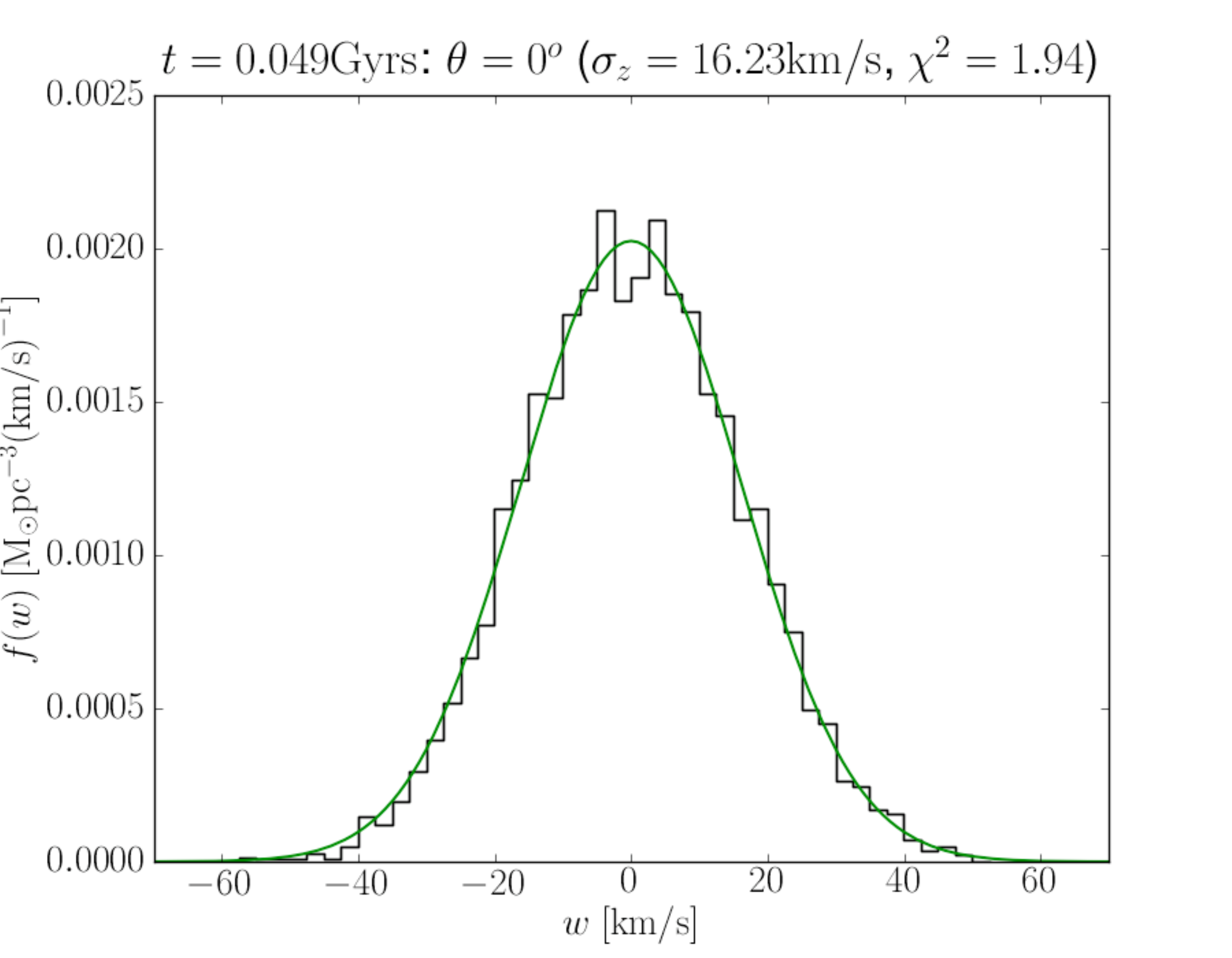}
\includegraphics[height=.4\textwidth]{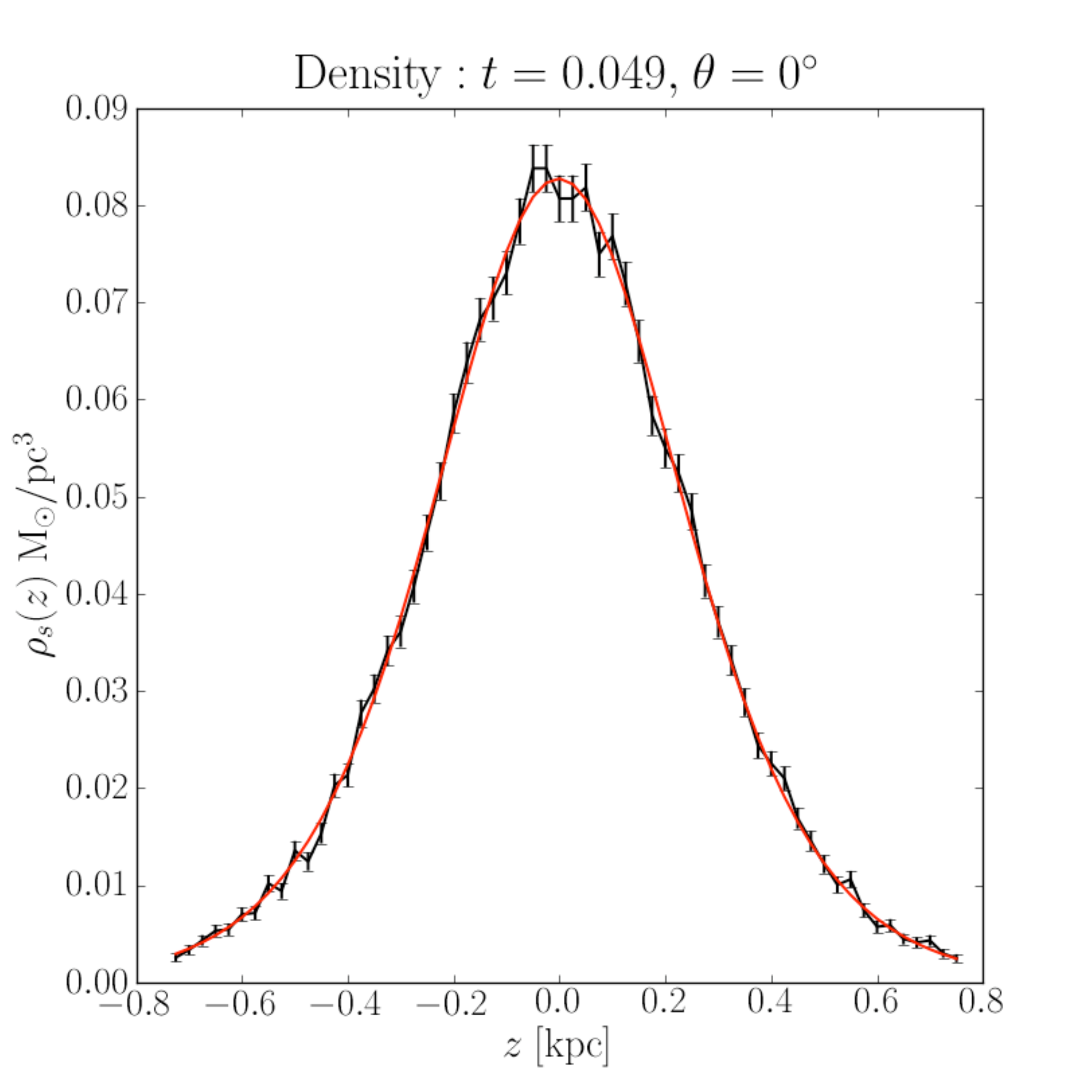}
\caption{Left panel: The vertical velocity distribution of star particles (in black) for one of the ``wedge'' patches at $R_\sun=8.5$\,kpc from the center of the galaxy. The green curve corresponds to the Gaussian fit. Right panel: The vertical density distribution of star particles (in black) for one of the ``wedge'' patches at $R_\sun=8.5$\,kpc from the center of the galaxy. The red curve corresponds to the prediction of the best-fitting mass model. The error bars represent Poisson noise.}
\label{den-fall-off}
\end{figure*}

We use a MCMC method based on a Metropolis algorithm \citep[e.g.][]{saha_principles_2003} to recover the local density. For the simulation data, we use the dark matter density (namely $\rhodm$ in equation \ref{dmeff}, adding the rotation curve term calculated for each volume) and the visible matter density $\rhos$ (which correspond to $\nu_{i,0}=m^*_i\nu_i(0)$ in equation \ref{new_rhodisc} or \ref{rhodisc}), as our input parameters. When we apply the HF method, we fit the distribution function at the midplane with a Gaussian (double Gaussian) for the unevolved (evolved) simulation. These fits are good for most of the volumes considered (an example is shown in the left panel of Figure \ref{den-fall-off}). When we adopt the MA method, we linearly interpolate the velocity dispersion of the tracers above the plane $\vztwoi(z)$, since this method is extremely sensitive to the velocity dispersion function adopted.

When we apply the two methods to the real data (see Section \ref{hf-data}) the situation is more complex. Firstly, we must fit a larger number of parameters: namely the local dark matter density $\rhodm$, the total visible density $\rhos$, the fraction of the different disc components ($\nu_{i,0}$), and their velocity dispersions in (and even potentially above) the plane ($\vztwoi$). Secondly, the data are magnitude rather than volume limited. We take this into account by drawing the observed stellar distribution from the model density fall-off using the observed luminosity function. The MCMC allows us to easily implement both these additional parameters and the sampling of the luminosity function. In addition, it is straightforward to model different tracer populations simultaneously, and apply constraints on the local {\it surface density} of the disc. Our full procedure is described in more detail in Section \ref{hf-data}.   Finally, with real data we cannot simply interpolate the velocity dispersion as a function of $z$, but we must consider the uncertainties on the velocities. Such uncertainties can be straightforwardly added to the MCMC and marginalised out (see Appendix \ref{app:cyl}).

We apply the MA and HF methods to our simulated Milky Ways in Section \ref{testsim}. We then apply the MA method to real data in Section \ref{hf-data}. For the simulation, we calculate the potential by modeling the visible matter in the disc as a single population. To simplify the calculation we introduce some dimensionless parameters described in Appendix \ref{adim} \citep{bahcall_distribution_1984,bahcall_k_1984,bahcall_self-consistent_1984}. This transforms equations \ref{new_nu}, \ref{nuform} and \ref{poisson} to \ref{nu-adim}, \ref{nu-adim2} and \ref{fin-adim}.

\section{Testing the methods}\label{testsim}
To test the MA and HF methods in Section \ref{theory} and evaluate the systematic errors, we apply both to a high resolution collisionless simulation of a Milky Way like galaxy.

We consider two different stages of the simulation: an unevolved one with an axisymmetric disc (shown in the left panel of Figure \ref{sim10-820}) and fulfilling all the hypotheses of the more restrictive HF method; and a more evolved stage (represented in Figure \ref{sim10-820}, right panel) presenting a bar, similar to the real Milky Way, that breaks many of the assumptions. The results for these two different stages of the simulation are described in Sections \ref{results} and \ref{inhom}, respectively.

\subsection{The simulation}\label{sim-sect}
We ran a simulation of a Milky Way like galaxy with the parallel tree code PkdGRAV \citep{stadel_cosmological_2001}, using the galaxy models of \cite{widrow_equilibrium_2005} for the initial conditions. These models are derived from a composite three-integral distribution function $f=f_\mathrm{disc}(E,E_z,L_z)+f_\mathrm{halo}(E)+f_\mathrm{bulge}(E)$ and provide near-equilibrium initial conditions.

The disc model has an exponential radial profile and a $\mathrm{sech}^2(z/z_\mathrm{s})$ vertical profile. Its distribution function applies in the epicyclic approximation with $\sigma_{R,\phi,z}\ll V_c$, so the vertical energy is an approximate integral
of motion: this leads to triaxial velocity ellipsoids in the disc models as seen in real spiral galaxies \citep{widrow_dynamical_2008}. The halo is modeled as a NFW profile. However, when its distribution function is combined with the disc one, the net halo density profile is slightly flattened along the $z$-axis near the centre, but preserving the $r^{-1}$ central cusp.

To have statistics comparable with the present data in the Solar Neighbourhood (e.g. \cite{holmberg_local_2000} considered $\sim2000$ A-stars in a cylindrical volume of radius $R=200$pc and height $|z|<200$pc centered on the Sun), we constructed a disc with $n_\mathrm{d}=30\times 10^6$ star particles. We chose the masses of the dark matter halo particles and the (star) bulge particles so that the heating time-scale for the disc is much larger than both the internal relaxation time-scale, and the time of the simulation ($\sim4$\,Gyr): $t_\mathrm{heat}\gg t_\mathrm{rel}\gg t_\mathrm{sim}$, where $ t_\mathrm{rel}$ is given by \citep{binney_galactic_2008}:
\begin{equation}
t_\mathrm{rel}=n_\mathrm{rel}t_\mathrm{cross}=\frac{n}{8\log\Lambda}\frac{b_\mathrm{max}}{v_\mathrm{typ}}
\end{equation}
where $v_\mathrm{typ}=\sqrt{GM/R_{\odot}}$ is the typical velocity at the Solar position $R_{\odot} = 8.5$\,kpc; $b_\mathrm{min}=2Gm_\mathrm{part}/v^2_\mathrm{typ}$, $b_\mathrm{max}=R_{\odot}$; and the Coulomb logarithm is $\log\Lambda=\log(b_\mathrm{max}/b_\mathrm{min})$. Given $n_\mathrm{d}=30\times 10^6$ total stars, the number enclosed within $R_\odot$ is $n=n_\mathrm{d}(R_\odot)\sim25\times 10^6$. Using this latter number, we find $t_\mathrm{rel}\simeq 1.17\times 10^{4}$\,Gyr. 

The heating time $t_\mathrm{heat}$ is given by \citep{lacey_massive_1985}:
\begin{equation}
t_\mathrm{heat}=\frac{\sigma_z^2V_\mathrm{h}}{8\pi G^2M_\mathrm{h}\rho_\mathrm{h}\log\Lambda_\mathrm{h}}
\end{equation}
where $\sigma_z$ is the vertical velocity dispersion of the disc particles; $M_\mathrm{h}$ the mass of the dark matter particles; $V_\mathrm{h}$ their typical velocity; $\rho_\mathrm{h}$ and $\log \Lambda_\mathrm{h}$ are the density and the Coulomb logarithm for the halo (a similar calculation can be done for the bulge particles). 

Using $t_\mathrm{heat} = k t_\mathrm{rel}$, with $k\sim10$, we find the following satisfy the above timescale constraints: $n_\mathrm{h}=15\times 10^6$ and $n_\mathrm{b}=0.5\times 10^6$ particles for the halo and the bulge respectively. 

The main features of the model we used are listed in Table \ref{sim-tab}. For comparison, some of the corresponding features of the real Milky Way are given in the Table \ref{mw-tab}. 
\begin{table}
\center
\caption{Parameters for the disc, dark matter halo and stellar bulge for the initial conditions of the simulation. From left to right columns show: the number of particles ($N$); the total mass ($M$); the softening length ($\varepsilon$); the half mass scale-length ($R_{1/2}$); and the half-mass scale height ($z_{1/2}$).}
\label{sim-tab}
\begin{tabular}{|c|c|c|c|c|c|}
\hline
&N & $M$& $\varepsilon$& $R_{1/2}$& $z_{1/2}$\\
& [$10^6$]& [$10^{10}$M$_\sun$] & [kpc] &  [kpc]  & [kpc]\\
\hline
Disc  & 30  & 5.30 & 0.015 & 4.99 & 0.17 \\
Bulge&0.5 & 0.83 & 0.012 & - & -\\
Halo &15 & 45.40 & 0.045 & - & - \\
\hline
\end{tabular}
\end{table}

\begin{table}
\center
\caption{The distinct components of the Milky Way. From left to right the columns show: the total mass ($M$); the half mass scale-length ($R_{1/2}$); and the half mass scale height ($z_{1/2}$). These values are compiled using the following relations: $z_{1/2}=0.55z_s=0.7z_0$ and $R_{1/2}=1.68 R_0$ (Read et al. 2008), where $z_s$ is the $\mathrm{sech}^2$ disc scale height, $z_0$ is the exponential disc scale height and $R_0$ is the exponential disc scale length.}  
\label{mw-tab}
\begin{tabular}{|c|c|c|c|c|}
\hline
&$M$  & $R_{1/2}$ & $z_{1/2}$ & Ref. \\
& [$10^{10}$M$_\sun$] & [kpc] & [kpc]&\\
\hline
Thin disc & $3.5-5.5^*$ & $3.35 - 9.24$ & $\sim 0.14-0.18$ & fl,o,fe,k \\
Thick disc & - & $5.04 - 7.56$ &  $0.49- 0.84$& o,n,s\\
Bulge & $\sim 1$ & - & - & d,fl \\
Halo & $\sim 40- 200$ & - & - & x,g\\
\hline
\end{tabular}
\small{References: fl=\protect\cite{flynn_mass--light_2006}; o=\protect\cite{ojha_radial_2001}; fe=\protect\cite{feast_local_2000}; k=\protect\cite{kuijken_mass_1989-2}; n=\protect\cite{ng_probing_1997}; s=\protect\cite{spagna_galactic_1996}; d=\protect\cite{dehnen_mass_1998}; x=\protect\cite{xue_milky_2008}; g=\protect\cite{guo_how_2009}.\\
 $^*$ total disc mass}
\end{table}

\begin{center}
\begin{figure*}
\includegraphics[width=0.45\textwidth]{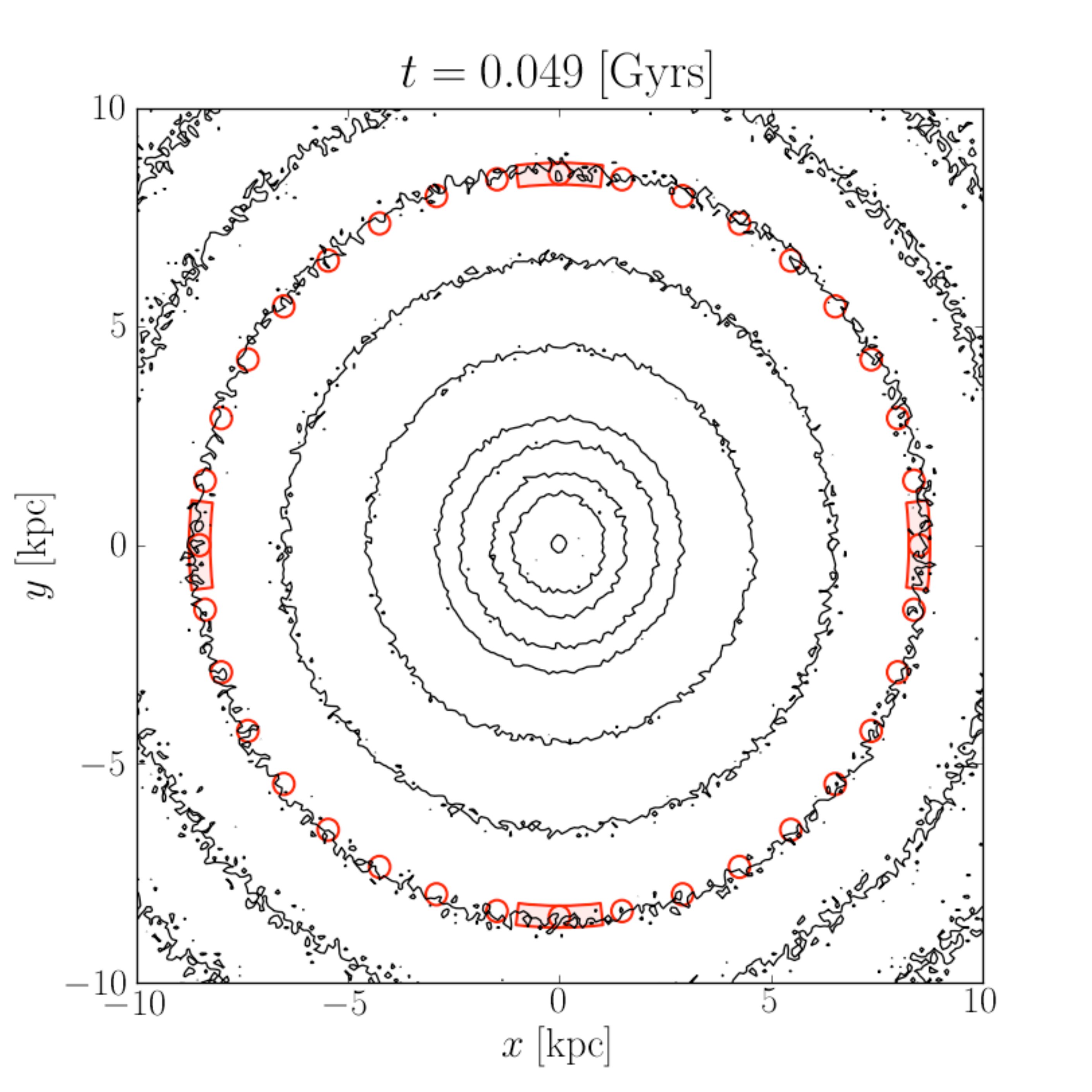}
\includegraphics[width=0.45\textwidth]{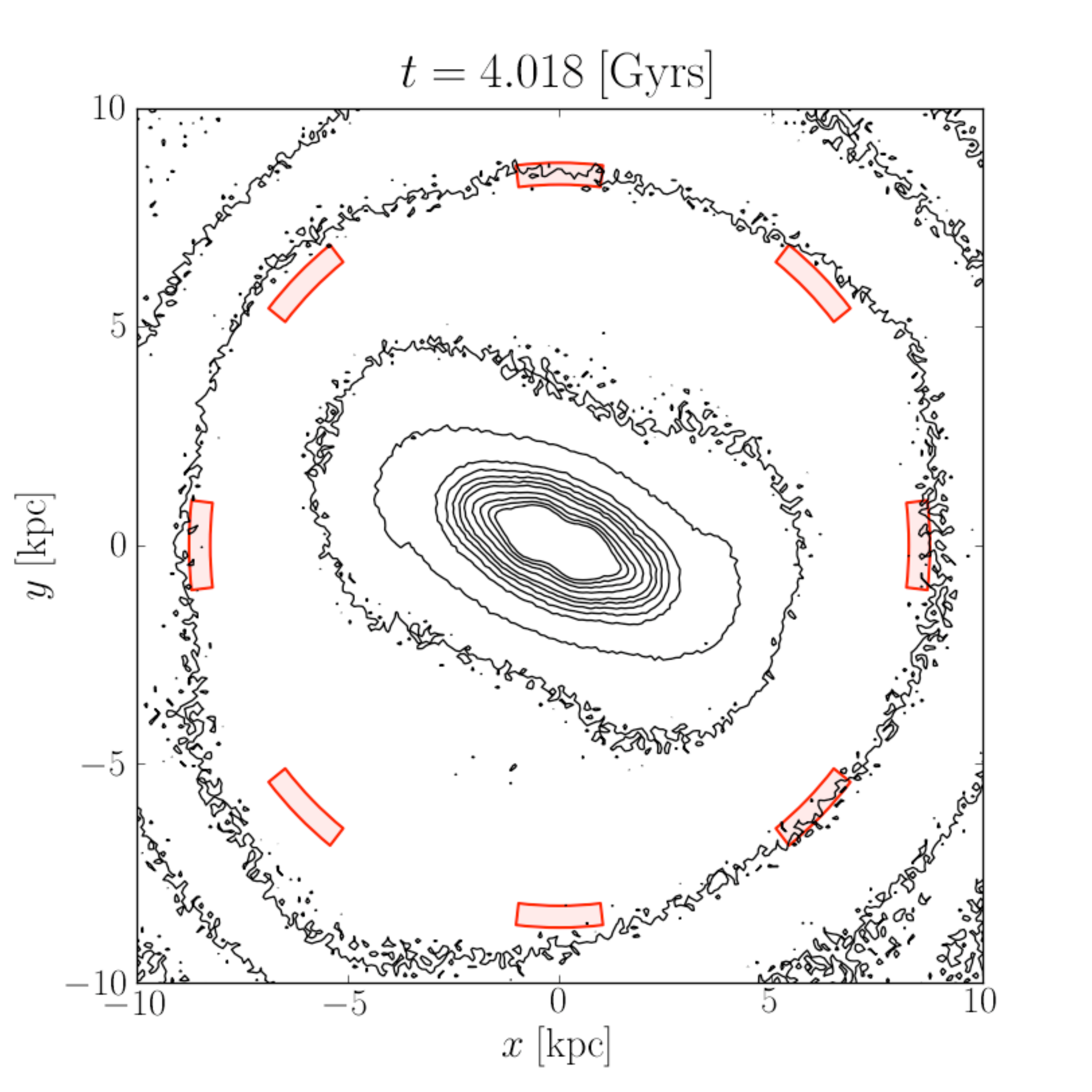}
\caption{Density contours viewed from top for the disc star particles. Left panel: an early time-step ($t\sim 0.05$\,Gyr) presenting an axisymmetric disc. Right panel: the evolved  simulation ($t\sim 4$\,Gyr), presenting a bar and spiral arms with inter-arm contrast $\rho_\mathrm{arm}/\rho_\mathrm{dip}\simeq 0.15$. The red circles correspond to the position of the cylindrical ``Solar Neighbourhood" patches, at a distance of $R_\sun=8.5$\,kpc from the Galactic Centre. The red shaded wedges represent the other volumes we used to compare the results of the analysis of the two stages of the simulation. We adopt patches of this shape to obtain better sampling. The angular position of the patches is calculated from $x,y=[R_\sun,0]$ anti-clockwise.}
\label{sim10-820}
\end{figure*}
\end{center}

In our analysis, we consider two different outputs of the simulation: an unevolved stage ($t\sim 50$Myrs) in which the disc is still axisymmetric, and an evolved one ($t\sim4$\,Gyr) which presents a bar similar to the real Milky Way. These two stages are shown in Figure \ref{sim10-820} (left and right panels, respectively). The unevolved disc is used to test the method in general, and to study what data are needed to recover the right value of the local density in the ideal case of data fulfilling all the assumptions. The evolved stage represents a more realistic situation and is used to test the effect of realistic disc inhomogeneities on the determination of the local density. The spiral arms -- that are the major driver of inhomogeneities at the Solar neighbourhood in the evolved disc -- are compatible with the Milky Way: our Galaxy has an inter-arm ratio of the spiral structure at the solar radius $R_\sun$ of $K\sim1.7$ \citep{drimmel_3d_2001}; the corresponding value for the simulation is $K\sim1.5$.

In the analysis of the simulation, we set the Solar Neighbourhood position at a Galactocentric distance of $R_\sun=8.5$\,kpc, in agreement with the IAU (International Astronomy Union) recommended value. We consider several small volumes at different angular position around the disc, represented as red circles and wedges in Figure \ref{sim10-820} (and see Section \ref{fitting}). For the unevolved (axisymmetric) disc, these different patches test the effect of sampling error on our derived $\rhodm$ and $\rho_s$; for the evolved disc, they examine the effect of disc inhomogeneities.

\subsection{How well does the simulation satisfy our assumptions} \label{test-hyp}
Both the MA and HF methods are based on several key assumptions, as outlined in Section \ref{fm-method} and Section \ref{hf_method}. To understand how well both methods can recover the local dark matter density, we first evaluate how well the two stages of the simulation fulfil these assumptions.

\subsubsection{Constant $\rhodm$ in the local volume}\label{dmc-sect}
\begin{center}
\begin{figure*}
\includegraphics[width=0.45\textwidth]{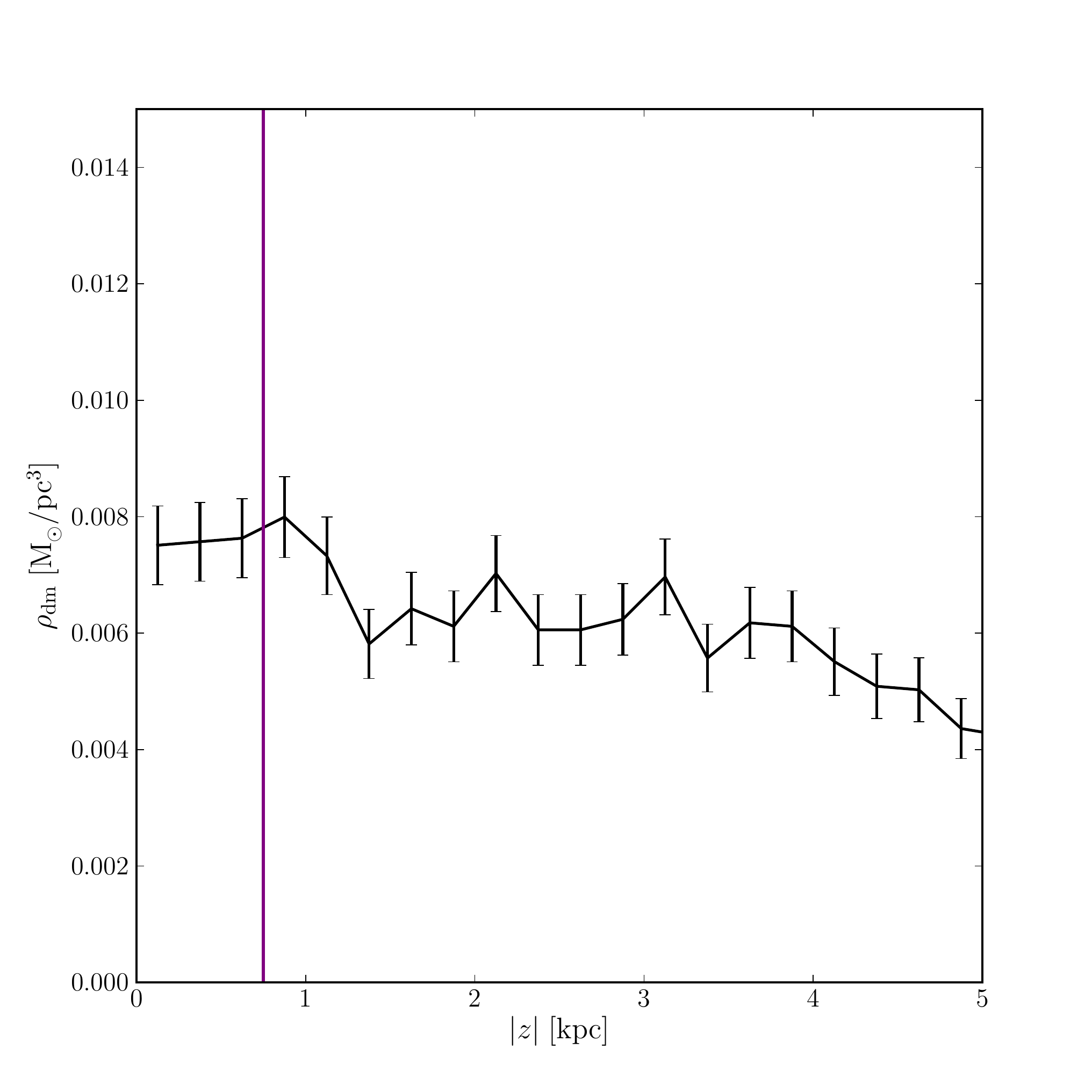}
\includegraphics[width=0.45\textwidth]{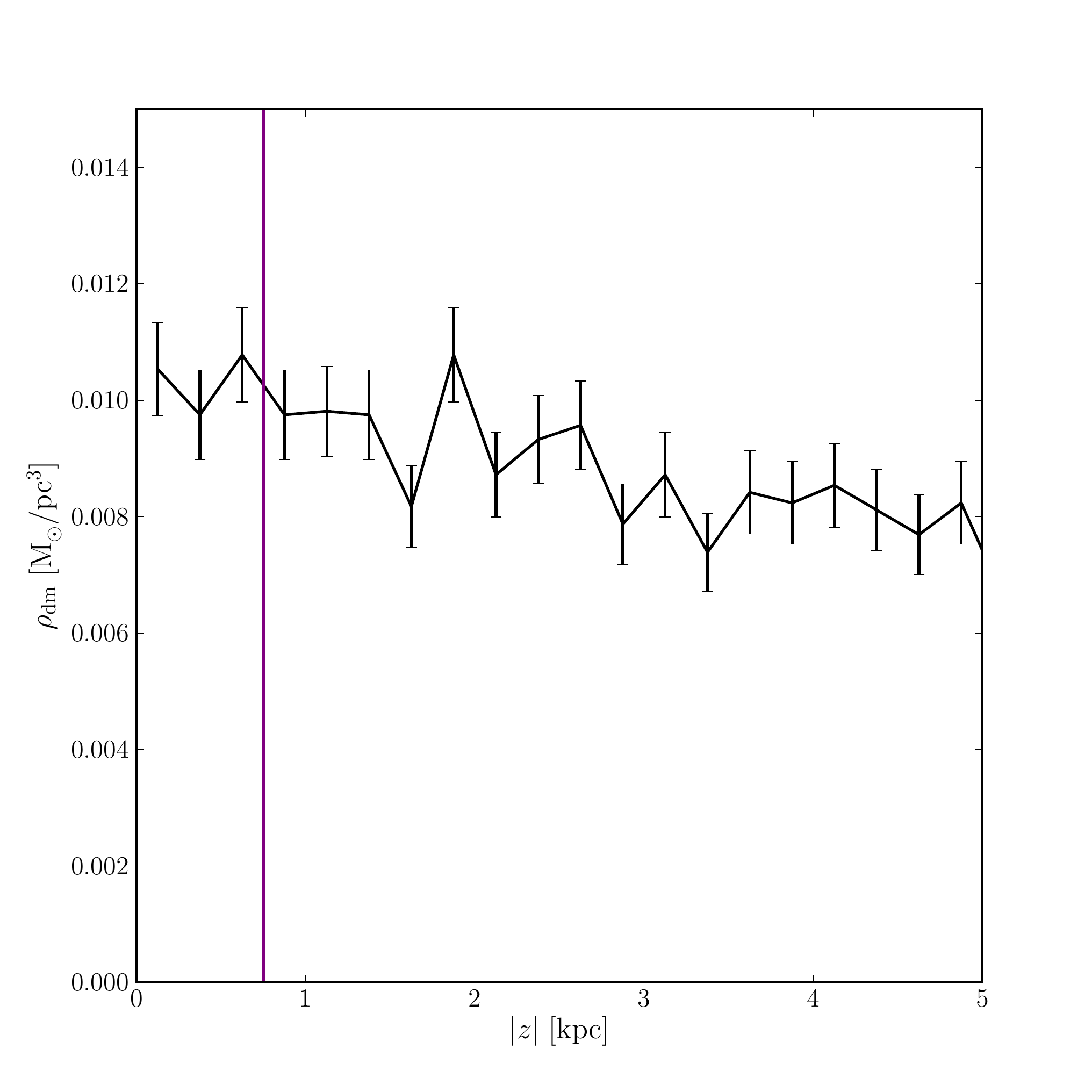}
\caption{The dark matter density as a function of $|z|$ for the the unevolved (left panel) and the evolved simulation (right panel). The purple line represents $z=0.75$\,kpc, i.e. the maximum height considered in our analysis. The errorbars correspond to the Poisson errors. The dark matter density is noisy owing to the large mass of the dark matter particles, but it is constant within the uncertainties for $|z|<0.75$\,kpc.}
\label{fig:dmc}
\end{figure*}
\end{center}

The hypothesis \ref{dm-hyp} of the MA method is well fulfilled as shown in Figure \ref{fig:dmc}, where we plot the dark matter density as a function of $z$ for the unevolved (left) and the evolved (right) simulation. The purple line represents $|z|=0.75$\,kpc, i.e. the maximum height considered in our analysis.

\subsubsection{Isothermality, tilt and equilibrium}\label{sec:testiso}

\begin{figure*}
\center
\includegraphics[width=\textwidth]{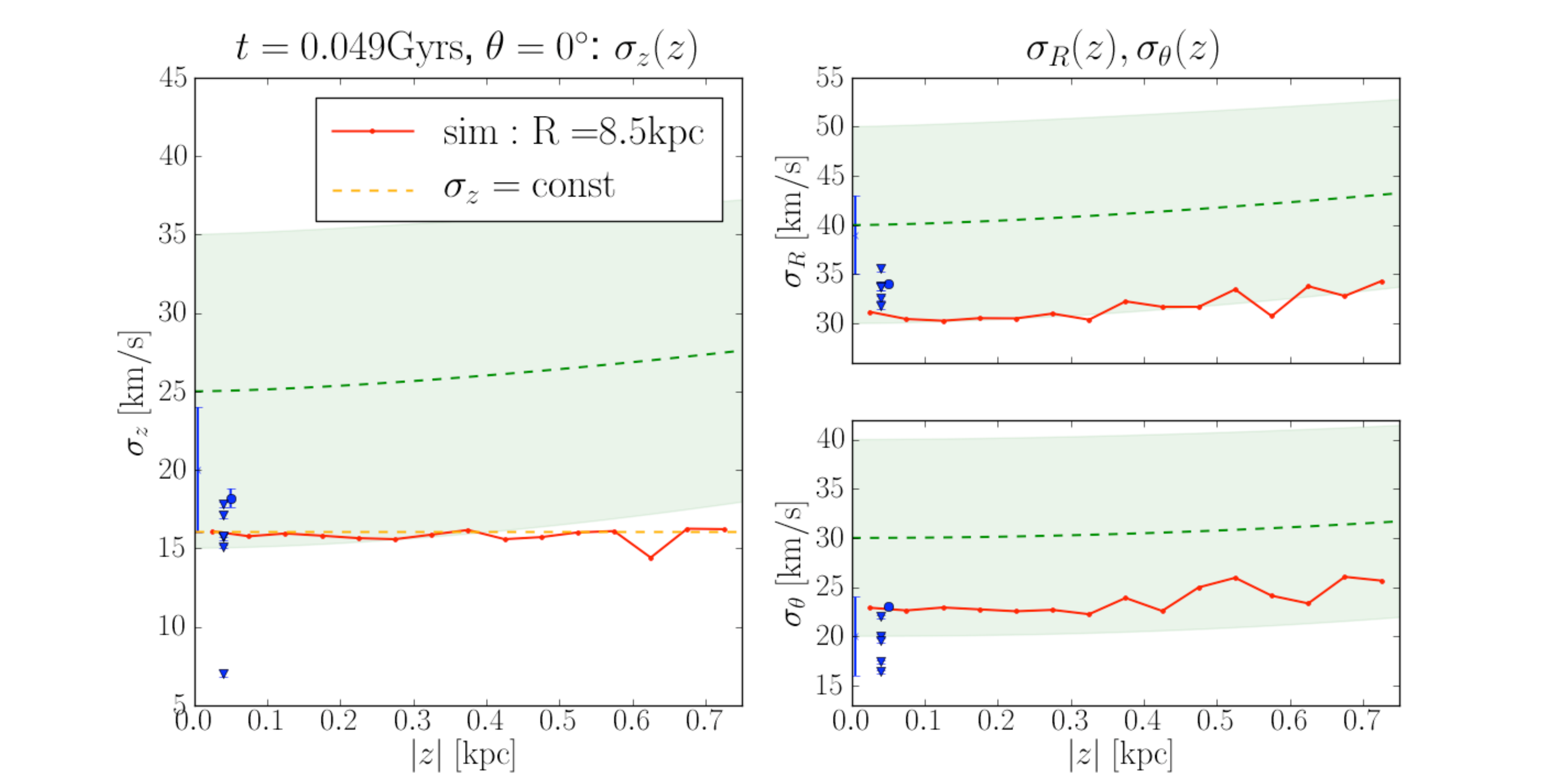}
\includegraphics[width=\textwidth]{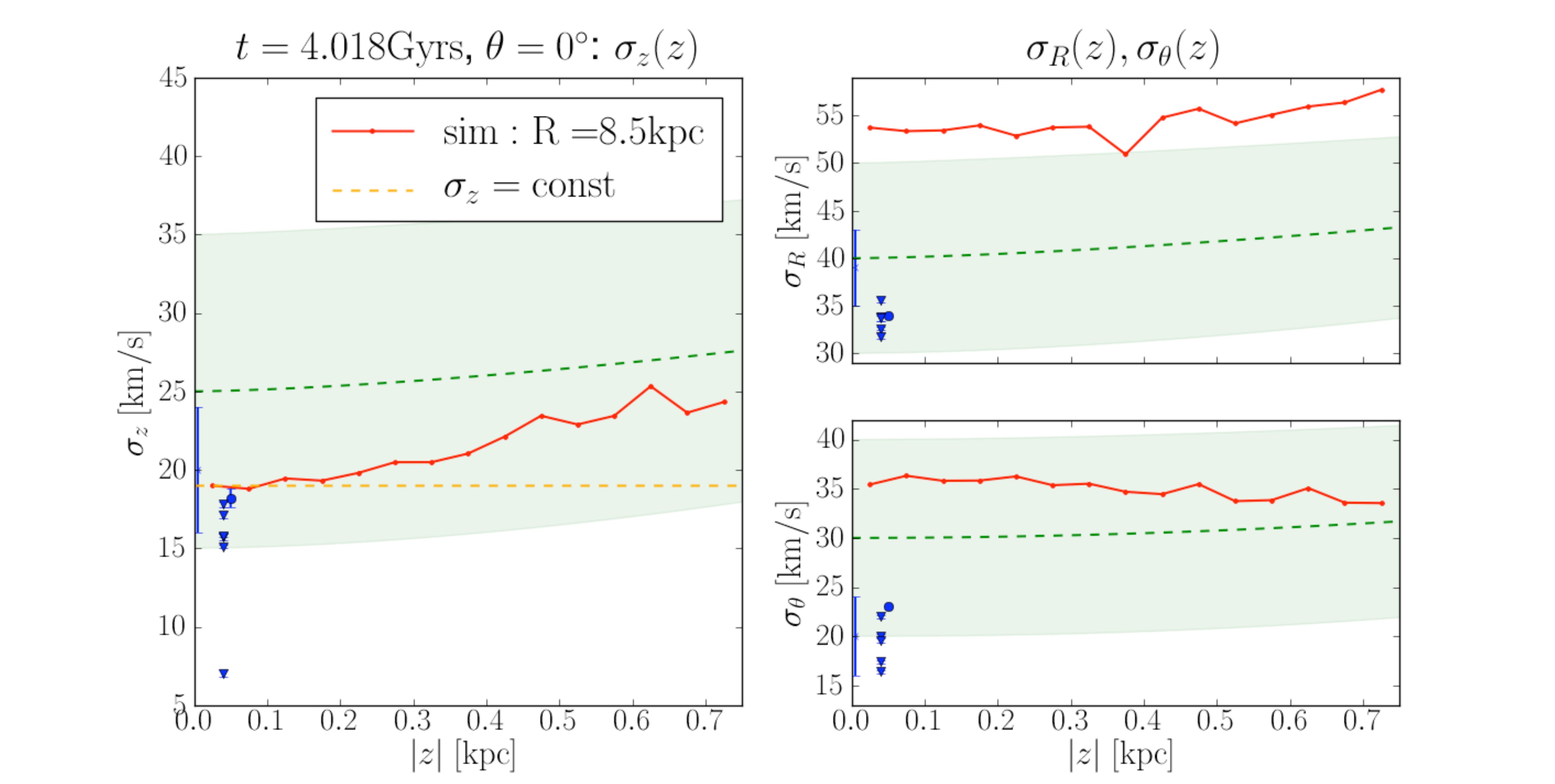}
\caption{Velocity dispersion gradients with $z$. Upper panel: unevolved simulation $t=0.049$\,Gyr. Lower panel: evolved simulation $t=4.018$\,Gyr. The dashed green line represent the best fit of the velocity dispersion by \citep{bond_milky_2009}, while the green shaded region shows the errors in the fitted parameters. The blue data points give the values of $\vztwo(z)$, $\vthetatwo(z)$ and $\vRtwo(z)$ taken from the literature \citep{holmberg_local_2004,seabroke_revisiting_2007}. The red points represent the values for our simulation at $R=8.5$\,kpc. The yellow and red dot-dashed lines in the $\vztwo(z)$ plot are lines of constant $\vztwo(z)$.} 
\label{isotherm}
\end{figure*}

The velocity dispersion $\vztwo$ as a function of $z$ should be constant, by definition, for an isothermal population. In the two left panels of Figure \ref{isotherm} the velocity dispersion $\vztwo(z)$ is represented for the two output times of the simulation considered ($t=0.049$\,Gyr in the upper panel and $t=4.018$\,Gyr in the lower one) at $R=8.5$\,kpc (in red). For comparison, the observational data for the Milky Way (blue data points), and the best fit $\vztwo(z)$ function determined by \cite{bond_milky_2009} (green dashed line: the light green shaded region represent the errors in the fit parameter) are shown. \cite{bond_milky_2009}'s fit are obtained from a sample of 53000 blue ($0.2 < g-r < 0.6$) disc stars from SDSS with radial velocity measurements, $b > 20^\circ$ and high metallicity ([Fe/H]$>-0.9$), up to $|z|<5$\,kpc. These stars are taken at high $z$ over the plane and are much hotter than the stars used in literature (A,F and K stars) to trace the local gravitational potential (blue dots). However, the fit does gives us information about the potential non-isothermality of the disc. 
The dashed yellow line is the isothermal line for 8.5\,kpc. These plots refer to a particular angular position in the disc ($\theta=0^\circ$), but the situation for $\vztwo$ is similar for the whole disc. 

The visible population in the disc for the unevolved stage ($t=0.049$\,Gyr) of the simulation is almost perfectly isothermal, while a significant deviation from isothermality is seen for the more evolved stage ($t=4.018$\,Gyr). 

When the disc species are not isothermal, the second term of the Jeans equation \ref{jeans} cannot be approximated as $ \vztwoi\partial \nu_i/\partial z$, but we must consider also the contribution of $z$-derivative of $\vztwoi(z)$. 

To quantify the effect of non-isothermality, we look at the the second and the third terms of the Jeans equation \ref{jeans} calculated for the two stages of our simulation. We compute these terms using a Smoothed Particle Hydrodynamics (SPH)-like method to determine smoothed quantities and gradients at the particle positions (for more details see Appendix \ref{SPH}).

In Figure \ref{jeans_terms}, the SPH calculated quantities are plotted for the two stages of the simulation considered $t=0.049$\,Gyr (left panel) and $t=4.018$\,Gyr (right panel) for $\theta=0^\circ$. The red line represents the potential term. The solid black and the dashed grey lines represent the sum of the two last terms of the Jeans equation in the non-isothermal ($r_\mathrm{NI}$) and in the isothermal ($r_\mathrm{I}$) case respectively, namely:
\beq
r_\mathrm{NI}=\nu_i\frac{\partial \Phi}{\partial z}+\frac{\partial (\vztwoi\nu_i)}{\partial z} \quad  \mathrm{(non-isothermal)}
\label{rni}
\eeq
 and 
\beq
r_\mathrm{I}=\nu_i\frac{\partial \Phi}{\partial z}+\vztwoi\frac{\partial \nu_i}{\partial z}
\quad \mathrm{(isothermal).}
\eeq
We see in this figure that, for $t=0.049$\,Gyr, the second term calculated as isothermal ($\vztwoi \partial \nu_i/\partial z$, dashed cyan line) and including non-isothermality ($\partial (\nu_i \vztwoi)/\partial z$ solid blue line) overlap almost perfectly, and that $r_\mathrm{NI}$ (black continuous line) and $r_\mathrm{I}$ (grey dashed line) are also very similar and close to zero. This is not surprising since the velocity dispersion $\vztwoi$ is almost constant with $z$ in the unevolved stage of the simulation.

As expected from Figure \ref{isotherm}, this is not the case for the simulation at $t=4.018$\,Gyr  where the isothermal (cyan) and the non-isothermal (blue) second term lines are clearly different. In this case $r_\mathrm{I}$ and $r_\mathrm{NI}$ are distinct and, while the non-isothermal residual averages to zero, the isothermal one presents a positive (negative) feature for $z<0$ ($z>0$). This suggests that using the isothermal approximation for the evolved stage of the simulation will introduce a bias that must be corrected. We show this in Section \ref{results}. 

Finally, notice that the sum of the second and third terms of the Jeans equation in Figure \ref{jeans_terms} is consistent with zero, excluding the presence of an important {\it tilt term} (hypothesis \ref{tilt-hyp} of the MA method) or significant non-equilibrium effects (hypothesis \ref{eq-hyp}).

\begin{figure*}
\center
\includegraphics[width=0.45\textwidth]{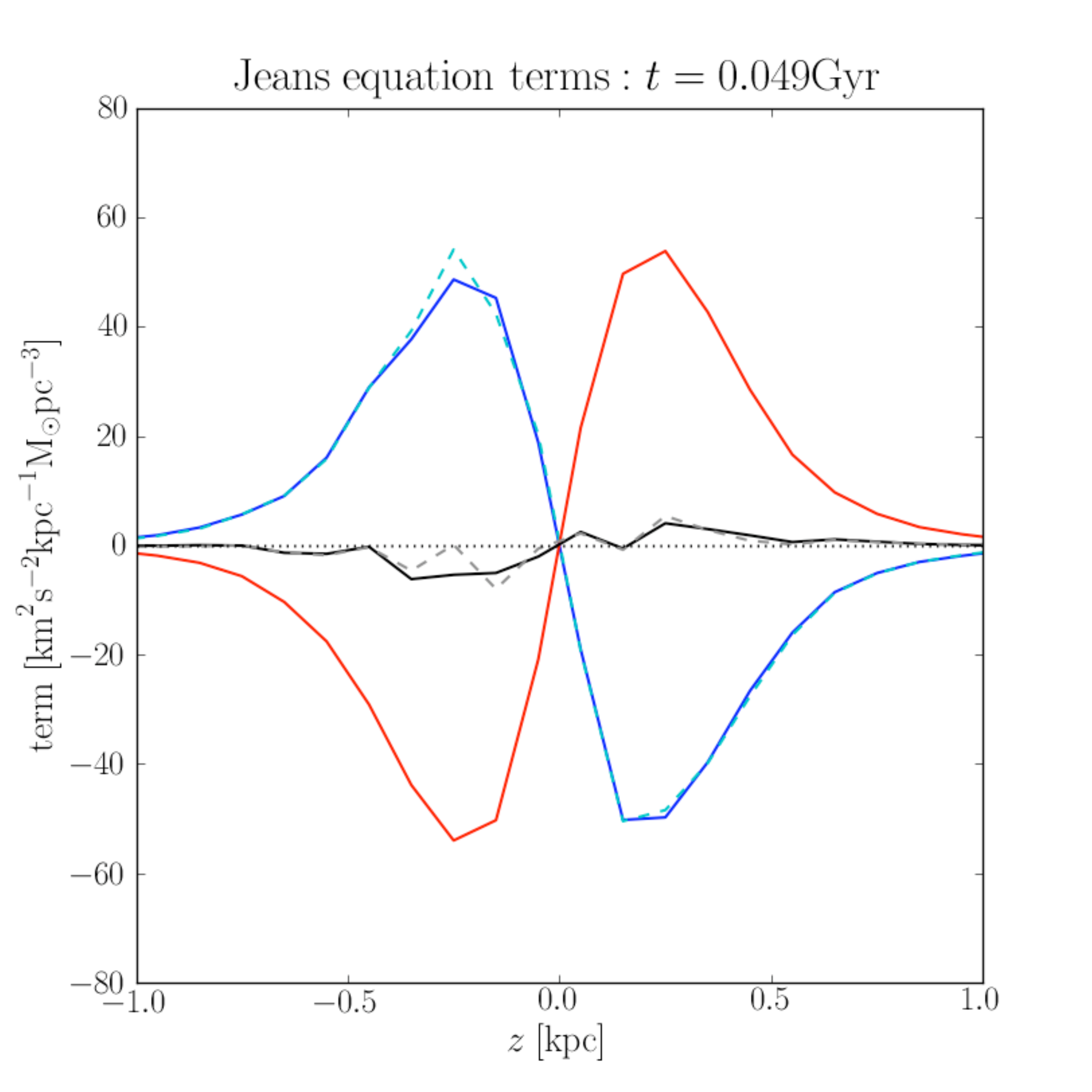}
\includegraphics[width=0.45\textwidth]{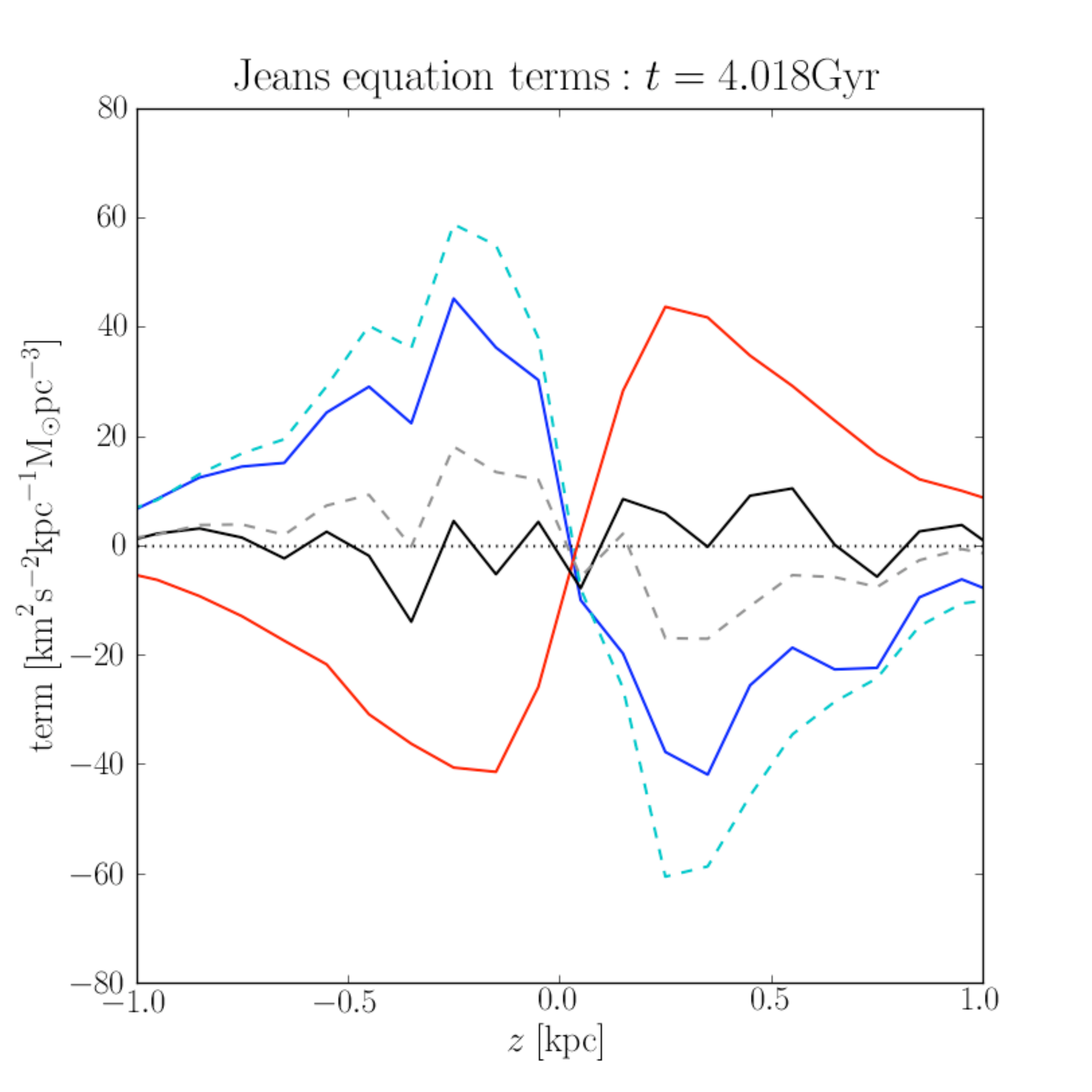}
\caption{The second and the third term of the Jeans equation \ref{jeans}, calculated for our simulation at $t=0.049$\,Gyr (left panel) and $t=4.018$\,Gyr (right panel) for $\theta=0^\circ$, with the SPH-like method. The different lines represent: dashed cyan: $\vztwoi\partial \nu_i/\partial z$ (2nd term: isothermal); solid blue: $\partial (\nu_i\vztwoi)/\partial z$ (2nd term: non-isothermal); solid red: $\nu_i\partial \Phi/\partial z$ (3rd term). The black and grey lines are the `residuals' given by the sum of the terms: solid black: $\nu_i\partial \Phi/\partial z+\partial (\vztwoi \nu_i)/\partial z$ (non-isothermal); dashed grey: $\nu_i\partial \Phi/\partial z+\vztwoi\partial \nu_i/\partial z$ (isothermal).}  
\label{jeans_terms}
\end{figure*}


\subsubsection{A flat rotation curve} \label{sec:rotcurve}

The second term of equation \ref{dmeff} is zero for flat rotation curves, i.e. for 
$V_c(R)=(Rd\Phi/dR)^{1/2}=\mathrm{constant}$. For a flat rotation curve the effective dark matter density corresponds to the halo mass density, $\rhodm^\mathrm{eff}=\rhodm(R)$, while the effect of a rising (falling) rotation curve is to give rise to a term of opposite (similar) sign to $\rhodm$, causing an underestimation (overestimation) of the dark matter density in the disc.

\begin{figure*}
\center
\includegraphics[width=0.45\textwidth]{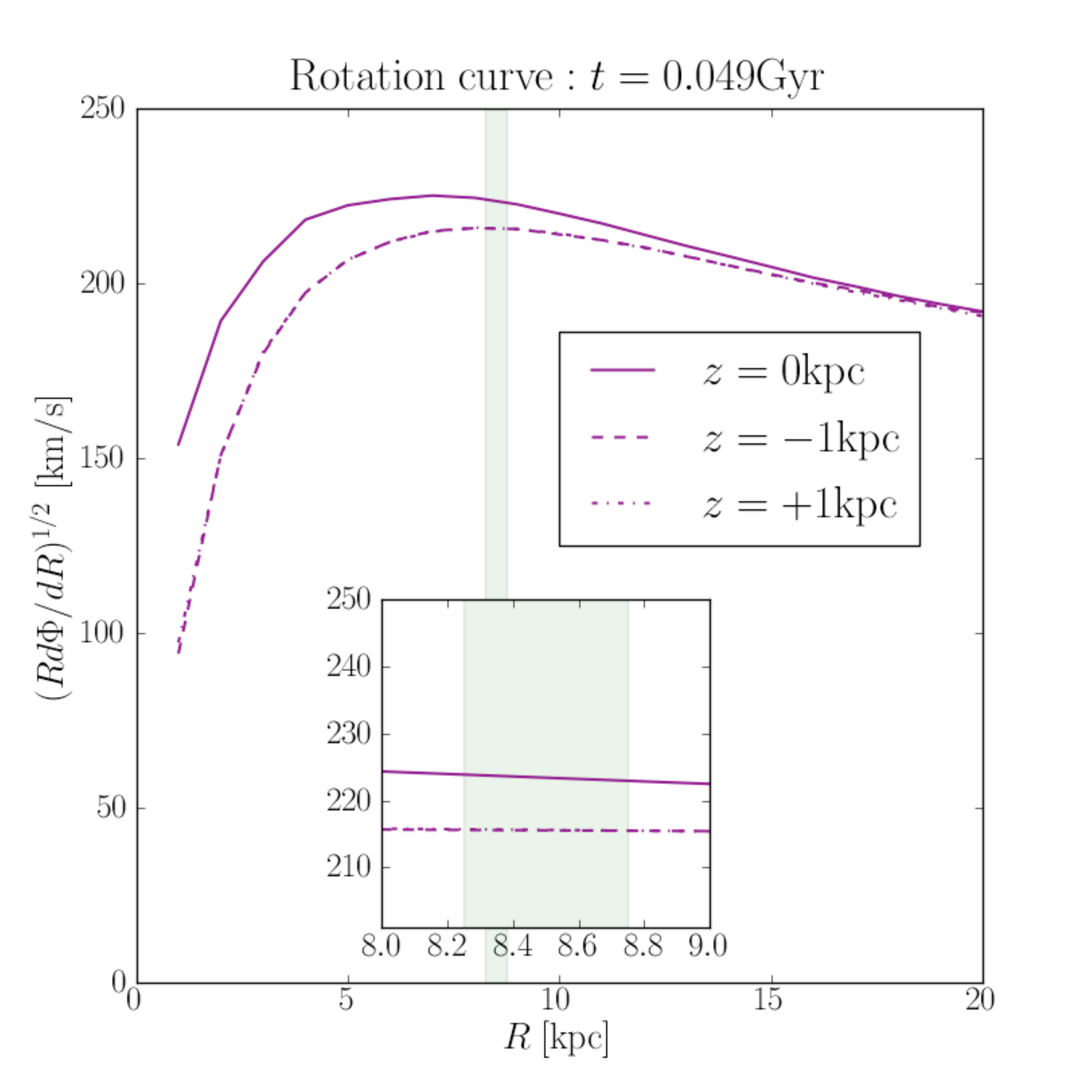}
\includegraphics[width=0.45\textwidth]{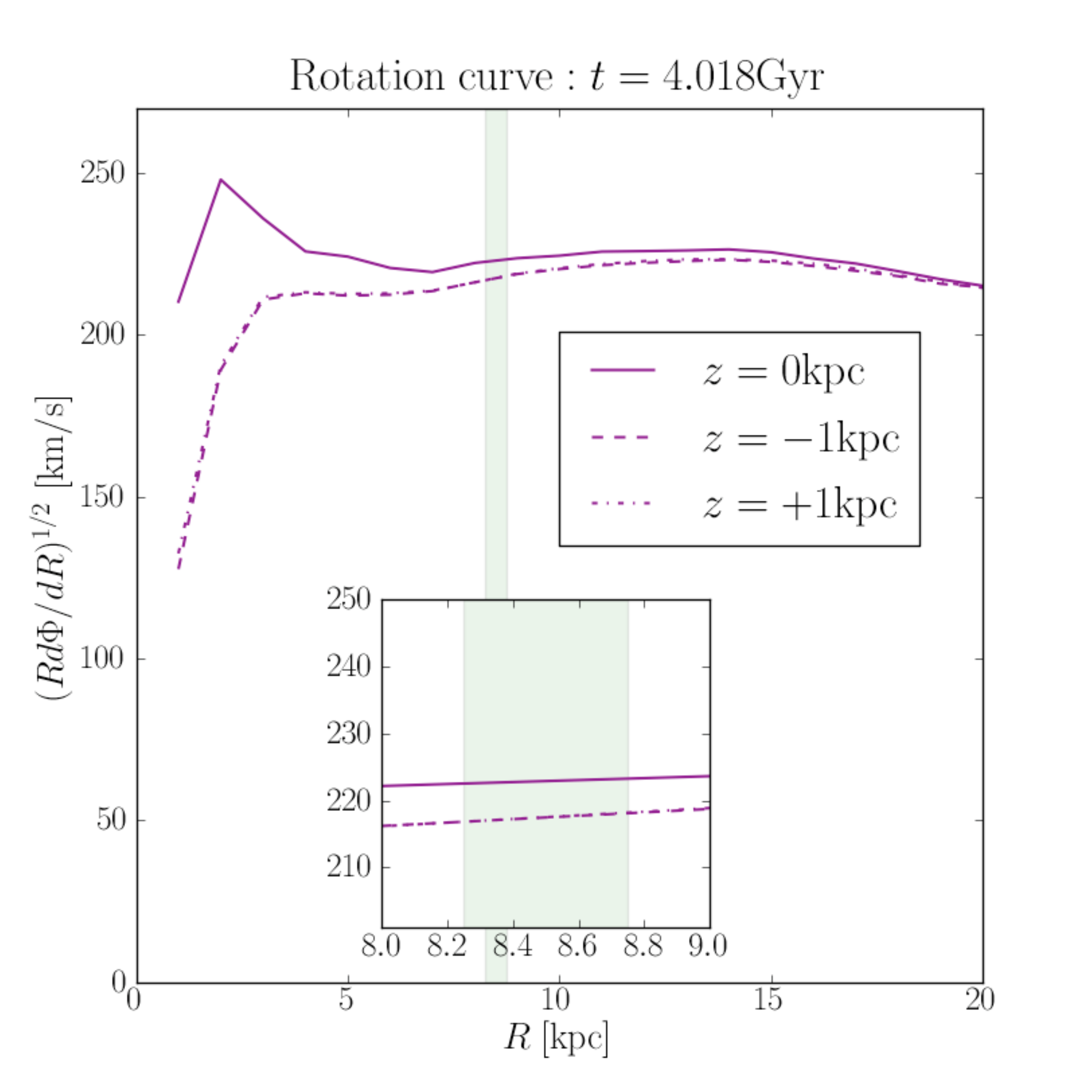}
\caption{Rotation curve for the unevolved stage of the simulation ($t=0.049$\,Gyr - left panel) and for the evolved one ($t=4.018$\,Gyr - right panel); it was calculated in large $R$-bins (1\,kpc) along a `slice' of the disc for each angular position considered using the SPH-method, here the patches at $\theta=0^\circ$ are shown. The solid line represents $V_c$ at the midplane, while the dashed and the dot-dashed line represent the rotation curve at $z=-1$\,kpc and $z=+1$\,kpc, respectively. The shaded green area is zoomed in the insert on the bottom of each plot and represents the radial position analysed in our work ($R=$8.5\,kpc).}
\label{fig:rotcurve}
\end{figure*}

In Figure \ref{fig:rotcurve} the rotation curves for the unevolved stage of the simulation ($t=0.049$\,Gyr) and for the evolved one ($t=4.018$\,Gyr) are plotted in the left and the right panel respectively. For the unevolved simulation the rotation curve is almost flat or slightly falling, while for the more evolved stage, in general, the rotation curve is usually slightly rising for $R=8.5$\,kpc; this means that we would expect a systematic underestimation of $\rhodm$ at $R=8.5$\,kpc for the evolved simulation.

\begin{figure*}
\center
\includegraphics[width=0.45\textwidth]{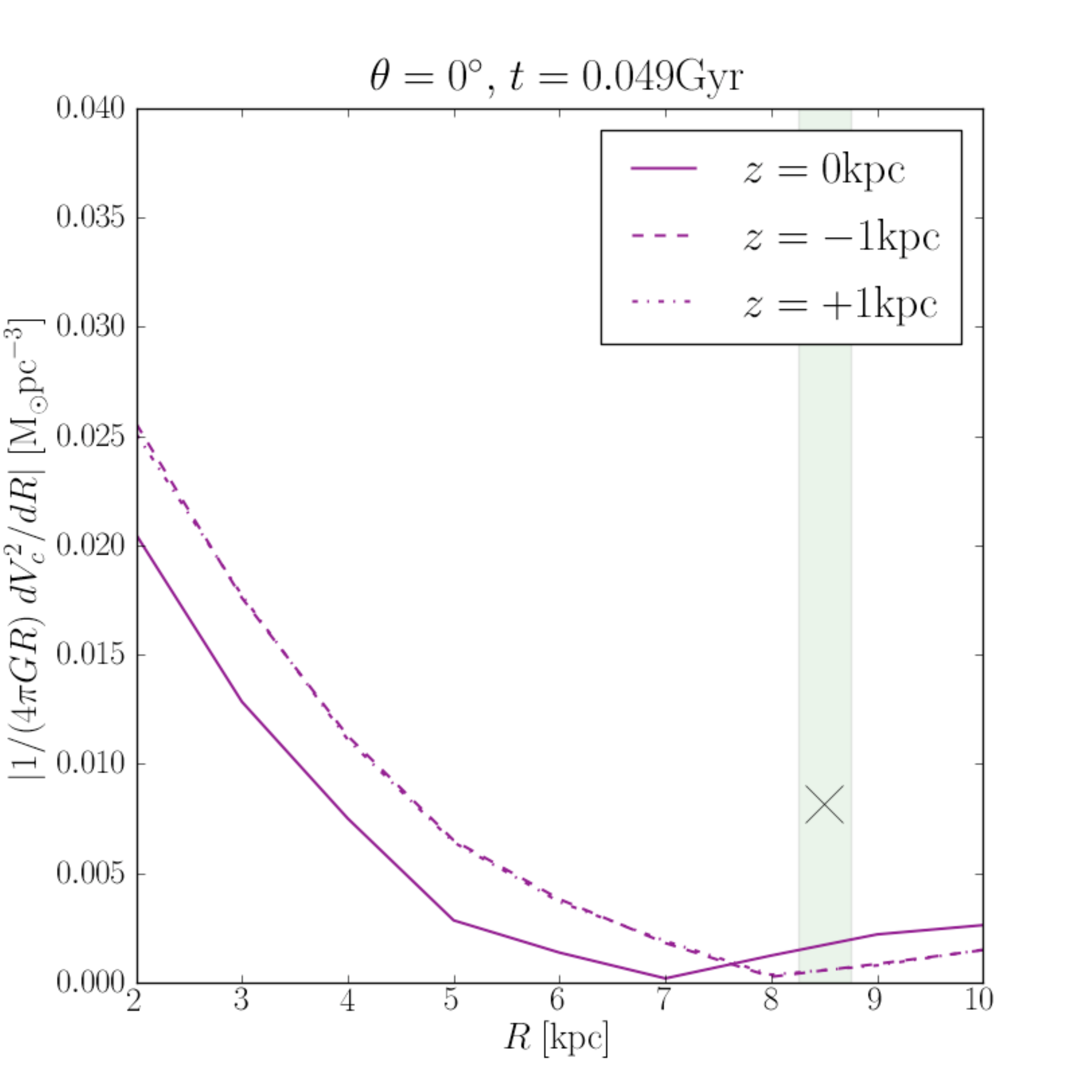}
\includegraphics[width=0.45\textwidth]{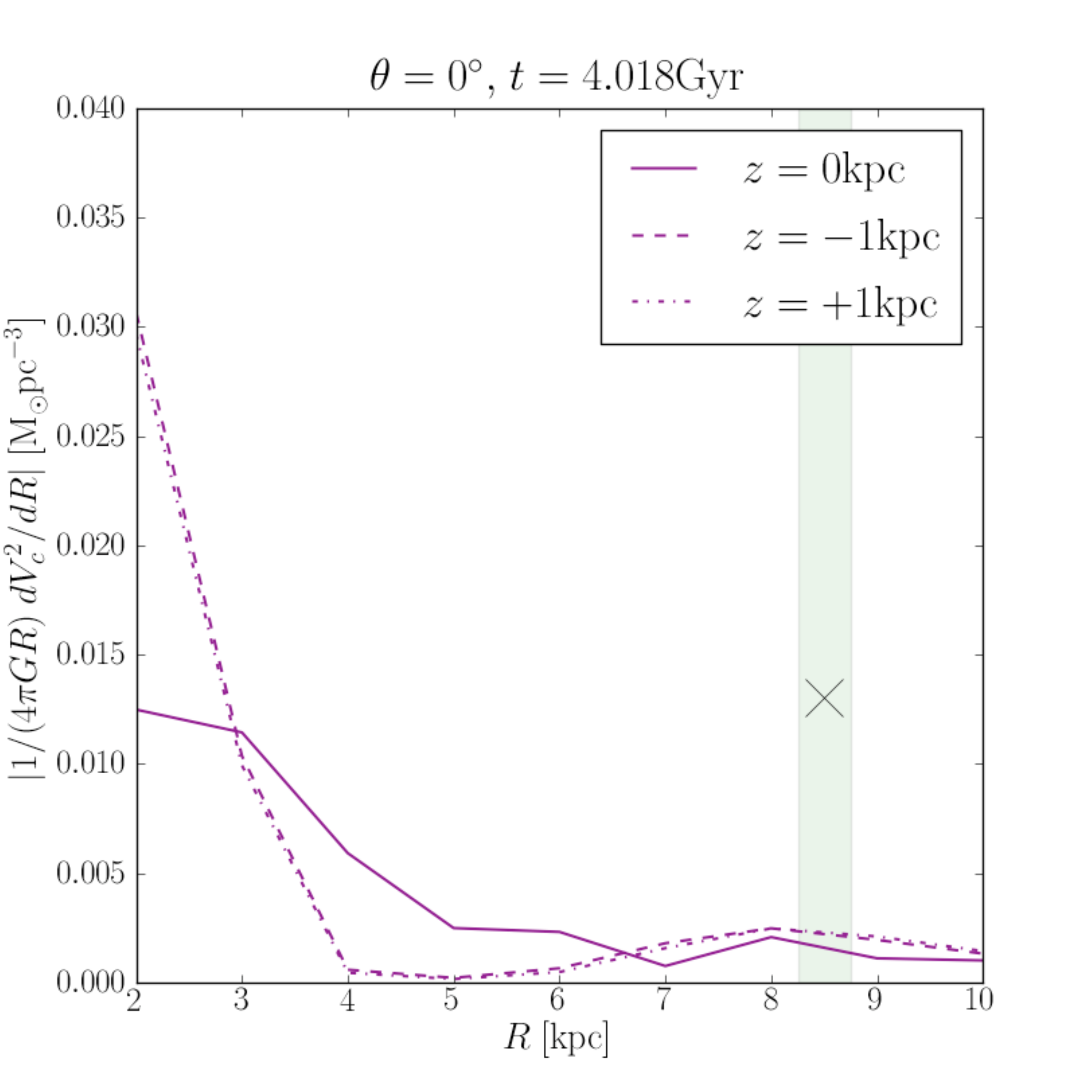}
\caption{Absolute value of the rotation curve term $|1/(4\pi G) dV_c^2/dR|$ for the unevolved stage of the simulation ($t=0.049$\,Gyr - left panel) and for the evolved one ($t=4.018$\,Gyr - right panel). The solid line represents the term calculated at the midplane, while the dashed and the dot-dashed line correspond to $z=-1$\,kpc and $z=+1$\,kpc, respectively. The shaded green area represents the radial position analysed in our work ($R=$8.5\,kpc), while the black cross gives the actual value of $\rhodm$ at $R=8.5$\,kpc.} 
\label{fig:rotterm}
\end{figure*}

To quantify the effect on the determination of $\rhodm$, we compute $V_c(R)=(Rd\Phi/dR)^{1/2}$ in large $R$-bins (1\,kpc) along a `slice' of the disc for each angular position considered using the SPH-method, then we calculate its $\partial/\partial R$ derivative to estimate the second term of equation \ref{dmeff}: $|(4\pi G R)^{-1} \partial V_c^2/\partial R|$. In Figure \ref{fig:rotterm} the absolute value of these terms are plotted for $\theta=0^\circ$ at $t=0.049$\,Gyr (left panel) and $t=4.018$\,Gyr (right panel). The black crosses show the values of $\rhodm$ at $R=8.5$\,kpc. 
For the evolved simulation, the shape of these plots is slightly different for the various angular positions at small $R$, due to the presence of the bar. However, at $R=8.5$\,kpc the contribution of the rotation curve term is between $10$ and $30$ per cent of $\rhodm$ (with positive sign). The shape of the rotation curve term with $R$ is always similar around the disc for the unevolved simulation and its contribution is $\sim 15-20$ per cent of $\rhodm$ at $R=8.5$\,kpc, always with negative sign.

For the real Milky Way, we can estimate the contribution of the rotation curve term from the Oort constants \citep{binney_galactic_1998}:
\beq
(4\pi G R)^{-1} \frac{\partial V_c^2}{\partial R}=\frac{B^2-A^2}{2\pi G}
\eeq

To determine the Oort constants, we must use stellar tracers that are well-mixed. As for the vertical potential determination, this means avoiding young stars. The most recent estimates using F giants  \citep{branham_oortF_2010} and K-M giants \citep{mignart_oorthipp_2000} from {\it Hipparcos} give $A=14.85 \pm 7.47$\,km\,s$^{-1}$\,kpc$^{-1}$ and $B=-10.85\pm 6.83$\,km\,s$^{-1}$\,kpc$^{-1}$ and $A = 14.5\pm 1.0$\,km\,s$^{-1}$\,kpc$^{-1}$ and $B =-11.5 \pm 1.0$\,km\,s$^{-1}$\,kpc$^{-1}$, respectively.  This is $\sim - 35$ per cent of the expected dark matter contribution as extrapolated from the rotation curve assuming spherical symmetry (see Section \ref{intro}), namely\footnote{this is just a simple average of the two cited values.} $-0.0033\pm0.0050$\msun\,pc$^{-3}$, leading to a slight overestimate of the dark matter density.

\subsubsection{Assuming that the $z$-motions are completely decoupled} 

The last assumption of the HF method is that the $z$ motion is decoupled so that the distribution function of the stars is only a function of $E_z$. If this is true, the distribution function of the stars in the midplane -- $f(E_z(0))=f(w_0)$ -- represents the distribution of the stars at all heights above the plane -- $f(E_z(z))=f(\sqrt{v_z^2+2\Phi(z)})$. Thus, it can be  integrated in $w_0=v_z(0)$ to predict the density fall-off. 

\begin{figure*}
\center
\includegraphics[width=0.33\textwidth]{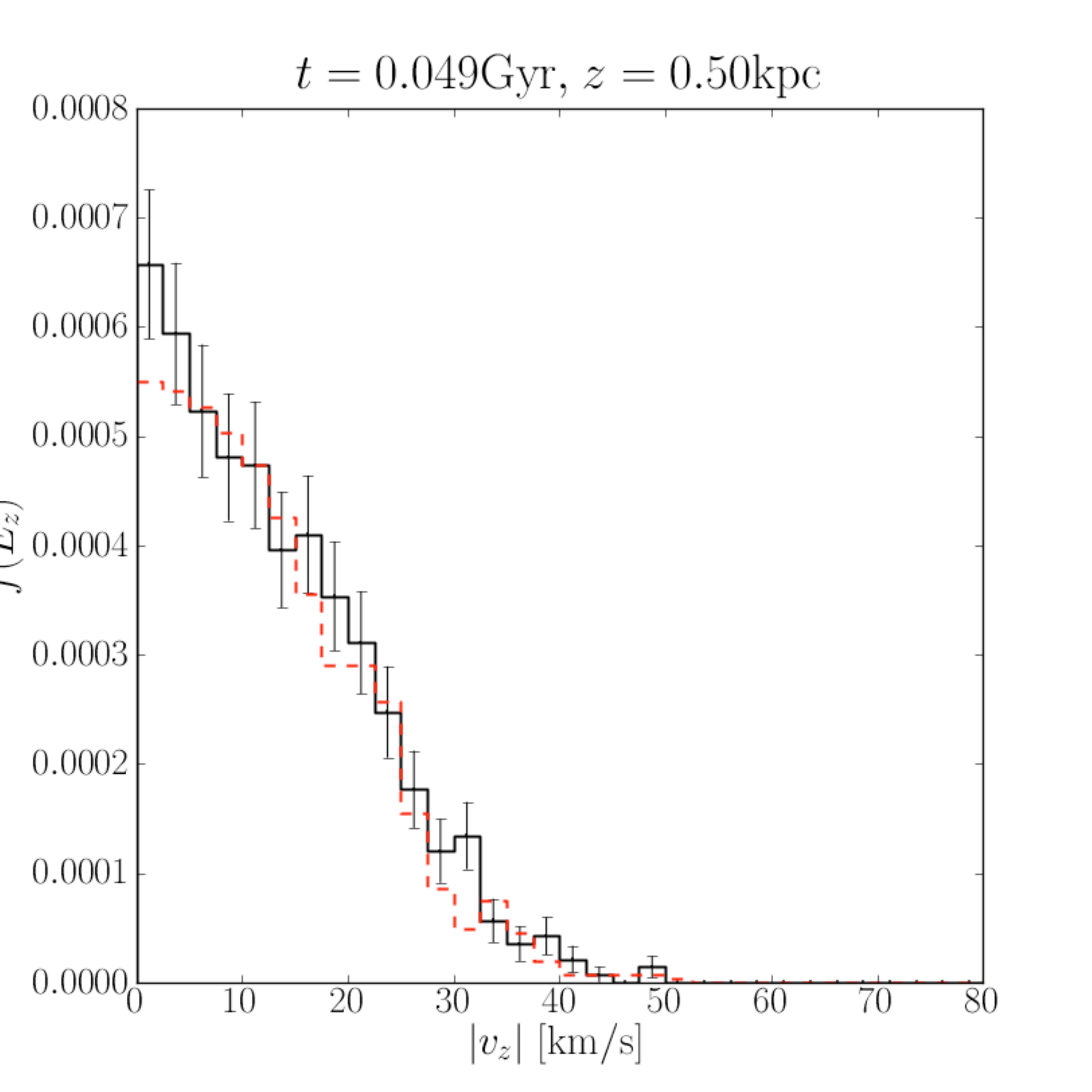}
\includegraphics[width=0.33\textwidth]{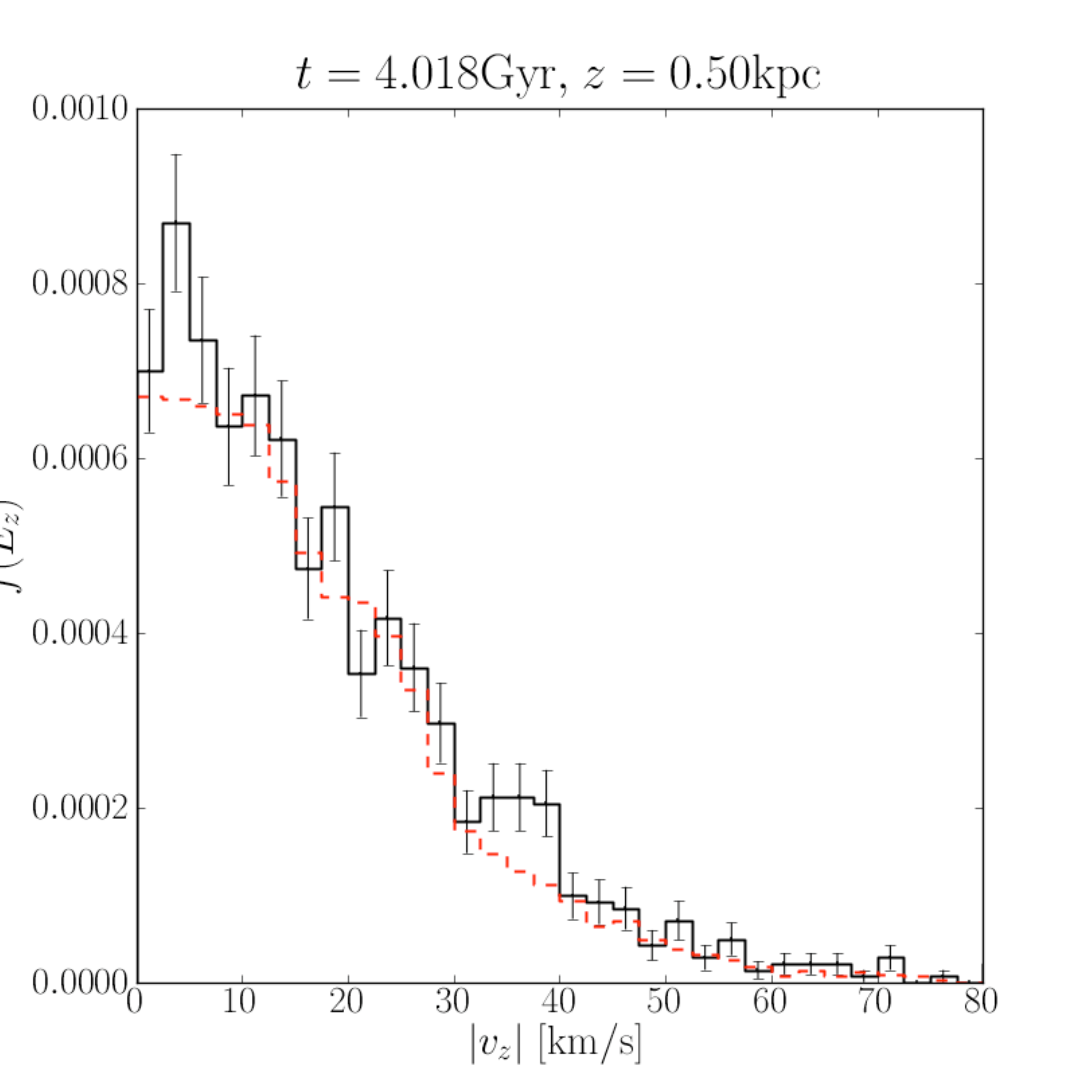}
\includegraphics[width=0.33\textwidth]{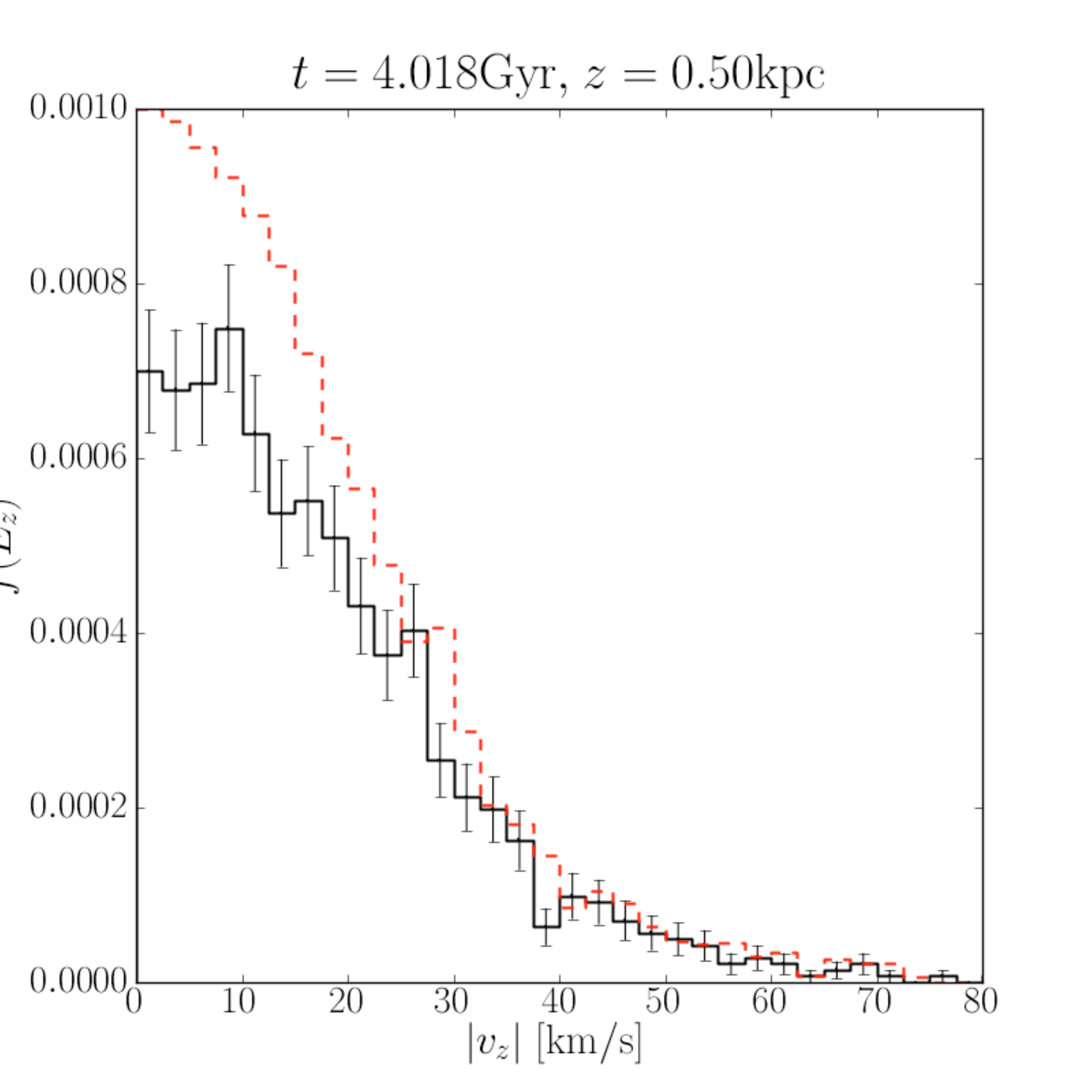}
\caption{Distribution function above the plane at $z=0.5$\,kpc (in black) compared with the one predicted from $f(E_z(0))$ (in red). The left panel represents the patch at $\theta=0^\circ$ for the unevolved simulation ($t=0.049$\,Gyr), while the centre and the right panels correspond to $\theta=45^\circ$ and $\theta=180^\circ$ in the evolved disc ($t=4.018$\,Gyr).} 
\label{fig:df}
\end{figure*}

In Figure \ref{fig:df}, we plot the distribution function at $z=0.5$\,kpc predicted from $f(w_0,0)$ for the unevolved simulation (first panel) and the evolved simulation (second and third panel representing two extreme cases at two different angular positions in the disc) in red. The actual distribution functions are over-plotted in black. As expected, while for the unevolved simulation the predicted distribution function is in good agreement with the actual one (left panel), the situation is different for the evolved stage. For most of the angular positions around the disc, the shape of the predicted distribution function is very different from the true one: the two volumes shown (at $\theta=45^\circ$ and $\theta=180^\circ$) in the second and third panel represent the best and the worst cases, respectively. From this analysis, we might expect the HF method to perform well on the $\theta=45^\circ$ patch, but poorly on the $\theta=180^\circ$ patch. We test this expectation in \ref{fitting}.

Note that \cite{statler_problems_1989} also considered this problem. Using St\"ackel potentials, he showed that, the $E_z$ is a good approximation to the Galactic third integral close to the midplane, but not above $z\approx 1$\,kpc. Two recent works by \cite{siebert_estimation_2008} and \cite{smith_tilt_2009} find that the tilt of the velocity ellipsoid for the Milky Way is indeed significant at $z\gtrsim 1$\,kpc, meaning that at such height, the motion is no longer separable. In our evolved simulation, we find important non-separability even at $z \sim 500$\,pc above the plane.

\begin{figure*}
\center
\includegraphics[width=0.33\textwidth]{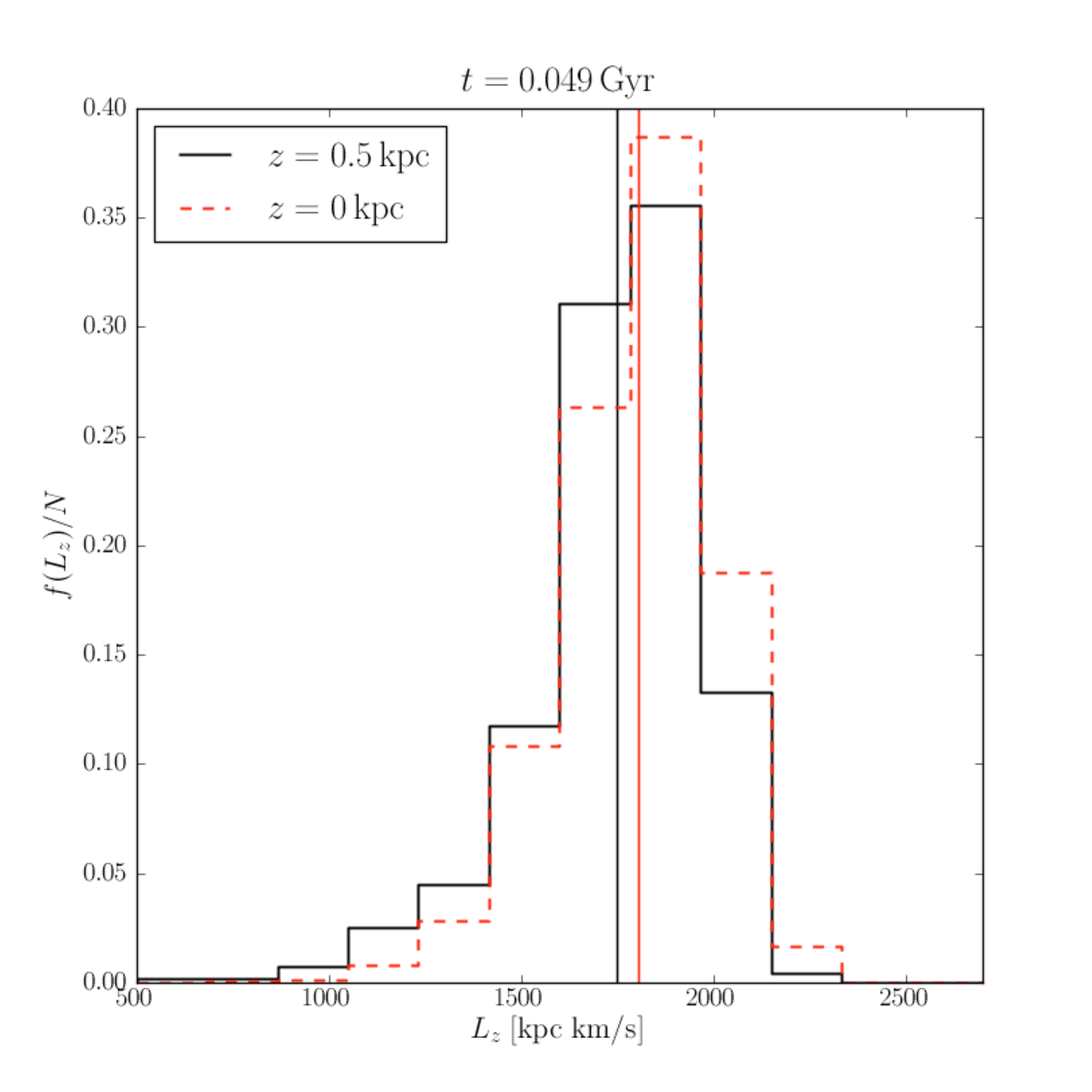}
\includegraphics[width=0.33\textwidth]{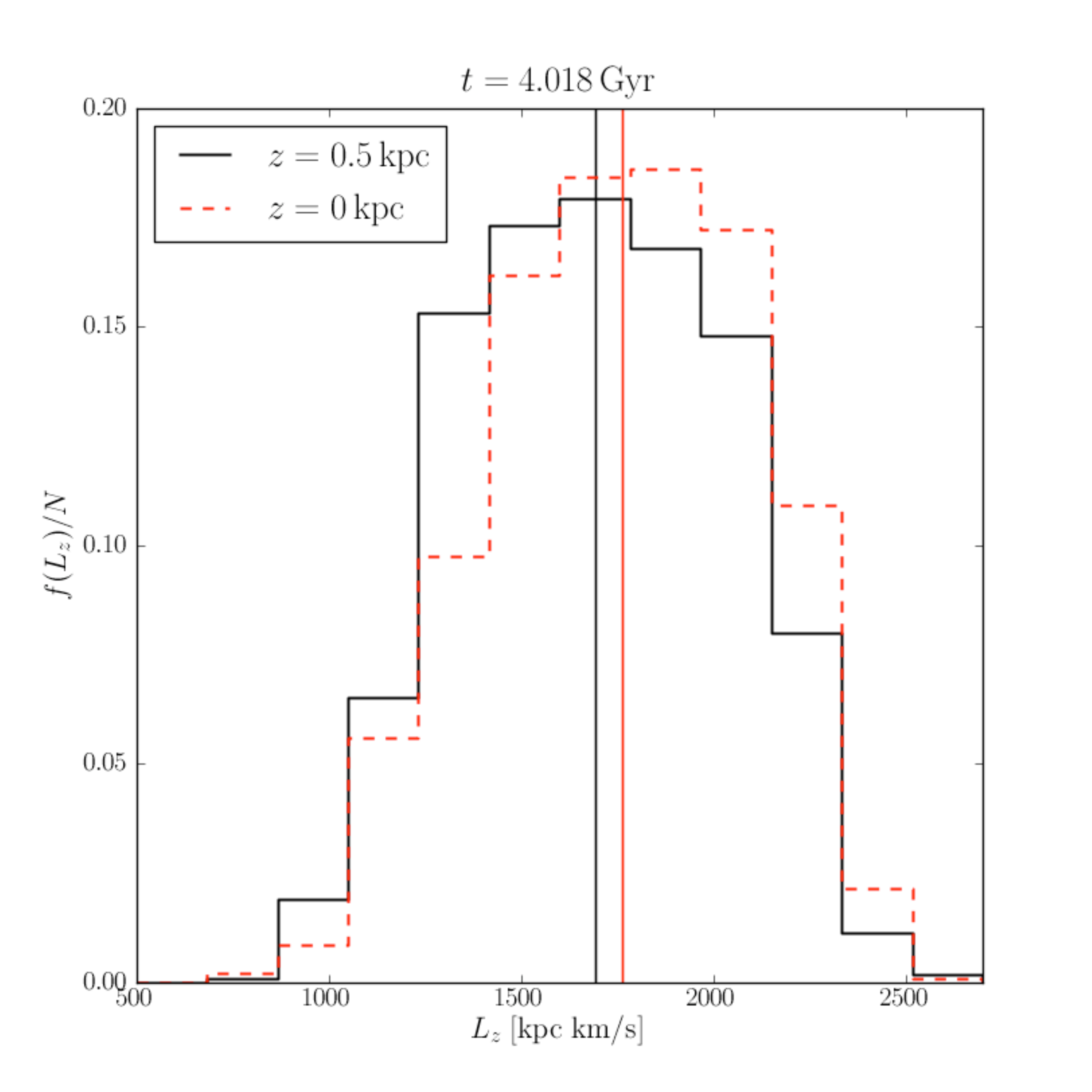}
\includegraphics[width=0.33\textwidth]{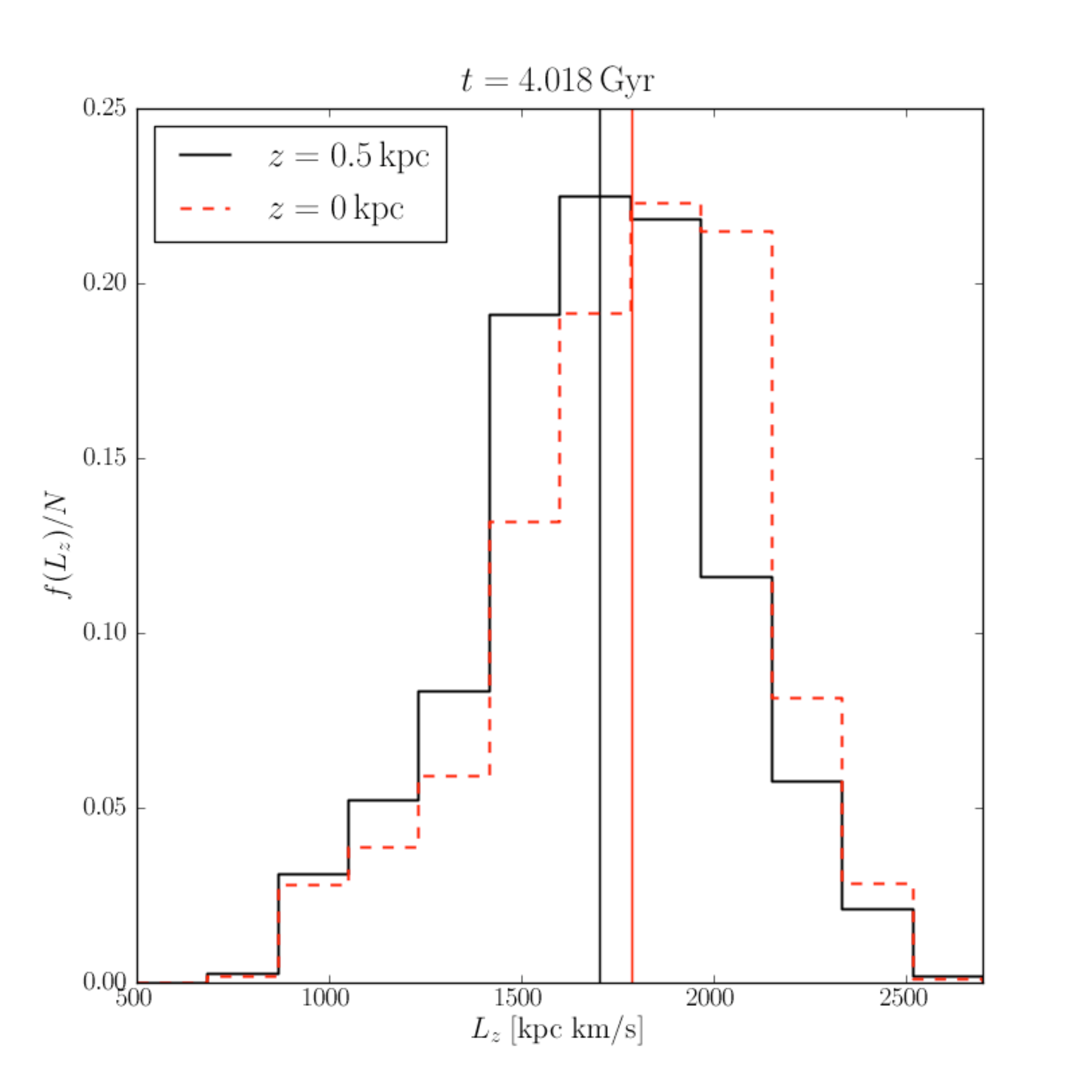}
\caption{Angular momentum distribution $g(L_z)$ -- normalized by the total number of stars $N$ -- at $z=0$\,kpc (dashed red histogram) and at $z=0.5$\,kpc (black histogram). The red and black vertical lines represent the median of the distributions at  $z=0$ and 0.5\,kpc respectively. The left panel represents the patch at $\theta=0^\circ$ for the unevolved simulation ($t=0.049$\,Gyr), the centre and the right panels correspond to $\theta=45^\circ$ and $\theta=180^\circ$ in the evolved disc ($t=4.018$\,Gyr). While the distributions have similar shape, the mean shifts by $\sim3$, 4 and 5 per cent for the unevolved simulation and the evolved simulation at $\theta=45^\circ$ and $\theta=180^\circ$, respectively. This means that the stars in the plane at 8.5\,kpc have a mean guiding center of $\sim8.0$\,kpc, while those at $z=0.5$\,kpc have a mean guiding center of 7.8, 7.8 and 7.7\,kpc for the unevolved simulation and the evolved simulation at $\theta=45^\circ$ and $\theta=180^\circ$, respectively.} 
\label{fig:Lz}
\end{figure*}

We also note that assuming the separability of the distribution function as $f = f(E_z) g(L_z)$, implies that $g(L_z)=\mathrm{const.}$ with height above the midplane.  We test this in Figure \ref{fig:Lz}, where we plot $g(L_z)$ at the midplane (dashed red histogram) and at $z=0.5$\,kpc (black histogram) for the unevolved (left panel) and for the evolved simulation (center and right panel). In the unevolved disc, $g(L_z)$ at the midplane and $z=0.5$\,kpc are similar. For the evolved simulation this is not always the case.  In accord with our analysis above, the situation is better for the `best case' $\theta=45^\circ$ patch than for the `worst case' $\theta=180^\circ$ patch.

\subsection{Results for the simulation}\label{fitting}
In this section, we test the MA and HF methods on our evolved and unevolved simulations. We define three different `Solar Neighbourhood' patches: 36 cylinders around the disc at angular separations of $10^\circ$ (represented as red circles in Figure \ref{sim10-820}); a `superpatch' that is the average of the 36 cylindrical patches; and 4 (or 8) wedges around the disc at angles: $\theta=0^\circ,90^\circ,180^\circ,270^\circ$ (and additionally $45^\circ,135^\circ,225^\circ,315^\circ$ for the evolved simulation which is not axisymmetric, to examine all the relevant positions with respect to the bar). All patches are represented as red shaded areas in Figure \ref{sim10-820}. The cylinders have sampling similar to the currently available {\it Hipparcos} data that we consider in Section \ref{hf-data}. The `superpatch' gives sampling equivalent to that expected from the GAIA mission (GAIA will obtain distances with an accuracy better than $0.1$ per cent for $\sim 100000$ stars within 80\,pc \citep{bailer-jones_what_2008}). However, we can only apply the superpatch to the unevolved simulation that is axisymmetric. For this reason we introduce also the wedges that contain approximately 5 times the number of stars in a cylinder; they are the best compromise to obtain larger sampling for a sufficiently local volume in the non-axisymmetric disc. Note that, for the unevolved disc, the cylinders and wedges tell us only about sampling errors since the disc is axisymmetric (the results for each patch should be statistically equivalent). For the evolved disc, however, the different patches explore the effect of spiral structure and disc inhomogeneities on our analysis.

We consider a single visible component to build the mass model for the disc, described by its density in the plane and its velocity dispersion. We set the Sun's position at $R_\sun=8.5$\,kpc. We let the local dark matter density $\rhodm$ vary in the range $[0,1]$\,M$_\sun$\,pc$^{-3}$, and the disc mass density $\rhos(0)$ vary in the range $\pm 0.014$ M$_\sun$\,pc$^{-3}$ around the actual value that we measure for the simulation. This range has a width comparable to the observational uncertainties for the data we consider in Section \ref{hf-data} (and see also Table \ref{mmodel}).

For the HF method, we need the distribution function in the midplane $f(w_0)$ to be used in equation \ref{den}. To compute this, we fit the velocity distribution of stars with $|z|\leq50$\,pc (see Section \ref{box}) with a Gaussian function for the unevolved simulation, and a double Gaussian for the evolved one (an example fit is shown in the left panel of Figure \ref{den-fall-off}).

\subsubsection{The unevovled simulation} \label{results}

We first consider the unevolved simulation ($t=0.049$\,Gyr) that fulfils the hypotheses of the methods. 


\paragraph{Maximum volume of the patch}\label{box}

We first consider the appropriate size of the volume for the MA method: it should be small enough in the radial direction (ideally infinitesimal) to average the potential and its derivatives over $R$ to solve the Jeans equation for a one-dimensional slab.  Of course, we need a large patch for the best possible sampling.  In this section, we use the unevolved simulation to measure how large our patch can be before systematic errors dominate over our sample error. For this, we use the `superpatch' described in Section \ref{fitting}, above. We consider the average of 36 cylinders around the disc at $R_\sun = 8.5$\,kpc with radius $R=150, 250,300,400,500$\,pc. 

In addition, the HF method requires measuring the distribution function in the midplane: $f(w_0)$. For this, we must choose a vertical scale to determine $f(w_0)$, and again there is a trade off between bias and sample error. To find the optimal height, we compute the velocity distribution considering star particles up to $|z|<50,75,100$\,pc. Note that for any patch size, there will be a bias error due to the finite volume considered. Here we find the largest patch size (for `GAIA' sampling; the `superpatch' described in Section \ref{fitting}) for which the bias error is small as compared to the sample error. If the sampling for a given volume is improved, then we will become more sensitive to bias. In this case, the optimal patch size will be smaller than that found here. 


For each choice of $R$ and $|z|$, we apply our MCMC method to explore the $\rhos$-$\rhodm$ parameter space and calculate the $\chi^2$ for each model. 

\begin{figure*}
\center
\includegraphics[width=0.45\textwidth]{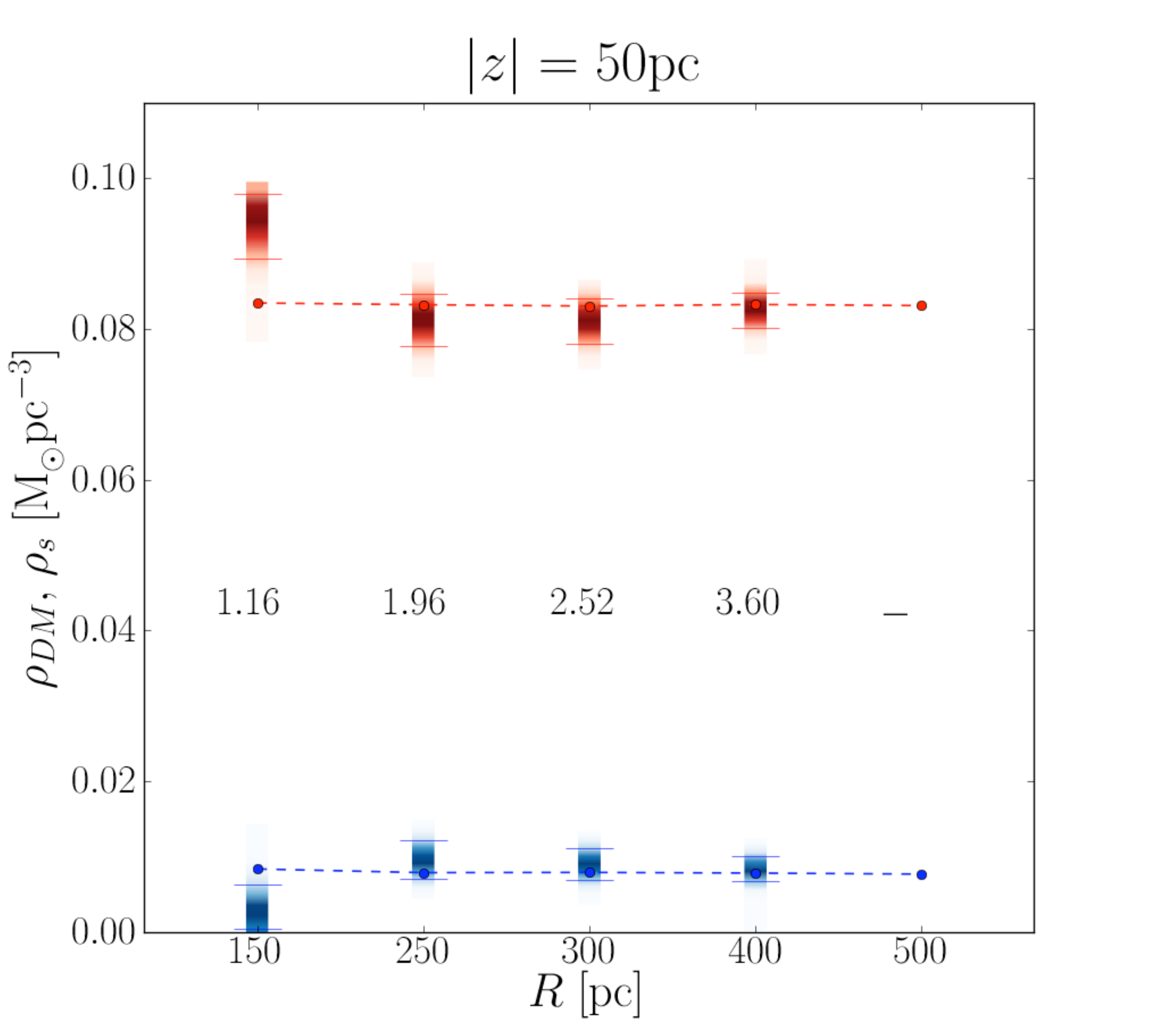}
\includegraphics[width=0.45\textwidth]{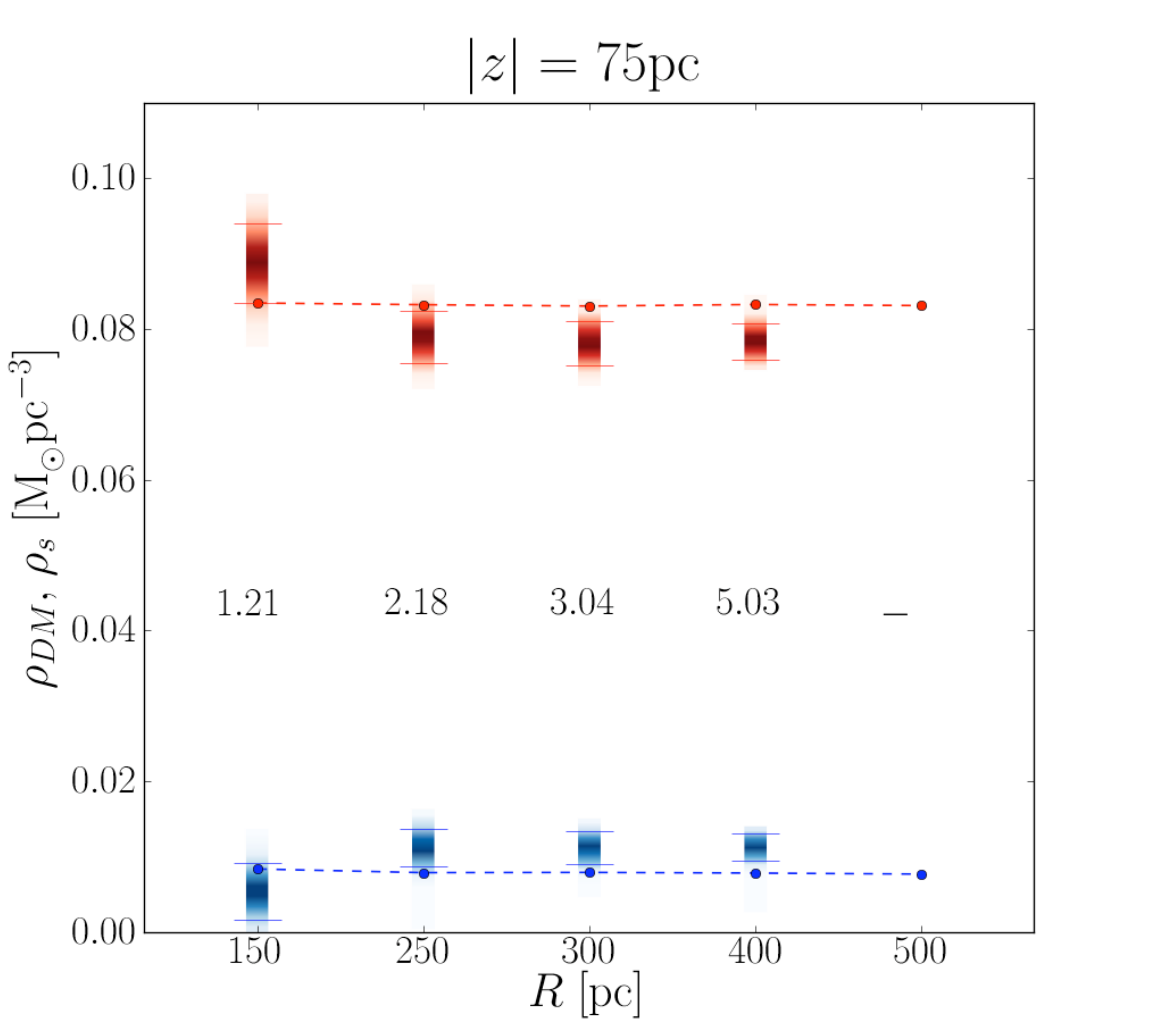}
\caption{Models explored by the MCMC for the HF method, using different size of the `local volume' box. The left (right) panel correspond to velocity distribution in the midplane constructed using stars with $|z|<50$\,pc ($|z|<75$\,pc). On the $x-$axis the different radial sizes are indicated. The blue (red) shaded rectangles represent the recovered dark (visible) matter density. The blue (red) dashed line and filled dots represent the actual value of $\rhodm$ ($\rhos$). The horizontal red (blue) segments represent the 90 per cent errors in the recovered value of $\rhos$ ($\rhodm$).}
\label{fig:box}
\end{figure*}

We first apply our MA method to test the optimal radial extent of a patch. For a cylinder of radius $R>500$\,pc the MCMC fails to find a solution indicating that the bias errors are dominant. For smaller patches, we recover the correct values of $\rhos$ and $\rhodm$ within our quoted errors, but find that the best $\chi^2$ shrinks with $R$. Next, we apply the HF method. In this case, the MCMC fails to find a solution if the midplane velocity distribution is averaged up to $|z|=100$\,pc. The best $\chi^2$ values for each case are reported in Table \ref{chi-tab} (the situation for the MA method is very similar to the first line). The recovered densities in the different volumes are shown in Figure \ref{fig:box}: for $R=250,300,400$\,pc and when we calculate the velocity distribution function in the midplane using stars with $|z|<50$\,pc, we always recover the correct answer even if the agreement between the predicted and the measured density fall-off of the tracers give rise to increasing $\chi^2$ value with $R$. For $R=150$\,pc the result is not as good, likely owing to the poorer sampling. Calculating the velocity distribution in the midplane from stars with $|z|<75$\,pc gives always slightly biased results.

Given the above results,  we will consistently use patches with $R=250$\,pc and average our midplane velocity distributions for stars with $|z|<50$\,pc. This volume is similar to that used by \cite{holmberg_local_2000} whose data we consider in Section \ref{hf-data}.

\begin{table}
\center
\caption{Best $\chi^2$ for different sizes of the `local volume' box; $|z|<50,75,100$\,pc is the height used to construct the midplane velocity distribution; $R=150,250,300,400,500$\,pc is the radial size of the cylindrical box. The dashes correspond to a failure of the MCMC in recovering the density, i.e. it can not find an acceptable value.}
\label{chi-tab}
\begin{tabular}{|c|c|c|c|c|c|}
\hline
&$R=150$\,pc & $250$\,pc & $300$\,pc & $400$\,pc &$500$\,pc\\
\hline
$|z|<50$\,pc & 1.16 & 1.96 & 2.52 & 3.60 & -\\
$|z|<75$\,pc & 1.21 & 2.18 & 3.04 & 5.03 & -\\
$|z|<100$\,pc & - & - & - & - & -\\
\hline
\end{tabular}
\end{table}

\paragraph{Degeneracy in $\rhos$ and $\rhodm$}\label{dege}

\begin{figure*}
\center
\includegraphics[width=0.45\textwidth]{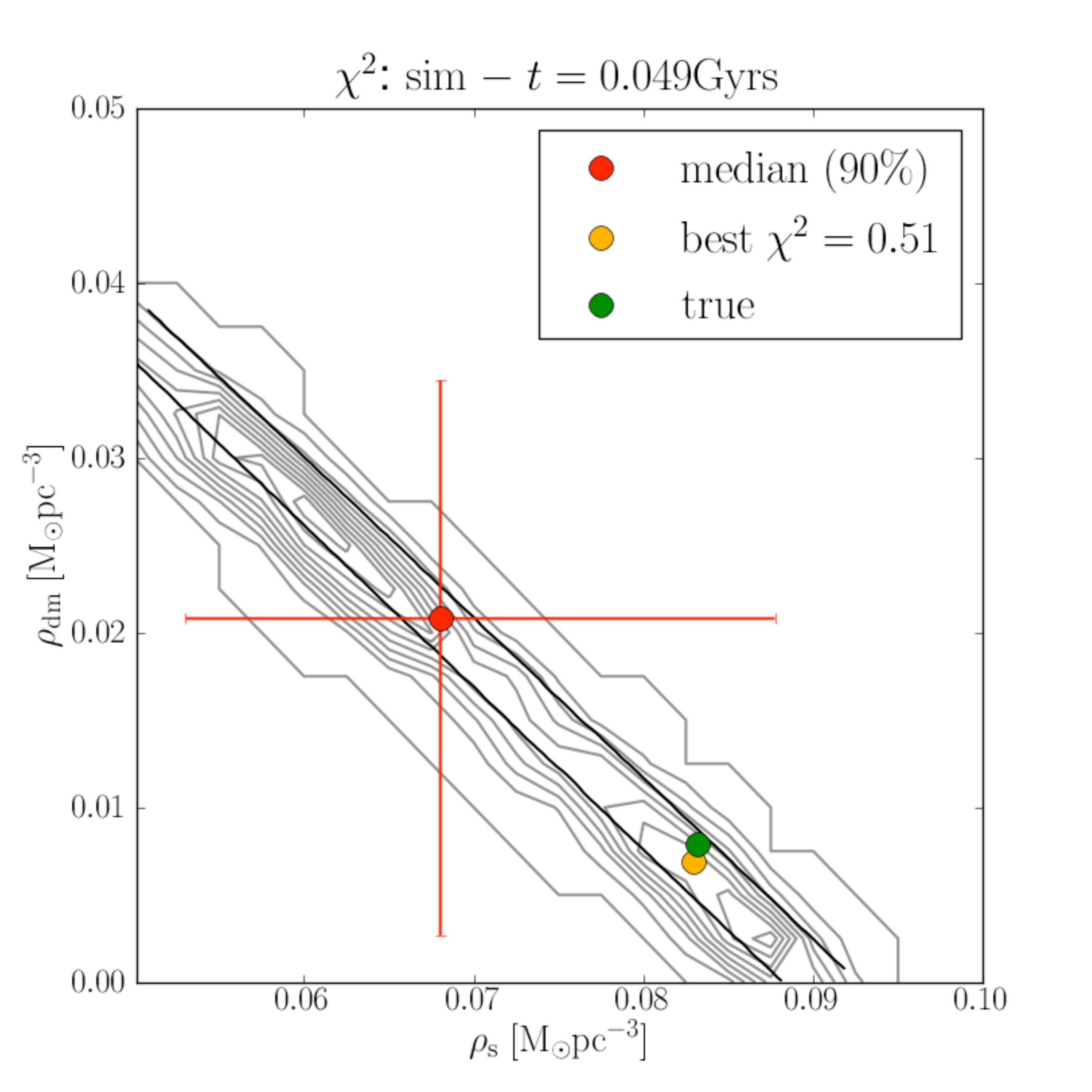}
\includegraphics[width=0.45\textwidth]{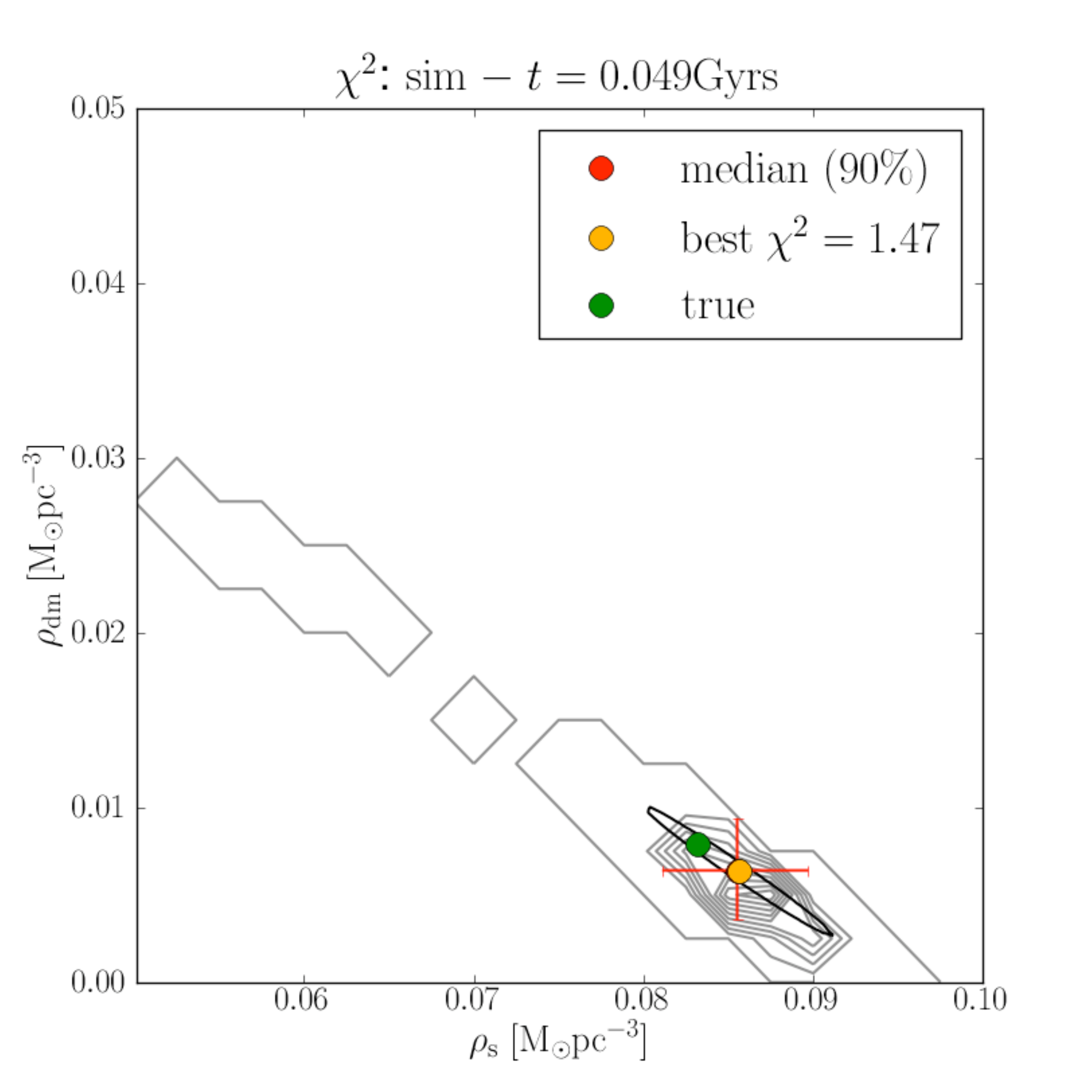}
\caption{MCMC models in $\rhos$-$\rhodm$ space for the `superpatch' applied to the unevolved simulation. The yellow dot corresponds to the best $\chi^2_\mathrm{best}$ model; the green dot corresponds to the true value; the red dot is the median of the distribution with $90$ per cent error bars; the black contours contain all models with $\chi^2\leq1.1\chi^2_\mathrm{best}$; and the grey contours represent the density of models explored by the MCMC. Left panel: fitting the density fall off up to $|z|=0.25$\,kpc ($\gtrsim  z_{1/2}$); Right panel: fitting up to $|z|=0.75$\,kpc ($\gtrsim 4 z_{1/2}$).}
\label{degeneracy}
\end{figure*}

In their work, \cite{holmberg_local_2000} fit the density fall off of the stellar tracers up to $0.1-0.2$\,kpc which approximately corresponds to the MW disc half mass scale height $z_{1/2}$. If we adopt the same criteria for our `superpatch', we see that the area of the $\rhos$-$\rhodm$ plane explored by the MCMC corresponds to a $45^o$ stripe with almost the same value of $\chi^2$ for all models. This means that we have a nearly flat distribution of models and a strong degeneracy between $\rhos$ and $\rhodm$. This is shown in the left panel of Figure \ref{degeneracy}. The grey contours represent the density of models explored by the MCMC, while the black contour contains all models with $\chi^2\leq1.1\chi^2_\mathrm{best}$.

This strong degeneracy means that we can only determine  the total density on the plane $\rho_{\mathrm{tot}}(0)=\rhos(0)+\rhodm(0)$, but not $\rhos$ and $\rhodm$ separately.  To break this degeneracy -- and obtain smaller error bars -- we must fit the tracers to higher $z$. This has been noted in earlier work. \cite{bahcall_self-consistent_1984} state that data up to $z=600$\,pc are required  to be sensitive to the SHM dark matter density.

In the right panel of Figure \ref{degeneracy}, we show our recovered $\rhos$ and $\rhodm$, but now fitting to $|z|=0.75$\,kpc ($\sim 4$ times $z_{1/2}$). This is sufficient to break the degeneracy and we recover the correct answer for both $\rhos$ and $\rhodm$ inside our $1\sigma$ error bars. We show results here for brevity only for the MA method, however the HF method produces similar results for this test. For the rest of our analysis we will fit the density fall-off of the tracers up to $0.75$\,kpc.

\paragraph{Introduction of realistic errors}\label{gaia}
As already stressed, the `superpatch' has statistics comparable to that expected for the GAIA mission. In this section, we consider the effect of realistic observational errors in the velocities and positions of the stars on the recovered stellar and dark matter densities.

We consider errors typical for current {\it Hipparcos} data (that we consider in Section \ref{hf-data}) and GAIA quality data. The {\it Hipparcos} mission provided $\sim 10^4$ stars out to $\sim 100$\,pc with proper motions and parallaxes accurate to $<10$ per cent \citep{dehnen_our_2002}. In \cite{holmberg_local_2000}, the (incomplete) radial velocity information from {\it Hipparcos} data were ignored and the velocity distribution was computed using only low latitude stars, whose motion is dominated by the proper motion. The confidence limits were estimated via a series of Monte Carlo simulations of observations drawn from synthetic {\it Hipparcos} survey catalogues, taking into account the {\it Hipparcos} magnitude limits and magnitude-dependent parallax and proper-motion errors. For the A and the F sample they found a $95$ per cent confidence limit of $\pm0.011$\msun\,pc$^{-3}$ and $\pm0.023$\msun\,pc$^{-3}$, respectively.

GAIA will determine distances for ~150 million stars with a accuracy better than $10$ per cent (within 8\,kpc) and some 100000 stars to better than $0.1$ per cent within 80\,pc \citep{bailer-jones_what_2008}. For an unreddened K giant at 6\,kpc, GAIA will measure the distance accurate to $2$ per cent and the transverse velocity with an accuracy of about 1\,km\,s$^{-1}$ \citep{bailer-jones_what_2008}.

To understand the impact of GAIA's accuracy, we introduce Gaussian errors in the velocity of $1$\,km\,s$^{-1}$ and an accuracy in the positions of $2$ per cent. We then run our MCMC chain on these input data with errors. We find that our recovered values for the density are unchanged, but the $\chi^2$ increases. We conclude that velocity-position errors are a perturbation on sample errors and model systematics. 

Here we included only uncorrelated errors on distances and velocities of the stars; correlated errors could be a concern when one calculates space velocity from proper motions. However, in the methods considered, only the vertical velocity of stars in a small volume (i.e. mostly high latitude stars) for which $v_z$ is mostly due to the radial velocity are considered. In addition, we show that the main uncertainties come from model rather than measurement uncertainties.

\paragraph{The importance of statistics}\label{stat}

In this section, we investigate the effect of sample size.  We considered a GAIA data quality mission with `superpatch' sampling. Now we consider smaller patches with sampling more similar to  {\it Hipparcos} data. Good statistics are particularly important for the HF method that requires the shape of the in-plane velocity distribution function rather than just its moments. 

We consider 4 cylindrical volumes around the disc with statistics comparable to {\it Hipparcos} data ($\sim 2000-3000$ within $|z|<200$\,pc), and 4 wedge-shaped larger volumes at the same angular positions, having the same radial and vertical size, but covering a larger azimuthal angle (and containing about 4-5 times more particles). 

\begin{figure*}
\center
\includegraphics[width=0.45\textwidth]{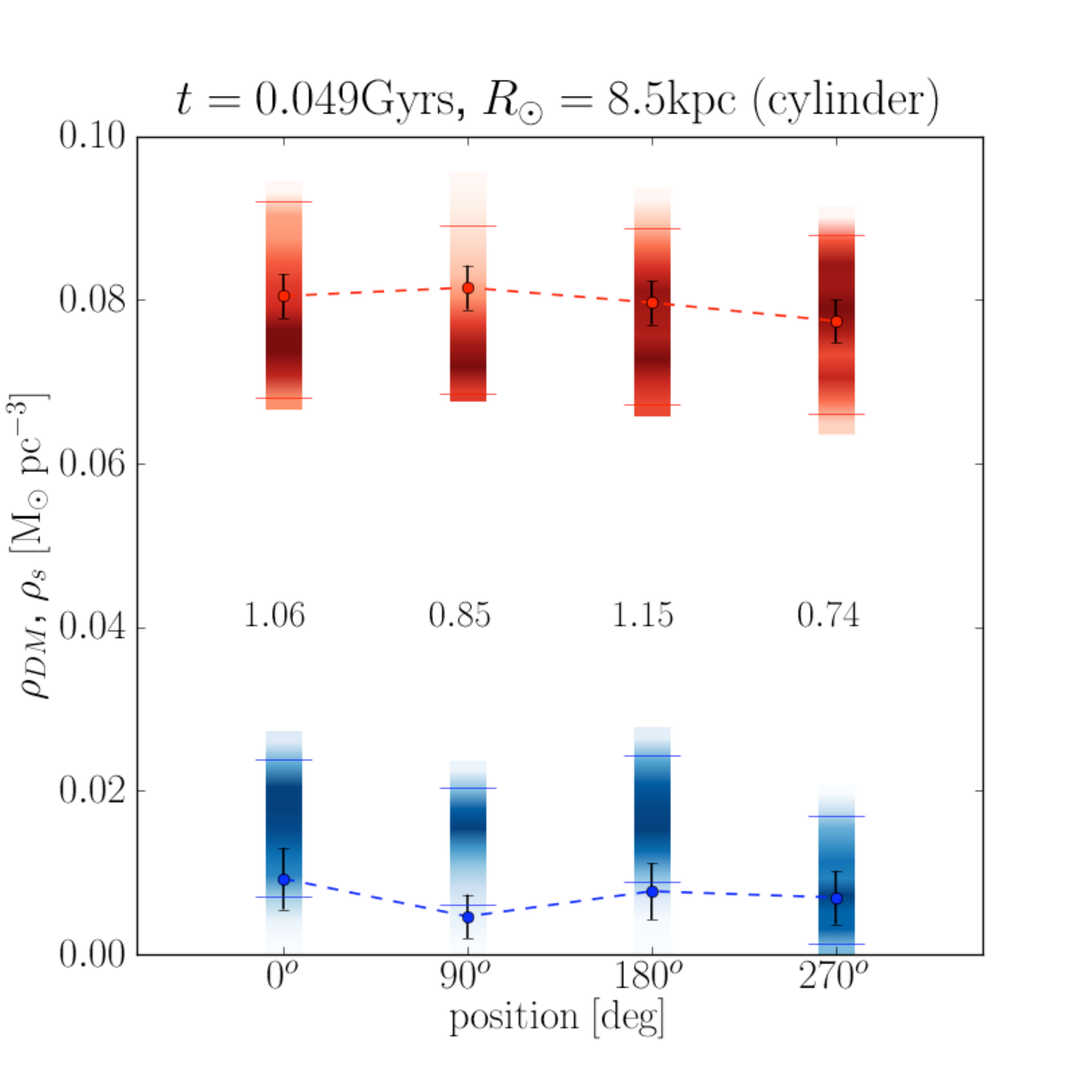}
\includegraphics[width=0.45\textwidth]{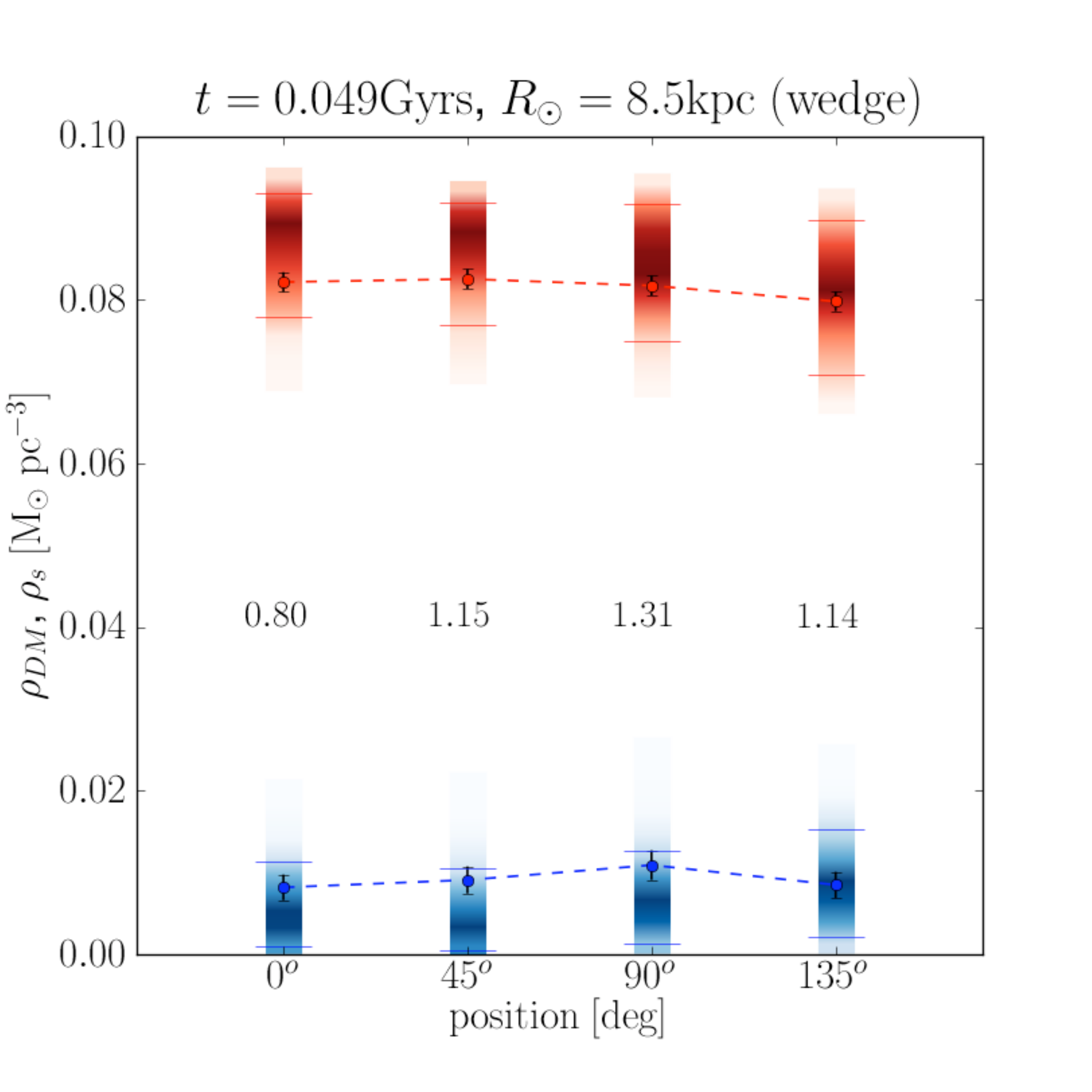}
\caption{Models explored by the MCMC for the MA method for the four cylinders (left panel) and the four wedges (right panel) represented as shaded areas of different colours. Blue corresponds to the recovered dark matter density $\rhodm$ and red to the recovered visible matter density $\rhos$ in the plane. The filled dots represent the corresponding actual values with Poisson errors. The red and blue horizontal segments show the 90 per cent errors on the recovered densities. The numbers correspond to the reduced $\chi^2$ values. Notice that the apparent fluctuations in the density at different angular positions are due to the sample noise.}
\label{fig:stat}
\end{figure*}

The results are reported in Figure \ref{fig:stat}, which shows the models explored by the MCMC for the MA method for the four cylinders (left panel) and the four wedges (right panel). In both cases, the method recovers the correct value of $\rhos$ and $\rhodm$ within our quoted errors, with the error bars shrinking with improved sampling as expected. The results are almost identical for the HF method for this early stage of the simulation.

\subsubsection{The evolved simulation}\label{inhom}

\paragraph{The HF method} 

In the previous section, we demonstrated that the MA and HF methods perform equivalently well when applied to the ideal situation of an isothermal axisymmetric disc, fulfilling all the standard assumptions. Both recover the local dark matter and midplane stellar densities within our quoted uncertainties. The situation is different when we consider the evolved stage of the simulation. The onset of spiral arms and a bar causes significant radial mixing that induces vertical non-isotropy and non-separability that violate key assumptions in the HF method. As such, we might expect its performance to degrade accordingly. 

\begin{figure}
\center
\includegraphics[width=0.5\textwidth]{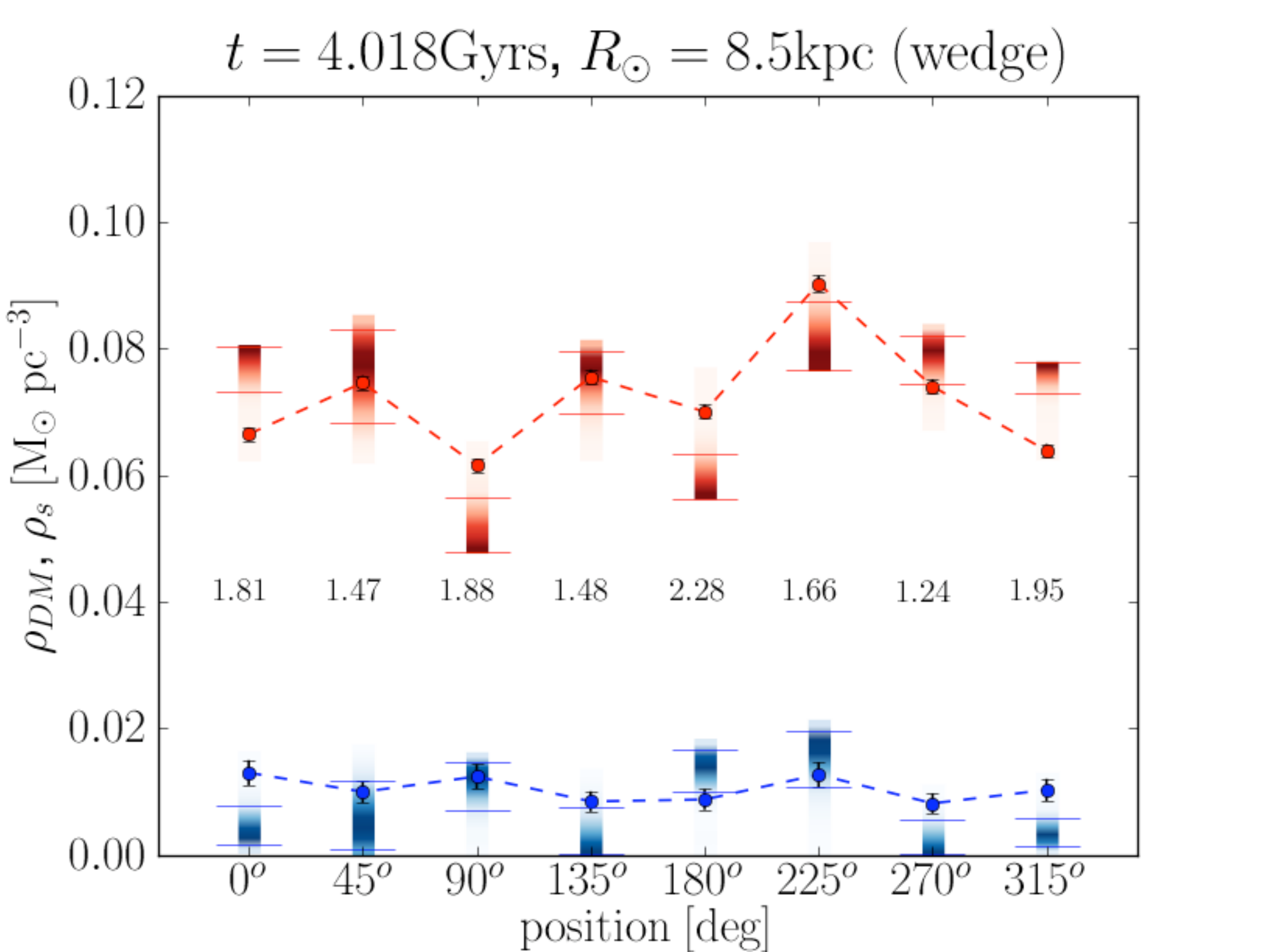}
\includegraphics[width=0.5\textwidth]{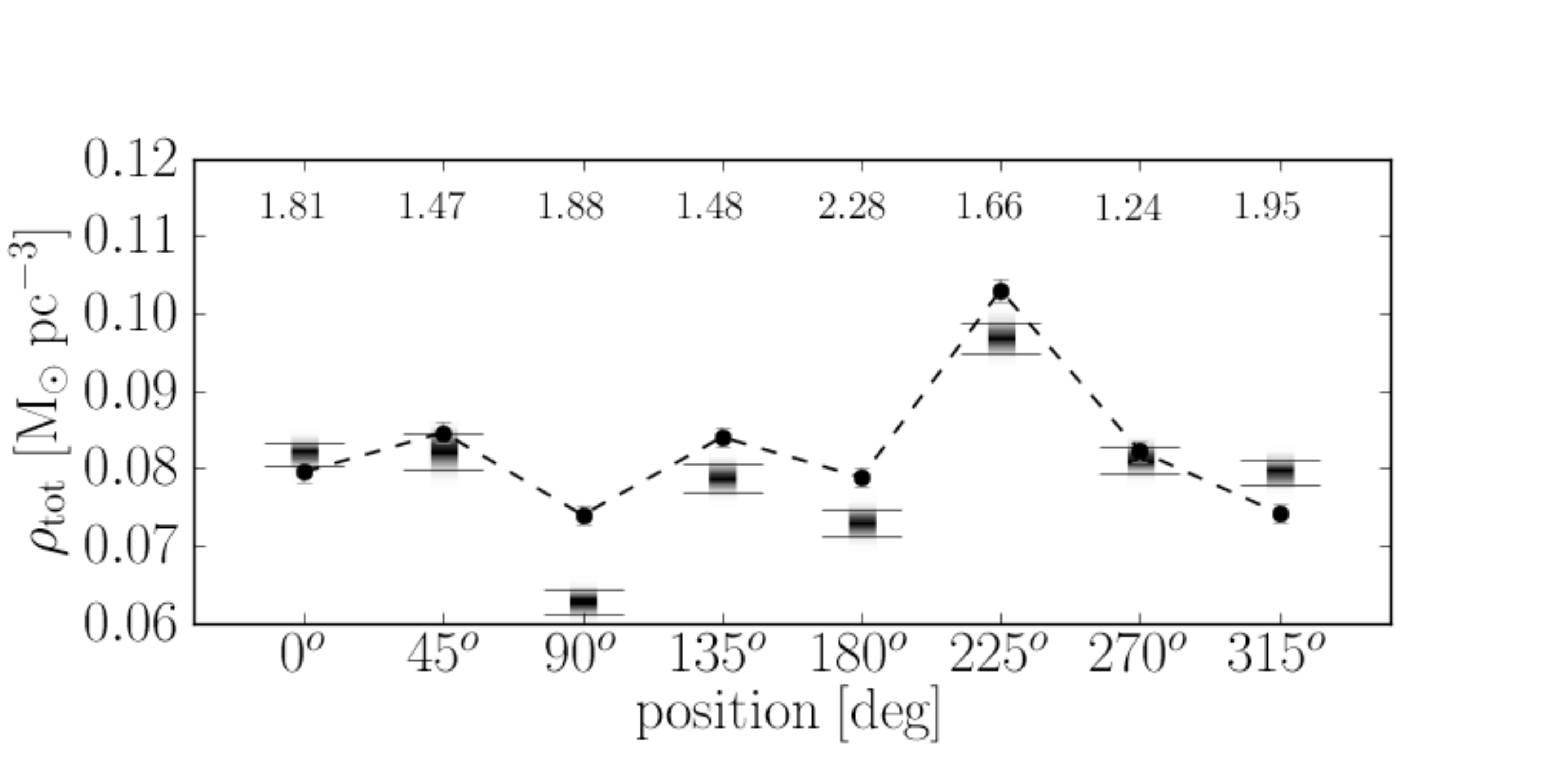}
\caption{Models explored by the MCMC for the HF method assuming isothermality of the disc population and using a double Gaussian fit of the velocity distribution for the 8 wedge-shaped ÔSolar neighbourhoodÕ volumes at $R=8.5$\,kpc. Upper panel: recovered dark and visible matter density (the symbols and colours are as in Figure \ref{fig:stat}). Lower panel: recovered total (dark+visible) matter density. The numbers under each stellar density are the reduced $\chi^2$ for the best-fitting model.}
\label{fig:820int_nonisot}
\end{figure}

We consider 8 different wedges\footnote{In order not to confuse sampling errors with systematic errors, we show the results for the evolved simulation only for the wedges. The results for the MA method applied to the cylinders are given in Appendix \ref{app:cyl}.} around the evolved disc  to sample patches that lie on/away from spiral/bar features. We first apply the HF method, assuming an isothermal disc mass model. The results are shown in Figure \ref{fig:820int_nonisot} (upper panel). As expected, we do not recover the correct value of the local stellar and dark matter densities for most of the volumes. The possible reasons are: the neglected non-isothermality of the disc; the unsatisfactory fit of the distribution function with a double Gaussian (at least for some of the volumes considered); and, at this stage of the simulation, that the distribution function of the stars above the plane is not well represented by the distribution in the midplane.

To test the first two possible sources of error, we correct for the non-isothermality of the disc population using equation \ref{new_rhodisc} instead of \ref{rhodisc}, and we interpolate linearly the distribution function instead of fitting it. The results are very similar; the reason for such a small change is that it is the non-isothermality of the {\it tracers} that really matters, not that of the whole disc model. (Recall that the HF method does not assume that the tracers are isothermal, but rather that their distribution function is a function only of the vertical energy $E_z$). Thus we can conclude that it is the assumption that $f=f(E_z)$ that leads to the systematic bias in the recovery of $\rhodm$ and $\rhos$ for the HF method applied to the evolved simulation. To see this, consider the wedges at $\theta=45^\circ$ and $\theta=180^\circ$. Recall from Section \ref{test-hyp} that for the former wedge, the velocity distribution at $z=0.5$\,kpc was well predicted from $f(w_0)$, while for the latter wedge the velocity distributions differed strongly. As might be expected, the $\theta=45^\circ$ gives an excellent recovery for $\rhodm$ and $\rhos$, while the $\theta=180^\circ$ wedge gives a very poor recovery. In the lower panel of Figure \ref{fig:820int_nonisot} the recovered total (visible+dark) matter density is shown: the HF method fails to recover the correct answer in many cases, even dramatically (e.g. see $\theta=90^\circ$ or $\theta=315^\circ$).

The above is a problem for the HF method -- and indeed any method that assumes that $f = f(E_z)$ -- if such methods are applied at heights larger than $\sim 1$ disc scale height. However, going to this height is necessary to break the degeneracy between $\rhodm$ and $\rhos$ (Section \ref{dege}). It may be possible to build an unbiased distribution function (or mixed) method that works at large height above the disc plane, by using more complex forms for $f$. This is beyond the scope of this present work.

\paragraph{The MA method} 

\begin{figure}
\center
\includegraphics[width=0.5\textwidth]{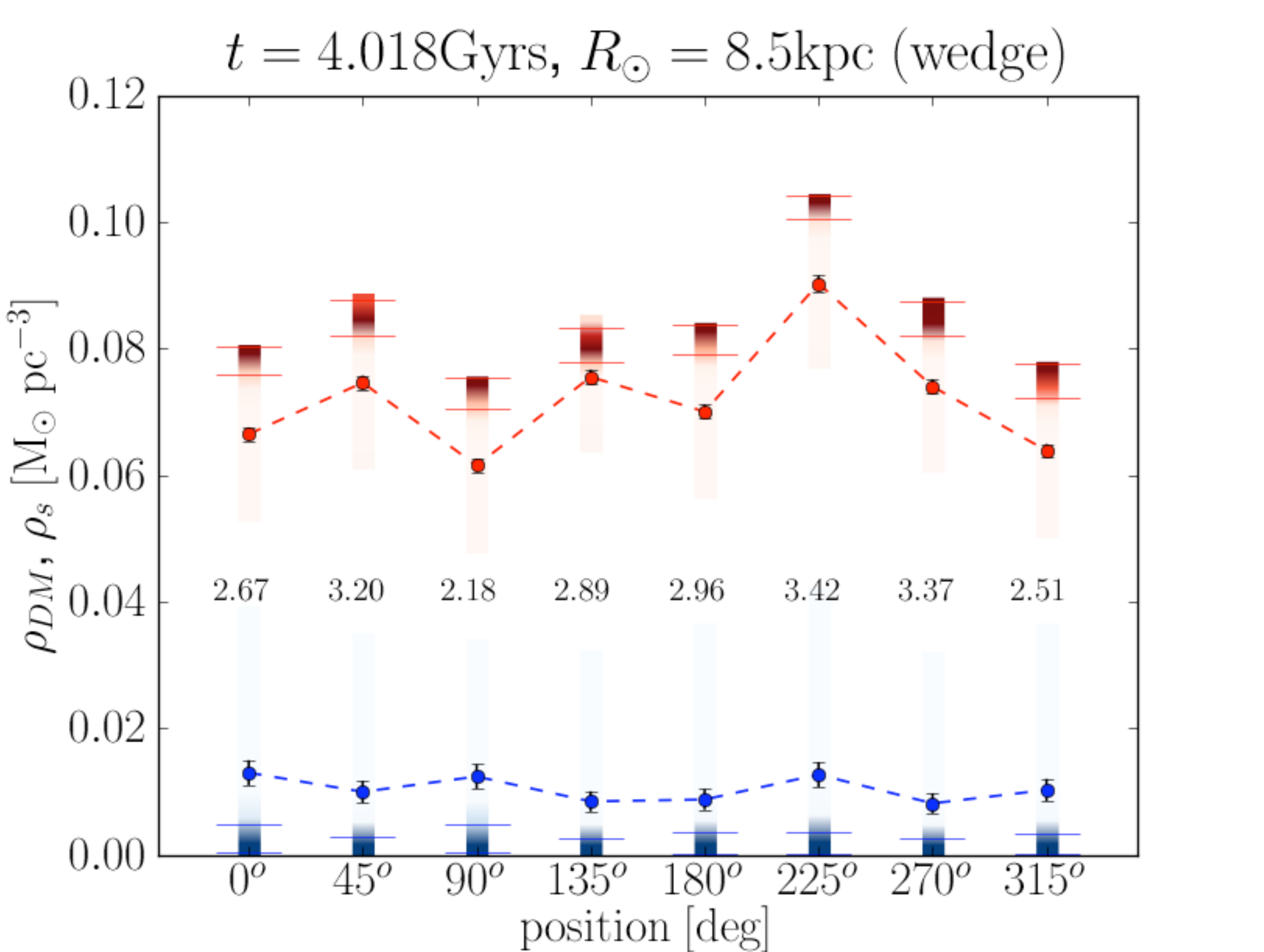}
\includegraphics[width=0.5\textwidth]{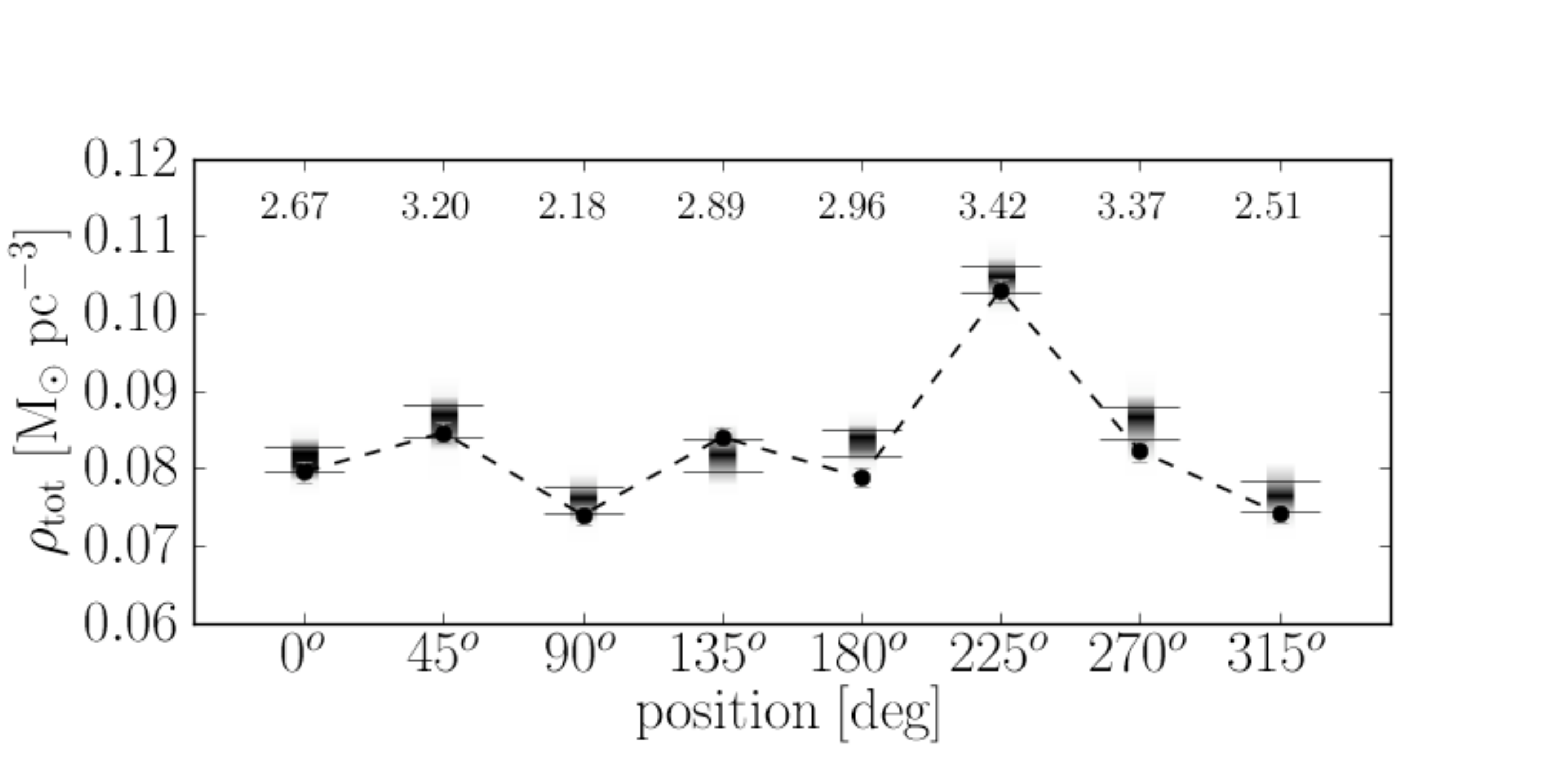}
\caption{Models explored by the MCMC for the MA method assuming isothermality for 8 wedge-shaped ÔSolar neighbourhoodÕ volumes at $R=8.5$\,kpc. The symbols and colours are as in Figure \ref{fig:820int_nonisot}. The numbers under each stellar density are the reduced $\chi^2$ for the best-fitting model.}
\label{fig:820int}
\end{figure}

We first apply the MA method assuming isothermality of the tracers to the 8 wedges. The results are shown in Figure \ref{fig:820int}. Notice that, similar to the HF method, the density recovery in all of the wedges is systematically biased and poor. The MCMC explores a very small area in the $\rhos$-$\rhodm$ parameter space, always pushing on the lower limit imposed for $\rhodm$. The error in this case has a particular direction: this probably owes to the deviation from zero of the sum of the second and third terms of the Jeans equation (represented as a grey line in Figure \ref{jeans_terms}). When we assume isothermality, this has a particular sign.

\begin{figure}
\center
\includegraphics[width=0.5\textwidth]{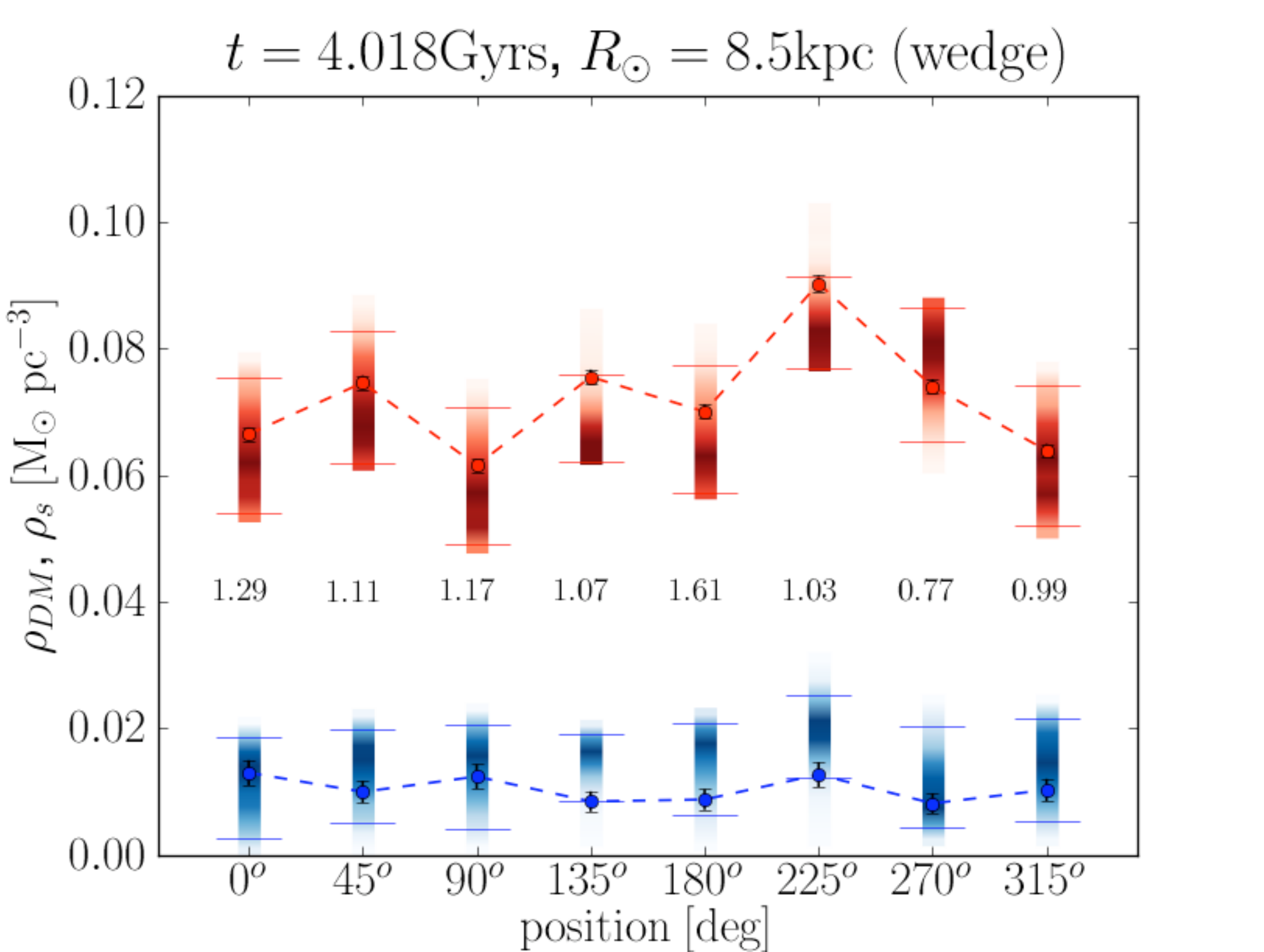}
\includegraphics[width=0.5\textwidth]{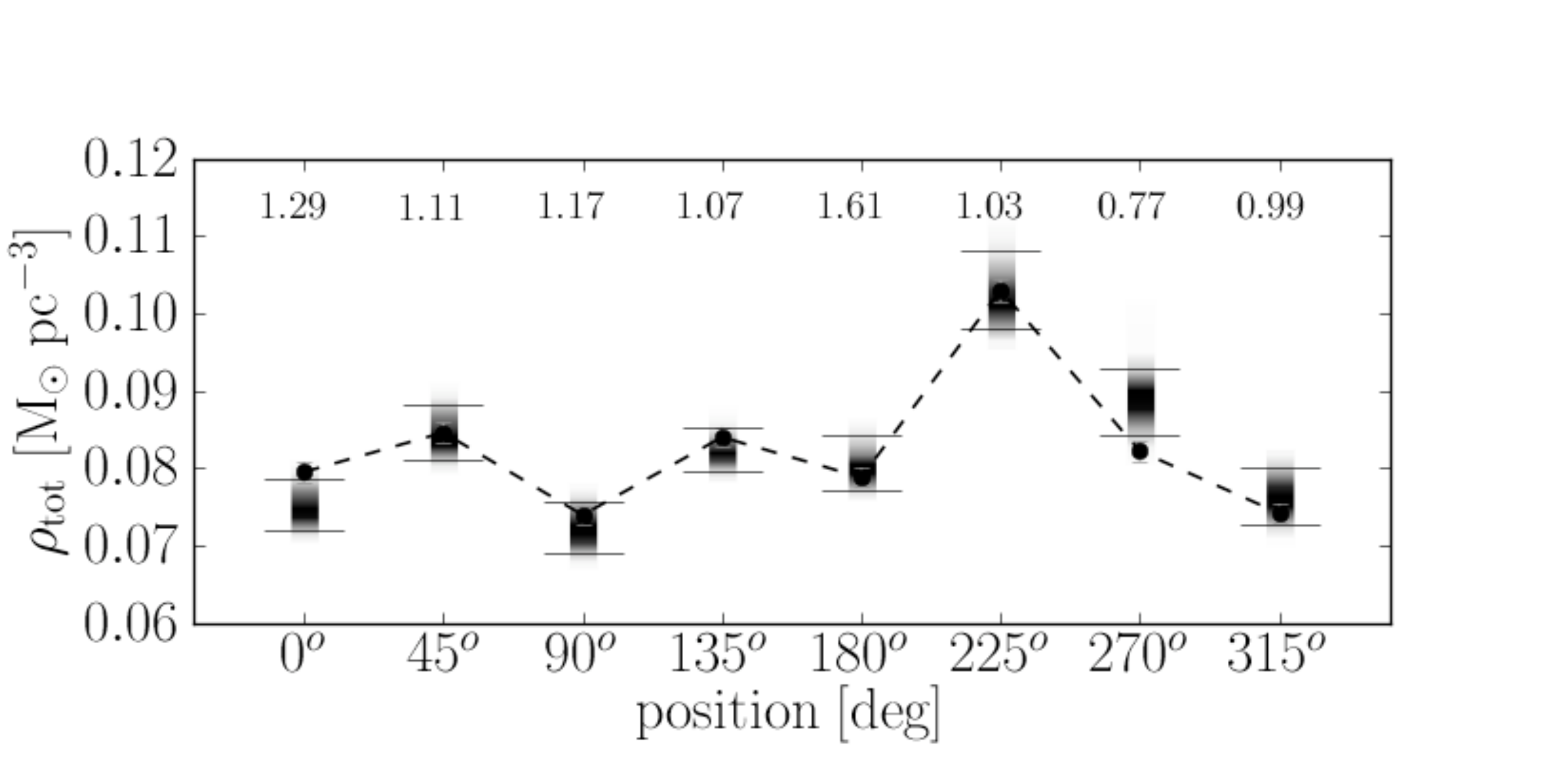}
\caption{Models explored by the MCMC for the MA method for 8 wedge-shaped ÔSolar neighbourhoodÕ volumes of the evolved simulation at $R=8.5$\,kpc. Upper panel: recovered values of dark and visible matter density. Lower panel: recovered values of the total (dark+visible) matter density. The symbols and colours are as in Figure \ref{fig:820int}.}
\label{fig:moment-met}
\end{figure}

Next, we include the non-isothermality of our tracers. The results are shown in Figure \ref{fig:moment-met}. Our results are now excellent for all patches, recovering the correct unbiased value for both $\rhodm$ and $\rhos$ (and the total matter density) within our quoted 90 per cent uncertainties. This emphasises the importance of knowing $\vztwoi(z)$ precisely for each tracer population. In fact, a small deviation from the actual velocity dispersion of the tracers is enough to lead to a wrong result; for this reason we linearly interpolate $\vztwoi(z)$. Note that this is possible for the simulation if we consider large enough wedges, so that the velocity dispersion is a quite smooth. For real data the situation is more complicated since we have to deal with velocity uncertainties and noisier velocity dispersions. In this case, we can use the MCMC to marginalise over such uncertainties. We demonstrate this for the evolved simulation in Appendix \ref{app:cyl}.

\begin{figure}
\center
\includegraphics[width=0.5\textwidth]{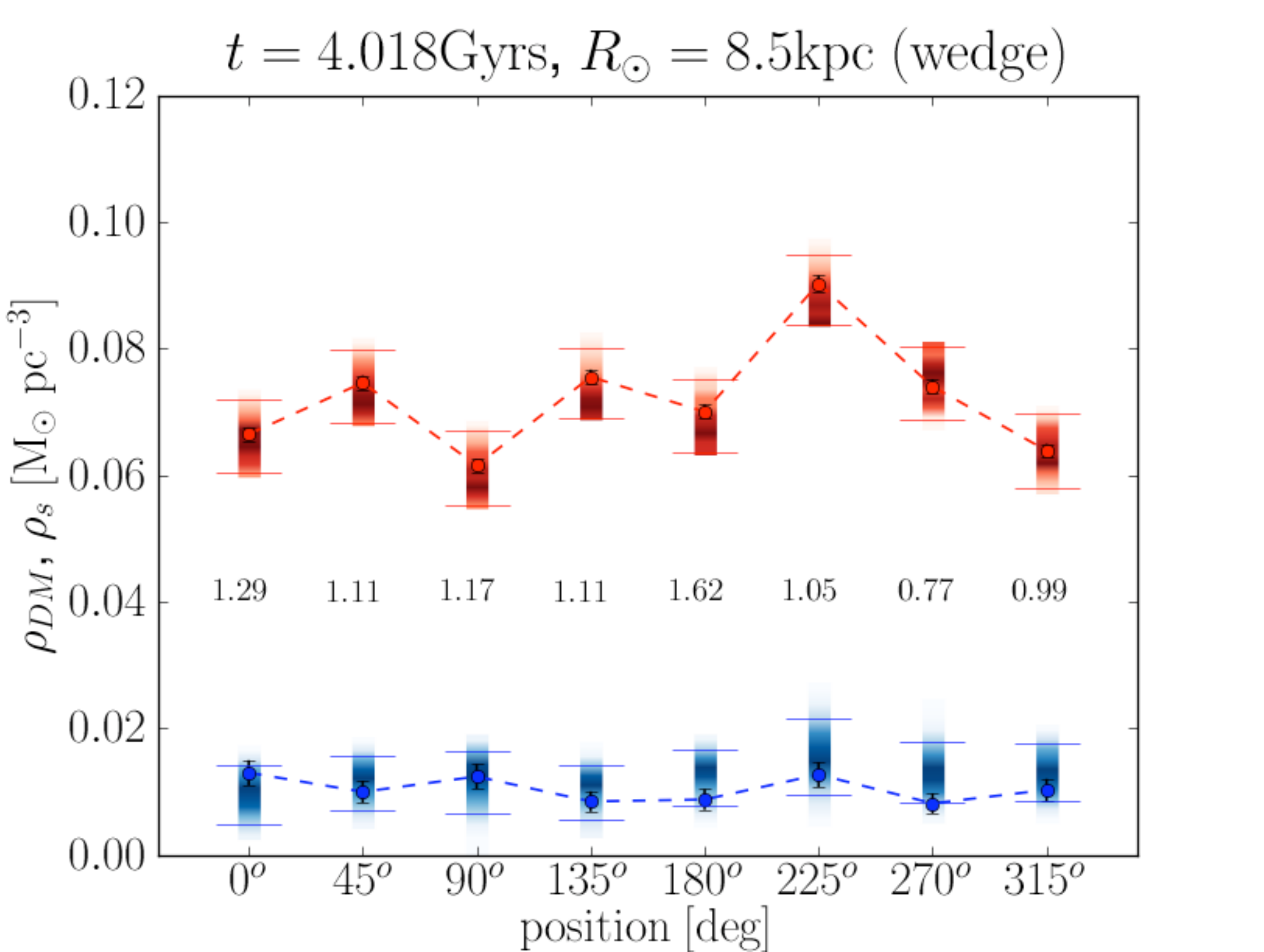}
\includegraphics[width=0.5\textwidth]{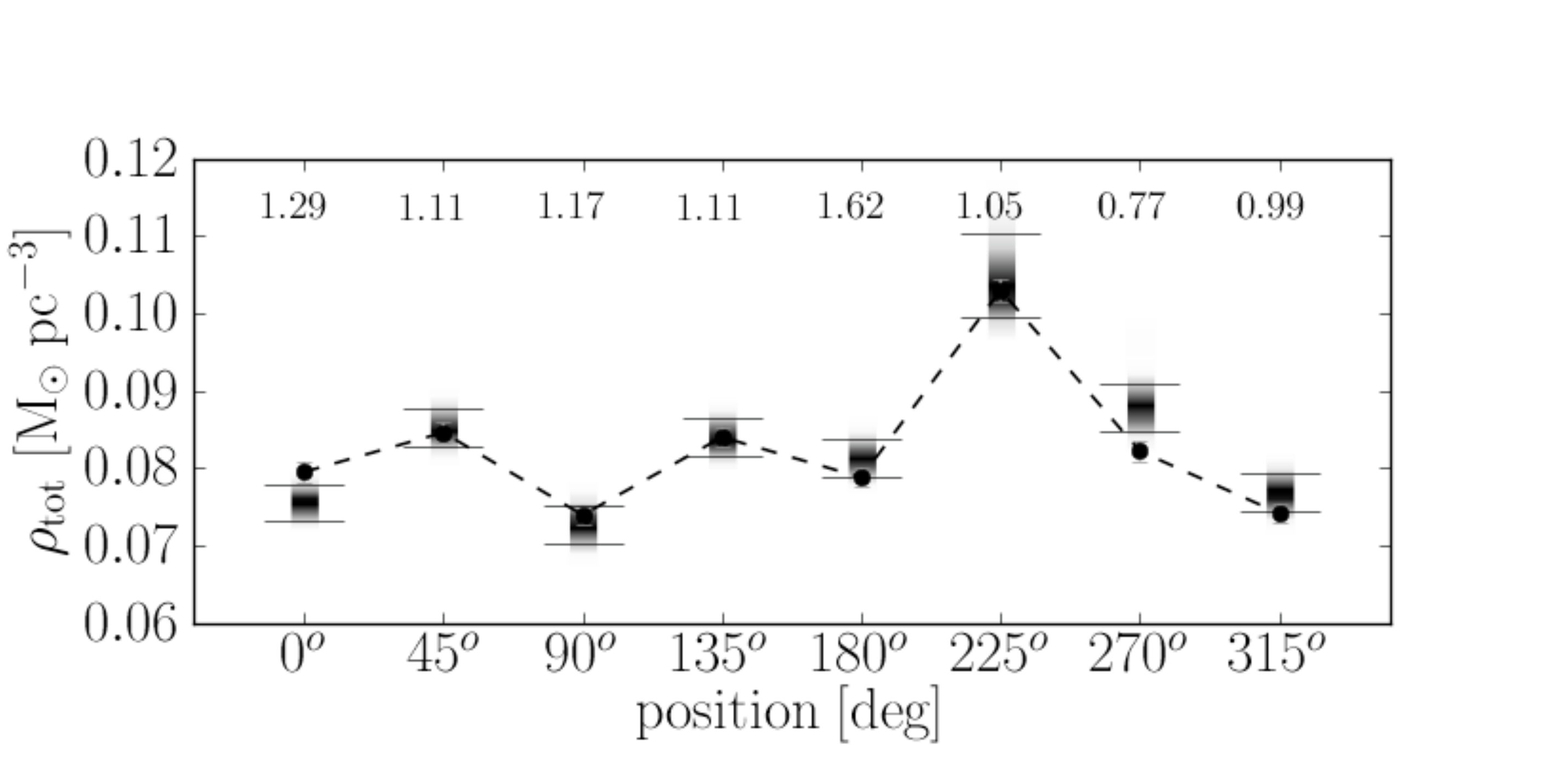}
\caption{Models explored by the MCMC for the MA method for 8 wedge-shaped ÔSolar neighbourhoodÕ volumes of the evolved simulation at $R=8.5$\,kpc. In this case tighter constraints on $\rhos$ are assumed (an error of $\pm0.007$\msun\,pc$^{-3}$ instead of $\pm0.014$\msun\,pc$^{-3}$). The symbols and colours are as in Figure \ref{fig:820int}.}
\label{fig:moment-met+constr}
\end{figure}

Note, however, that the errors are still large even though the relative amount of dark matter in the simulation is larger than we expect in the Milky Way. We can further improve on this if the errors on $\rhos(0)$ can be reduced. We explore this in Figure \ref{fig:moment-met+constr} where we assume that $\rhos(0)$ is known to an accuracy of $\pm0.007$\msun\,pc$^{-3}$ instead of $\pm0.014$\msun\,pc$^{-3}$ as previously assumed. The results are correspondingly improved, as expected. This suggests that the key limiting factors to determining $\rhodm$ are a good measure of the non-isothermality of the tracer population, and an accurate determination of the local visible matter density. 

\section{Application to real data}\label{hf-data}

In this section, we illustrate the power of our new minimal assumption (MA) method by applying it to the {\it Hipparcos} data used by \cite{holmberg_local_2004} to calculate the local surface density up to $z=0.7$\,kpc. As we demonstrated in Section \ref{dege}, fitting the density fall-off up to large $z$ is required to break the degeneracy between $\rhos$ and $\rhodm$.

\subsection{The data} 

We use the raw data of the `HD sample' \cite{holmberg_local_2004} from Chris Flynn (private communication) consisting of 139 K-giants from \cite{flynn_catalog_1993}'s catalogue in a cone pointing towards the South Galactic Pole with an aperture of 430deg$^2$, having a limiting visual magnitude of $V=9.2$, a magnitude range of $0.0<M_V<2.0$ and a colour range of $1.0<B-V<1.5$ (see figure 11, upper panel in \cite{holmberg_local_2004}). \cite{holmberg_local_2004} compute the velocity distribution of the tracers using a volume complete (to 100\,pc) sample of 395 K-stars from the {\it Hipparcos} catalogue with radial velocity information (in the same colour and absolute magnitude ranges). Because of the nature of those data, the analysis is more complicated and uses the {\it Hipparcos} luminosity function for K-giants (figure 2 in \cite{holmberg_local_2004}). A further complication as compared to our simulation data is the mass model for the real Milky Way which has several gas and stellar components, each with its local density and velocity dispersion. The density in the midplane $\nu_{i,0}$ and the velocity dispersion $\vztwoi(0)$ of the various visible components \cite{flynn_mass--light_2006} are listed in Table \ref{mmodel}.

The HD sample contains very few stars, so we also include additional constraints from the literature. This illustrates the power of our MA technique coupled to the MCMC since additional constraints are straightforward to add. As additional data, we include the two volume complete samples of stars from {\it Hipparcos} data employed by \cite{holmberg_local_2000} in their calculation of the local density: the $A$ star sample (including B5 to A5 stars) which contains 2026 stars in a cylinder with radius and height of 200\,pc, and the $F$ sample (A0 to F5) which comprises 3080 stars within 100\,pc. We also ensure that the surface density calculated for each model explored by the MCMC agrees with the observational constraints. In the second column of Table \ref{mmodel}, the current observational constraints for the surface densities of the different visible components are listed. From the values in this table, we adopt a total visible surface density for the disc of $\Sigma_{vis}$(R$_\sun$)=$49.4\pm 4.6$\,M$_\sun$ pc$^{-2}$. For each model explored by the MCMC we then calculate the expected surface density as:
\beq
\Sigma^\mathrm{exp}_\mathrm{s}=2\int_0^\infty \rhos(z)dz=2\int_0^\infty \sum_i\nu_{i,0}\exp\left(-\frac{\Phi(z)}{\vztwoi}\right)dz ,
\eeq
where $\Phi(z)$ is the potential computed according to the parameters of the model. We then compare this with $\Sigma_{vis}$(R$_\sun$), including the result in our determination of the $\chi^2$ for each model. 

\begin{table}
\center
\caption{The disc mass model taken from Flynn et al. 2006. Each component in the table gives the local mass density in the midplane $\rho(0)$ in \msun pc$^{-3}$, the total column density $\Sigma$ in \msun pc$^{-2}$, and the vertical velocity dispersion $\vztwoi(0)^{1/2}$ in km\,s$^{-1}$. Uncertainties on the densities are of order $50$ per cent for all the gas components (indicated with $^*$) and $10-20$ per cent for all the stellar components. For the thick disc, the column density is rather well known, while the velocity dispersion and the volume density are poorly known such that they should have larger error bars. However, these two quantities are essentially nuisance parameters for our analysis here. Since they anti-correlate and -- as pointed out by \protect\cite{kuijken_mass_1989} -- the local gravitational potential is mainly constrained by the column density, we simply assume small errors for both here such that the integrated column agrees with the observed value.}
\label{mmodel}
\begin{tabular}{|c|c|c|c|}
\hline
Component &$\nu_{i,0}(0)$& $\Sigma_i$ &  $\vztwoi(0)^{1/2}$ \\
& [\msun\,pc$^{-3}$] & [\msun pc$^{-2}$] & [km\,s$^{-1}$] \\
\hline
H$_2^*$ & 0.021 & 3.0 & $4.0\pm 1.0$ \\
HI(1)$^*$ & 0.016 & 4.1& $7.0\pm 1.0$ \\
HI(2)$^*$  & 0.012  & 4.1& $9.0\pm1.0$ \\
Warm gas$^*$ & 0.0009 & 2.0 & $40.0\pm1.0$\\
Giants & 0.0006  & 0.4 & $20.0\pm 2.0$ \\
$M_V<2.5$ & 0.0031 &0.9 &$7.5\pm 2.0$ \\
$2.5 < M_V < 3.0$ &  0.0015  & 0.6 & $10.5\pm 2.0$ \\
$3.0 < M_V < 4.0$ &  0.0020  & 1.1 & $14.0\pm 2.0$  \\
$4.0 < M_V < 5.0$ &  0.0022 & 1.7 & $18.0\pm 2.0$  \\
$5.0 < M_V < 8.0$ &  0.007 & 5.7 & $18.5\pm 2.0$\\
$M_V > 8.0$ & 0.0135  & 10.9&$18.5\pm 2.0$ \\
White dwarfs &	0.006  & 5.4&$20.0\pm 5.0$ \\
Brown dwarfs & 0.002 & 1.8&$20.0\pm 5.0$ \\
Thick disc& 0.0035  & 7.0&$37.0\pm 5.0$  \\
Stellar halo & 0.0001 & 0.6&$100.0\pm 10.0$  \\
\hline
\end{tabular}
\end{table}

As parameters to fit in the MCMC, we use the local dark matter density $\rhodm$; the total visible density in the midplane $\rhos(0)$; the relative fractions of the visible components $\nu_{i,0}/\rhos(0)$; their velocity dispersions in the midplane $\vztwoi(0)^{1/2}$; the velocity dispersion as a function of $z$ of the tracers; and the normalisation of the density fall-off of the tracers. We allow the densities and the velocity dispersions of the different components to vary within their measured uncertainties (the errors for each component are given in Table \ref{mmodel}). We let the total visible density in the plane $\rhos(0)$ vary within its observed range: $\rhos(0)=0.0914\pm0.0140$\msun\,pc$^{-3}$; and we let the dark matter density vary between 0 and 0.5\msun\,pc$^{-3}$. The velocity dispersion of the tracers in the midplane is given by the Gaussian fit of the velocity distribution calculated by \cite{holmberg_local_2004}, namely $\vztwoi(0)^{1/2}=18.3\pm0.6$\,km\,s$^{-1}$ for the HD sample, and by \cite{holmberg_local_2000}, i.e. $\vztwoi(0)^{1/2}=5.7\pm0.2$\,km\,s$^{-1}$ for the A sample and $\vztwoi(0)^{1/2}=8.3\pm0.3$\,km\,s$^{-1}$ for the F sample.

After computing the expected density fall-off for the tracers of the (magnitude limited) HD sample through (\ref{new_nu}), we apply the {\it Hipparcos} luminosity function and the magnitude cut $V<9.2$, to compare it to the observed number of stars in the cone. The $A$ and the $F$ samples from \citep{holmberg_local_2000} are easier to fit, since they are volume complete.

\begin{figure}
\center
\includegraphics[width=0.45\textwidth]{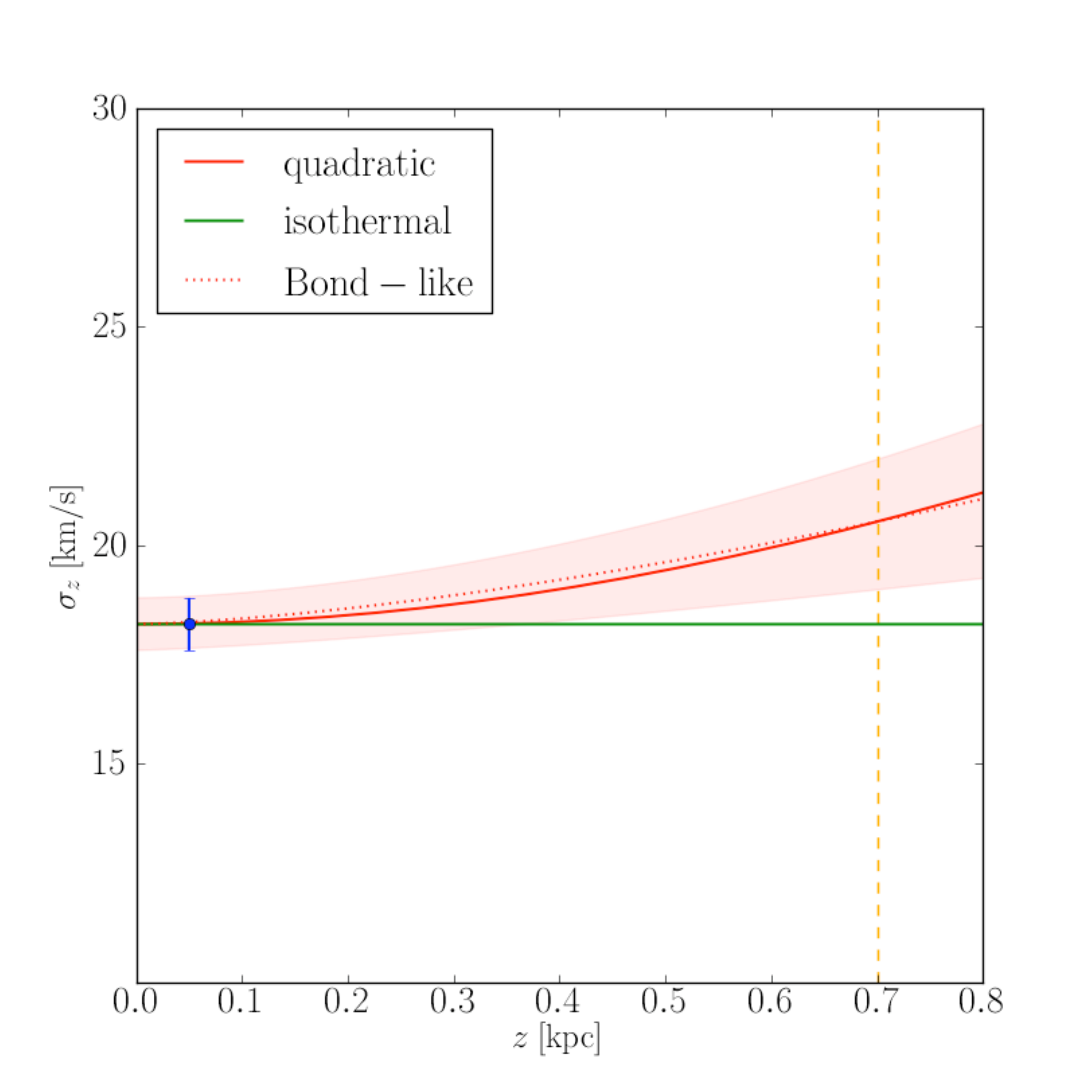}
\caption{Bond-like (dotted red line), quadratic (red solid line) and isothermal (solid green line) velocity dispersion functions for the `HD sample'. The shaded red area represents the errors in the Bond-like function. The blue dot represents the measured velocity dispersion in the local sample ($|z|<100$\,pc) and the dashed orange line is at $z=0.7$\,kpc (the upper $z$-limit for the HD sample).}
\label{fig:bondHF}
\end{figure}

Unfortunately, we do not have much information about the velocity dispersion above the plane of the different disc components included in the mass model. As such, we consider two extreme assumptions: one in which all of the visible components of the disc and the tracers are isothermal; and another in which the tracers and  all of the visible components of the disc are non-isothermal. We model the non-isothermality of the stars in this second case assuming a behaviour similar to the fit by \cite{bond_milky_2009} to blue disc stars. We proceed in the following way:\begin{enumerate}
\item We use the velocity dispersion in the plane of each component $\vztwoi(0)^{1/2}$ with its error bar (see Table \ref{mmodel}) and the constants $A=4\pm 0.8$ and $b=1.5\pm 0.2$ calculated by Bond to compute the velocity dispersion for that particular species at the maximum fitted height $z_\mathrm{max}$ with the help of Bond's fitting function:
\beq
\vztwoi(z_\mathrm{max})^{1/2}=\vztwoi(0)^{1/2}+A|z_\mathrm{max}/\mathrm{kpc}|^b.
\eeq
\item Since the function fitted by Bond is discontinuous in $z=0$, we use a quadratic function:
\beq
\vztwoi(z)=\vztwoi(0)(1+C|z|^2),\label{fit-vdisp}
\eeq
We choose the parameter $C$ of equation (\ref{fit-vdisp}) so that the quadratic pass through $\vztwoi(0)$ and the value of $\vztwo(z_\mathrm{max})$.
\end{enumerate}
In Figure \ref{fig:bondHF}, the quadratic curve (red solid line) and the Bond-like fit (red dotted line) for the HD tracers are shown. The shaded red area represent the uncertainties on the Bond's fit due to the errors in $A$ and $b$ calculated by \cite{bond_milky_2009} and the uncertainties in $\vztwoi(0)^{1/2}$ (blue point). Notice that the quadratic function obtained is very close to the Bond's fit and lies inside its quoted uncertainties.

We stress that the velocity dispersion law from \cite{bond_milky_2009} refers to different types of stars that are hotter than the A, F and K stars we consider here. However, recall that our goal is simply to explore the effect of varying the functional form of $\vztwoi$.

To summarise, our approach is as follows: (i) we use the mass model of table \ref{mmodel} (with a constant dark matter contribution) to calculate the potential; (ii) we use this potential and an isothermal/Bond-like velocity dispersion law (separately normalised for each tracer population) to predict the density fall-off of the three tracer populations; (iii) we simultaneously predict the total visible surface density; (iv) from the comparison of the three predicted and observed density laws (and the predicted and observed visible surface density) we accept or discard the initial guess for the potential at each iteration of the MCMC.

\begin{figure*}
\center
\includegraphics[width=0.45\textwidth]{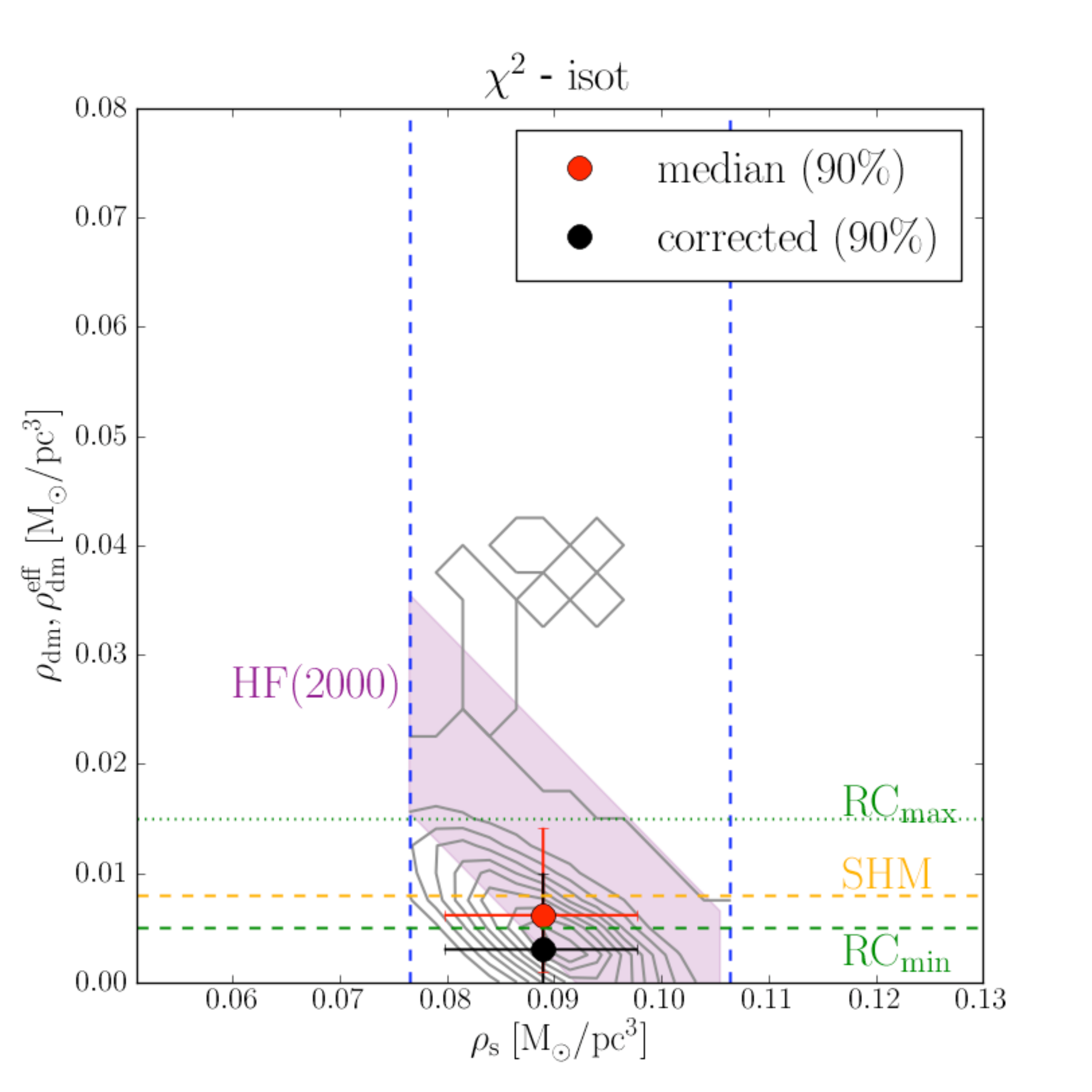}
\includegraphics[width=0.45\textwidth]{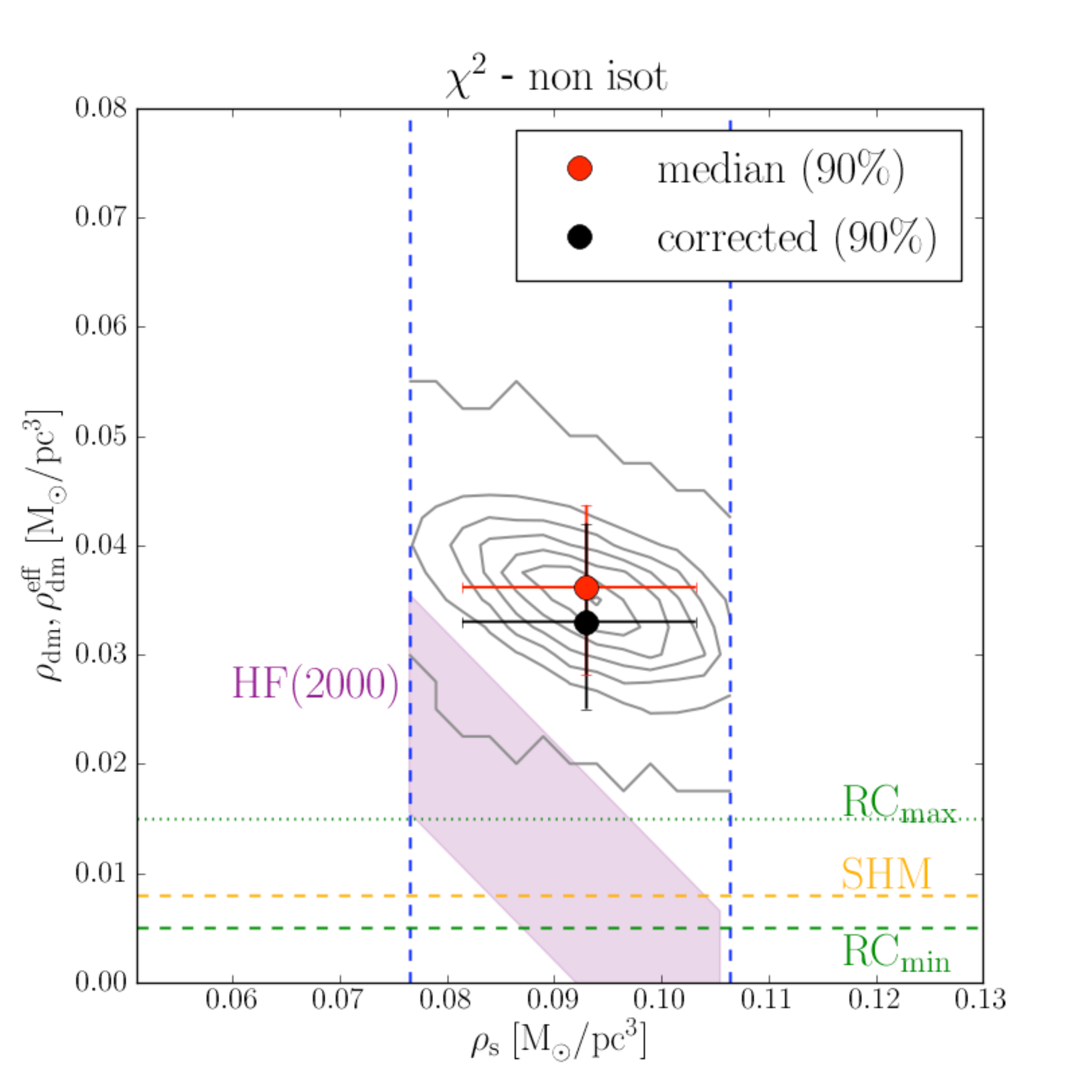}
\caption{The recovered visible and dark matter density calculated with the MA method, assuming isothermality (left panel) and non-isothermality of all the stellar populations (right panel) for the real data. The grey contours represent the density of models explored by the MCMC, the red dot represents the median values of $\rhos$ and $\rhoeff$ (see equation \ref{dmeff}); the red error bars correspond to the $90$ per cent confidence interval of the distribution. The black dot is the result corrected for the rotation curve term calculated from the Oort contants (see Section \ref{sec:rotcurve})}. The purple area represents the values estimated by Holmberg and Flynn (2000). The blue dashed lines show the imposed priors on $\rhos$ and $\rhodm$. The green and the yellow lines represent the minimum value and the maximum value of $\rhodm$ measured using rotation curves in the literature and the SHM value, respectively.
\label{fig:isot-nonisot}
\end{figure*}

The application of the MA method assuming isothermal or Bond-like velocity dispersion profiles leads to very different results for the recovered visible and dark matter density, but with a very similar value of $\chi^2$. The results are given in Figure \ref{fig:isot-nonisot}. The recovered visible and dark matter density calculated with the MA method assuming isothermality (upper panel) and a Bond-like non-isothermality (lower panel) are shown. The red dot represents the median of the distribution of the models explored by the MCMC in the $\rhos$-$\rhodm$ plane within a $90$ per cent confidence interval. The blue dashed lines correspond to the priors imposed on $\rhos$; the purple stripe shows the result by \cite{holmberg_local_2000}; and the green and yellow horizontal dashed lines represent the lower limit of the local dark matter density ($\simeq 0.005$\msun\,pc$^{-3}$) as extrapolated from the Milky Way's rotation curve (a summary of these values is given in Table \ref{dmden}) and the `Standard Halo Model' (SHM) canonical value ($\simeq 0.008$\msun\,pc$^{-3}$), respectively.

If all of the stellar tracers are assumed to be isothermal, we obtain a fit similar to \cite{holmberg_local_2000} with a dark matter density of $0.006^{+0.008}_{-0.005}$\msun\,pc$^{-3}$. By contrast, if we assume instead a `Bond-like' non-isothermality for the stellar populations in the disc, the recovered dark matter density is much larger ($0.036^{+0.007}_{-0.008}$\msun\,pc$^{-3}$); the measured local dark matter densities, corrected for the rotation curve using the Oort constants (see section \ref{sec:rotcurve}), are: $0.003^{+0.009}_{-0.007}$\msun\,pc$^{-3}$ (for the isothermal tracers) and $0.033^{+0.008}_{-0.009}$\msun\,pc$^{-3}$ (for non-isothermal tracers). They are represented as black dots in Figure \ref{fig:isot-nonisot}. 

\begin{figure*}
\center
\includegraphics[width=0.45\textwidth]{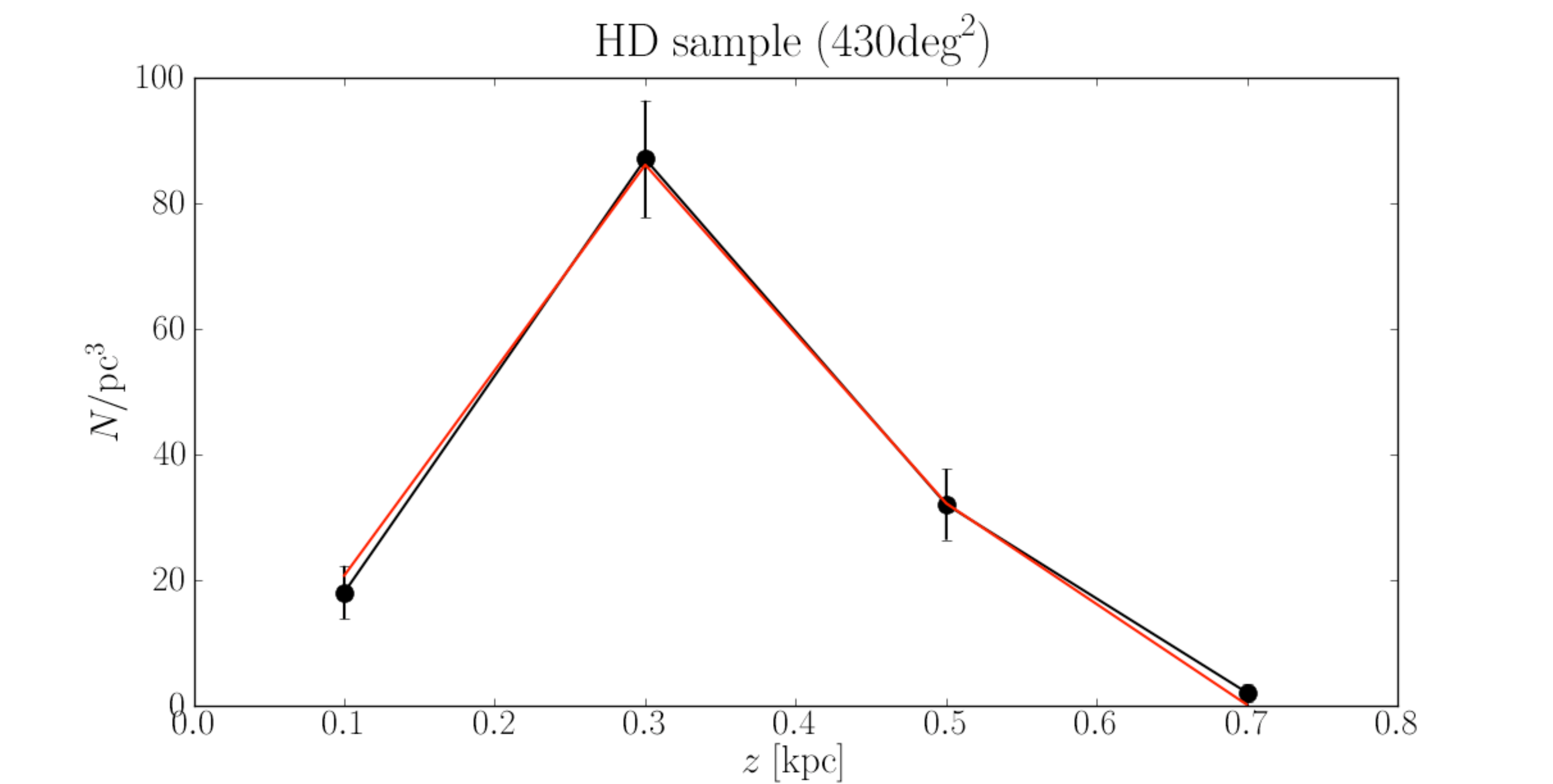}
\includegraphics[width=0.45\textwidth]{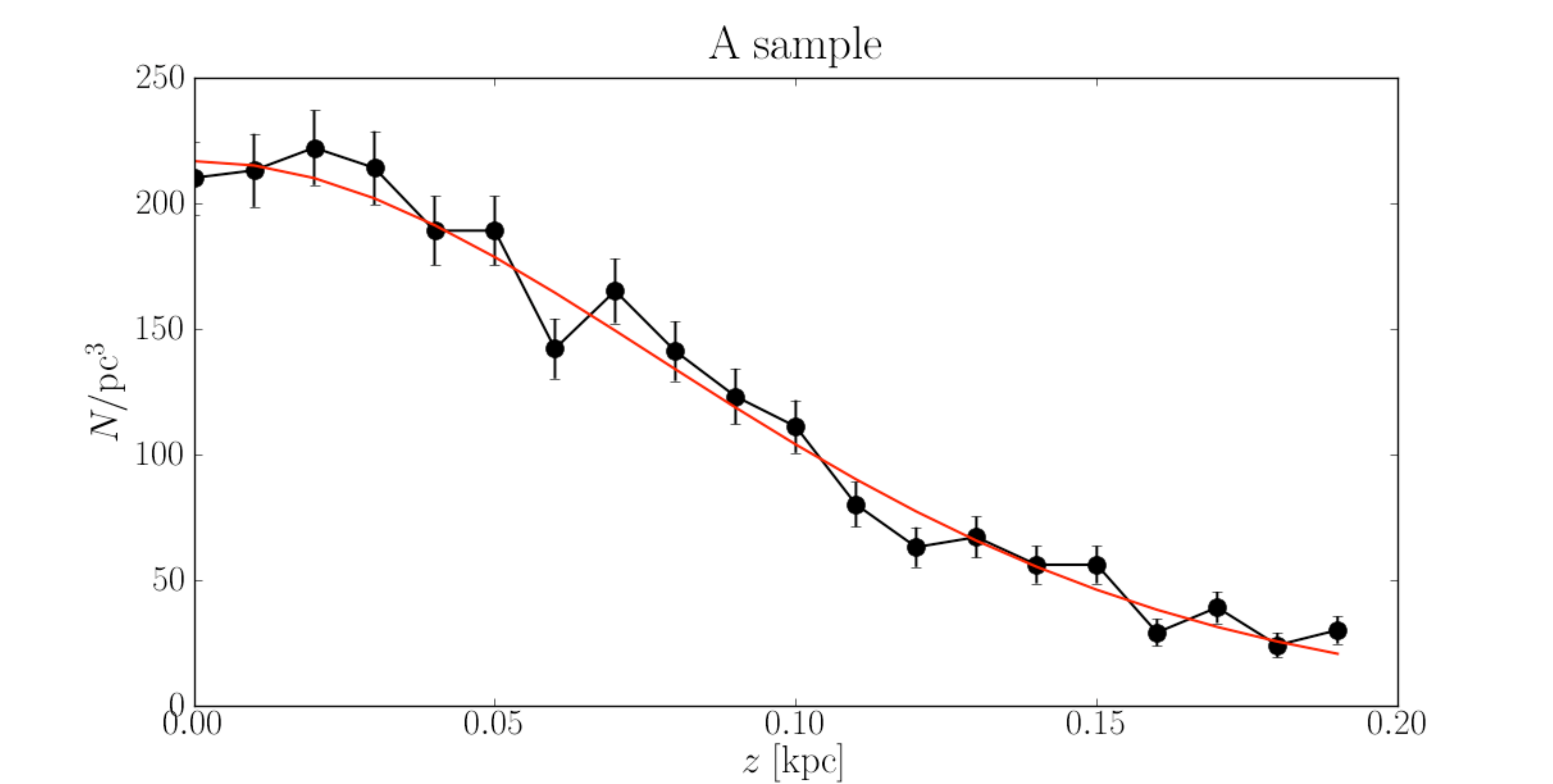}
\includegraphics[width=0.45\textwidth]{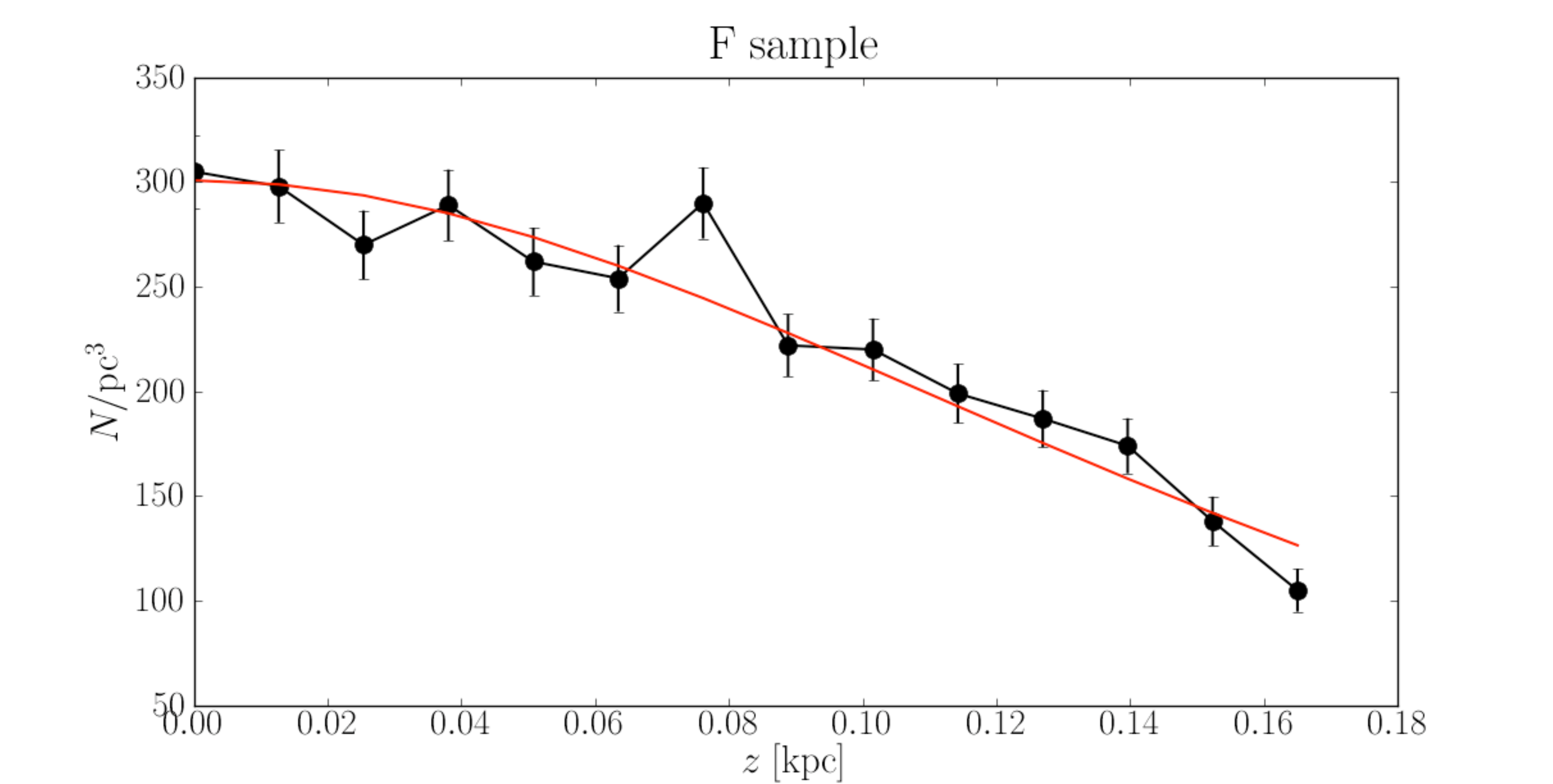}
\caption{The recovered density fall-off for the three tracers considered, assuming isothermality of all the disc populations: the HD sample from \protect\cite{holmberg_local_2004} (first panel); and the A and F star samples from \protect\cite{holmberg_local_2000} (second and third panels). Similarly good fits were obtained for the maximally non-isothermal model.}
\label{fig:hf-fit}
\end{figure*}

Yet the (non-reduced) $\chi^2$ for both models is comparable: $\chi^2=41.5$ for the fully isothermal model and $\chi^2=42.3$ for the non-isothermal model. This means that, for the data we consider here, we cannot discriminate between these two scenarios. Note that our $\chi^2$ values seem rather high (similar to those for the model fits in \cite{holmberg_local_2000}). The number of fitted parameters is 38, using 39 data points and two additional constraints (the total visible density and the surface density). This latter constraint is non-linear and so we cannot simply compute a reduced $\chi^2$. However, assuming that this constraint enters linearly, this gives a remaining 3 degrees of freedom and a reduced $\chi^2$ of $13.8$ for the isothermal model and $14.1$ for the non-isothermal model. This is still high, suggesting that our models are a poor representation of the data, despite the apparent goodness of the fits (shown in Figure \ref{fig:hf-fit}). The reason for this is that our method leads by construction to a smooth density fall-off which cannot account for the (statistically significant) wiggles present in the analysed samples. 

Finally, we repeated our analysis using the isothermal mass model of Table \ref{mmodel}, but still assuming a Bond-like non-isothermal velocity dispersion for the tracers. We found that the result was almost unchanged. This means that the method is very sensitive to the velocity dispersion of the tracer population that must be known accurately. However, the visible components of the mass model are less important. This is not surprising: the velocity dispersion of the tracers enters in equation \ref{new_nu} and thus directly affects the tracer density fall-off. By contrast, the mass-model velocity dispersion profiles appear only in equation \ref{poisson} (through equation \ref{new_rhodisc}), and uncertainties in these profiles are marginalised out when we calculate $\rhodm$ and $\rhos$. Thus, it is vital to obtain an accurate determination of $\vztwoi(z)$ for our tracers, but not crucial to know the precise form of the mass model. 

\begin{table}
\center
\caption{Extrapolated values of the local dark matter density using other methods from the literature. From these, we can place a reasonable lower limit on $\rhodm$ of $0.005$\msun\,pc$^{-3}$ ($\sim0.20$\,GeV\,cm$^{-3}$).}
\label{dmden}
\begin{tabular}{|c|c|c|}
\hline
$\rhodm(R_\sun)$  & Method & Reference\\
$[$\,GeV\,cm$^{-3}]$ & & \\
\hline
$0.519^{+0.021}_{-0.017}$ & microlensing+ & \cite{gates_local_1995}\\ 
 &mass modeling&\\
$0.385\pm0.027$&Bayesan approach +&\cite{catena_novel_2009}\\
 &Einasto profile&\\
$0.364$& rotation curve +& \cite{sofue_unified_2008}\\ 
&spherical halo&\\
$0.20\div0.52^a$& rotation curve +  & \cite{weber_determination_2010}\\
&mass modeling$^b$&\\
\hline
\end{tabular}\\
\small{$^a$ range for the different mass models considered.\\
$^b$ this is a lower limit calculated considering smooth DM halo; substructures can only enhance the local density.}
\end{table}

\section{Conclusions}\label{conclusion}

We have revisited systematic problems in determining the local matter densities from stellar motions. We used a high resolution N-body simulation of a Milky Way like Galaxy to test different methods in the literature and the systematic errors potentially introduced by their assumptions. We introduced a new method -- the minimal assumption (MA) method -- based on moments of the Jeans equations, combined with a Monte Carlo Markov Chain technique to marginalise over the unknown parameters. Given sufficiently good data, we showed that our MA method can recover the correct local dark matter density even in the face of disc inhomogeneities, non-isothermal tracers and non-separability of the $z$-motion. Finally, we illustrated the power of our approach by applying it to {\it Hipparcos} data from the literature. 

Our key results are as follows:
\begin{enumerate}
\item As noted previously by \cite{bahcall_self-consistent_1984}, data up to high $z$ ($|z| \sim 0.6$\,kpc -- i.e. significantly larger than the Milky Way disc scale height) are required to break a degeneracy between the local dark matter density $\rhodm$, and the local visible matter density $\rhos$. 

\item Methods that assume that the distribution function of a tracer population is a function only of the vertical energy $f=f(E_z)$ become systematically biased if the motion of the tracers is not truly separable in $z$. This effect becomes important when fitting to data that extend to heights larger than the disc scale height -- as is necessary to break the $\rhodm$-$\rhos$ degeneracy (c.f. point (i), above). The initial conditions in our simulation were separable, but as the disc evolves and reaches a true equilibrium, the distribution function is no longer separable. If we assume that $f=f(E_z)$, then this introduces a systematic error that we have no way to correct. For this reason, we favour moment based methods that assume nothing about the form of $f$. 

\item We introduced a new minimal assumption (MA) method for recovering the local matter and dark matter densities $\rhotot$ and $\rhodm$. Our method is based on solving the combined Jeans-Poisson equations using an MCMC technique to marginalise over the unknown parameters. We showed that our MA method can correctly recover both $\rhodm$ and $\rhos$ even in the face of disc inhomogeneities, non-separability of the $z$-motion, and vertical non-isothermality of the tracers, provided that the run of dispersion with height of the tracers $\vztwoi(z)$ is known. 

\item Our derived MA method is very sensitive to the precise form of $\vztwoi(z)$ for the tracers. For this reason, we interpolate the measured data (marginalising out any velocity uncertainties), rather than assuming a functional form. By contrast, the form of $\vztwoi(z)$ for the other disc components in the mass model is not important; we may safely assume that these are isothermal.  

\item We applied our new MA method to recent data from \cite{holmberg_local_2000, holmberg_local_2004}. We first made the assumption that the star tracer populations (A, F, K stars) were isothermal. This recovered $\rhodm$ = $0.003^{+0.009}_{-0.007}$\msun\,pc$^{-3}$  ($90$ per cent confidence), consistent with previous determinations. If, however, we assume instead a non-isothermal profile similar to the blue disc stars from SDSS DR-7 \citep{abazajian_seventh_2009} measured by \cite{bond_milky_2009}, we obtain a fit with a very similar $\chi^2$ value, but with $\rhodm$ = $0.033^{+0.008}_{-0.009}$\msun\,pc$^{-3}$ ($90$ per cent confidence). This illustrates the importance of measuring $\vztwoi(z)$ for the tracers. 

\item A combination of good statistics, precise knowledge of the local amount of visible matter, and a good measure of $\vztwoi(z)$ for the tracers is crucial for obtaining an accurate unbiased measure of $\rhotot$ and $\rhodm$. This will become possible with future generation Galactic surveys. 
\end{enumerate}

\appendix
\section{Introduction of dimensionless variables}\label{adim}
In Section \ref{fm-method} and \ref{hf_method} we presented the basic equations used to calculate the potential in the MA and HF method. In this Appendix we re-write these equation (namely equations \ref{new_nu}, \ref{nuform} and \ref{poisson}) using dimensionless variable to simplify the calculations \citep{bahcall_distribution_1984,bahcall_k_1984,bahcall_self-consistent_1984}.

The Poisson equation \ref{poisson} can be rewritten as:
\begin{equation}
\frac{\partial^2\Phi}{\partial z^2}=4\pi G\nu_{0,1}\left(\sum_{i=1}^N\frac{\nu_i(z)}{\nu_{0,1}}+\epsilon\right)
\end{equation}
with $\epsilon=\rhodm/\nu_{0,1}$ ($i=1$ indicates the population with the largest scale height). 

The following dimensionless variables can then be introduced:
\beq
\phi=\frac{\Phi}{\vztwo_{,1}}
\eeq
\beq
z_1=\sqrt{\frac{\vztwo_{,1}}{2\pi G \nu_{0,1}}}
\eeq
\beq
x=z/z_1
\eeq
\beq
\alpha_i=\vztwo_{,1}/\vztwo_{,i}
\eeq
\beq
\xi_i=\nu_{0,i}/\nu_{0,1}
\eeq
\beq
\epsilon=\rhodm/\nu_{0,1}
\eeq
and the solution to equation \ref{boltz} becomes:
\begin{equation}
\nu_i(z)=\nu_{0,i}\exp[-\alpha_i\phi(z)]
\label{nu-adim}
\end{equation}
Using this and the above dimensionless quantities we can write:
\begin{equation}
\frac{d^2\phi}{d x^2}=2\sum_{i=1}^N\xi_i\exp(-\alpha_i\phi)+2\epsilon
\label{fin-adim}
\end{equation}
with $\phi(0)=0$ and $d\phi(0)/dx=0$. For a specified ratio of the mass densities in the plane ($\xi_i$) and the velocity dispersions ($\alpha_i^{1/2}$), equation \ref{fin-adim} can then be integrated numerically for any $\epsilon$.

Finally, for the minimal assumption (MA) method, we must define an additional dimensionless variable:
\beq
\alpha_{i,z}=\vztwo_{,1}(0)/\vztwo_{,i}(z)
\eeq

In this way we can write the solution to equation \ref{boltz} and \ref{poisson} as:
\begin{equation}
\nu_i(z)=\xi_i\frac{\alpha_{i,z}}{\alpha_i}\exp\left(-\int_0^x \alpha_{i,z}\frac{d\phi}{dx}dx\right)
\label{nu-adim2}
\end{equation}
\beq
\frac{d^2\phi}{d x^2}=2\left[\sum_{i=1}^N\xi_i\frac{\alpha_{i,z}}{\alpha_i}\exp\left(-\int_0^x \alpha_{i,z}\frac{d\phi}{dx}dx\right)+\epsilon\right].
\eeq

\section{The SPH analysis method}\label{SPH}

The local density, velocity dispersion and derivatives for the Jeans equation terms are extracted from the simulation using weighted sums over the particles as in Smoothed Particle Hydrodynamics \citep{gingold_smoothed_1977,lucy_numerical_1977,monaghan_smoothed_1992}.

The density is given by: 
\beq
\nu_i = \sum_j^N m_j W(|\urij|,h_i)
\label{eqn:sphcont}
\eeq
where $h_{i}$ and $m_j$ are the smoothing length and mass of particle
$i$ and $j$, respectively; we define $\urij = \uri - \urj$ and similarly for
other vectors; and $W$ is a symmetric kernel that obeys the
normalisation condition:
\beq
\int_{V} W(|\ur-\ur'|,h) \dthr' = 1
\label{eqn:normw}
\eeq
and the property:
\beq
\lim_{h\rightarrow 0} W(|\ur-\ur'|,h) = \delta(|\ur-\ur'|)
\eeq
In the limit $N\rightarrow\infty, h\rightarrow 0$ (and using
$m_j/\nu_j\rightarrow \dthr'$) equation \ref{eqn:sphcont} recovers the
continuum density.

The smoothing lengths $h_i$ were adapted to ensure a fixed enclosed mass $M_\mathrm{SPH} = m N_\mathrm{SPH}$ where $m$ is the mass of a particle and $N_\mathrm{SPH} = 128$ is the neighbour number. We used the standard cubic spline smoothing kernel for $W$ \citep{monaghan_smoothed_1992}.

The velocity dispersion tensor is given by: 
\beq
\sigma_{ab,i} = \frac{1}{\nu_i} \sum_j^N m_j v_{a,j} v_{b,j} W(|\urij|,h_i)
\eeq
where $a,b = [0,1,2]$ give the index of the velocity vector and velocity dispersion tensor, respectively. 

Apart from the gradient of the gravitational potential that was taken directly from the tree (this is just the acceleration), gradients were calculated using a second order accurate polynomial reconstruction at each point in the collisionless fluid, as in \cite{maron_gradient_2003} and references therein. Briefly, assuming that the fluid is smooth (and therefore differentiable), we can perform a polynomial expansion at second order about a point, $i$: 

\begin{eqnarray}
q_{ij}  & = & a_0 + a_1 x_{ij} + a_2 y_{ij} + a_3 z_{ij} + a_4 x_{ij}^2 + a_5 y_{ij}^2 + a_6 z_{ij}^2 + \nonumber \\
& & a_7 x_{ij} y_{ij} + a_8 x_{ij}z_{ij} + a_9 y_{ij} z_{ij}  + O(h^3)
\end{eqnarray}
where ${\bf x_{ij}} = \urij / h_i = [x_{ij}, y_{ij}, z_{ij}]$ and $q_i$ is the quantity we wish to differentiate at particle $i$ (e.g. the density). 

The coefficients of this expansion can then be determined by inverting the following $10\times10$ matrix equation:

\begin{equation}
{\bf M} {\bf a} = {\bf q} 
\end{equation}  
where:
\beq
{\bf a}^\mathrm{T} = [a_0, a_1, a_2, a_3, a_4, a_5, a_6, a_7, a_8, a_9] 
\eeq
\begin{eqnarray}
{\bf q}^\mathrm{T} & = \sum_j^N m_j q_j \overline{W}_{ij} & \left[1, x_{ij}, y_{ij}, z_{ij},x_{ij}^2,y_{ij}^2,z_{ij}^2, \right . \nonumber \\
& & \left . x_{ij}y_{ij},x_{ij}z_{ij},y_{ij}z_{ij}\right]
\end{eqnarray}
\begin{eqnarray}
{\bf M} & = & \sum_j^N m_j \overline{W}_{ij}
\left(\begin{array}{rrr}
1 & \dx & \dy \cdots \\
\dx & \dx^2 & \dx\dy \cdots \\
\dy  & \dy \dx & \dy^2 \cdots \\
\dz & \dz \dx & \dz \dy \cdots \\
\dx^2 & \dx^3 & \dx^2\dy \cdots \\
\dy^2 & \dy^2 \dx & \dy^3 \cdots \\
\dz^2  & \dz^2 \dx & \dz^2 \dy \cdots \\
\dx\dy & \dx^2\dy & \dx\dy^2 \cdots \\
\dx\dz  & \dz\dx^2 & \dx\dz\dy \cdots \\
\dy\dz & \dy\dz \dx & \dz \dy^2 \cdots \\
 \end{array} \right . \nonumber \\ 
& &  \left . \begin{array}{lrrr}
 \cdots \dz & \dx^2 & \dy^2 & \dz^2 \cdots \\
\cdots  \dx\dz & \dx^3 & \dx\dy^2 & \dx\dz^2 \cdots\\
\cdots \dy\dz & \dy\dx^2 & \dy^3 & \dy\dz^2  \cdots\\
\cdots \dz^2 & \dz \dx^2 & \dz \dy^2 & \dz^3 \cdots\\
\cdots \dx^2\dz & \dx^4 & \dx^2\dy^2 & \dx^2\dz^2 \cdots \\
\cdots \dy^2 \dz & \dy^2 \dx^2 & \dy^4 & \dy^2 \dz^2 \cdots\\
\cdots \dz^3 & \dz^2 \dx^2 & \dz^2 \dy^2 & \dz^4  \cdots\\
\cdots \dx\dy\dz & \dy\dx^3 & \dx\dy^3 & \dx\dy\dz^2  \cdots\\
\cdots \dx\dz^2 & \dz\dx^3 & \dx\dz\dy^2 & \dx\dz^3 \cdots \\
\cdots \dy \dz^2 & \dy\dz \dx^2 & \dz \dy^3 & \dy \dz^3  \cdots\\
 \end{array} \right . \nonumber \\ 
& &  \left . \begin{array}{lrr}
\cdots \dx \dy & \dx \dz & \dy \dz \\
\cdots \dx^2 \dy & \dx^2 \dz & \dx \dy \dz \\
\cdots \dx \dy^2 & \dy\dx \dz & \dy^2 \dz \\
\cdots \dz \dx \dy & \dx \dz^2 & \dy \dz^2 \\
\cdots \dx^3 \dy & \dx^3 \dz & \dx^2\dy \dz \\
\cdots \dx \dy^3 & \dy^2 \dx \dz & \dy^3 \dz \\
\cdots \dz^2 \dx \dy & \dx \dz^3 & \dy \dz^3 \\
\cdots \dx^2 \dy^2 & \dy\dx^2 \dz & \dx\dy^2 \dz \\
\cdots \dz\dx^2 \dy & \dx^2 \dz^2 & \dx\dy \dz^2 \\
\cdots \dz \dx \dy^2 & \dy \dx \dz^2 & \dy^2 \dz^2 \\
\end{array} \right)
\end{eqnarray}
and $\overline{W}_{ij} = \frac{1}{2}[W_{ij}(h_i) + W_{ij}(h_j)]$ is the symmetrised smoothing kernel (the superscript $^\mathrm{T}$ means transpose). 

Having determined all of the coefficients of ${\bf a}$ (by solving ${\bf a} = {\bf M}^{-1} {\bf q}$), the gradients of $q$ evaluated at $i$ then simply follow as: 

\beq
\frac{\partial q_i}{\partial x} = a_1 ; \frac{\partial q_i}{\partial y} = a_2 ;\frac{\partial q_i}{\partial z} = a_3
\eeq
\section{Results for the evolved simulation (cylinders)}\label{app:cyl}
In Section \ref{inhom} we applied the HF and the MA method to the evolved simulation, considering several wedge shaped volumes at a Galactocentic distance $R=8.5$\,kpc around the disc. These wedge-shaped volumes allowed us to sample the star particles sufficiently well that we could study systematic errors on the recovery of the local density, without being affected by sampling errors.

\begin{figure}
\center
\includegraphics[width=0.5\textwidth]{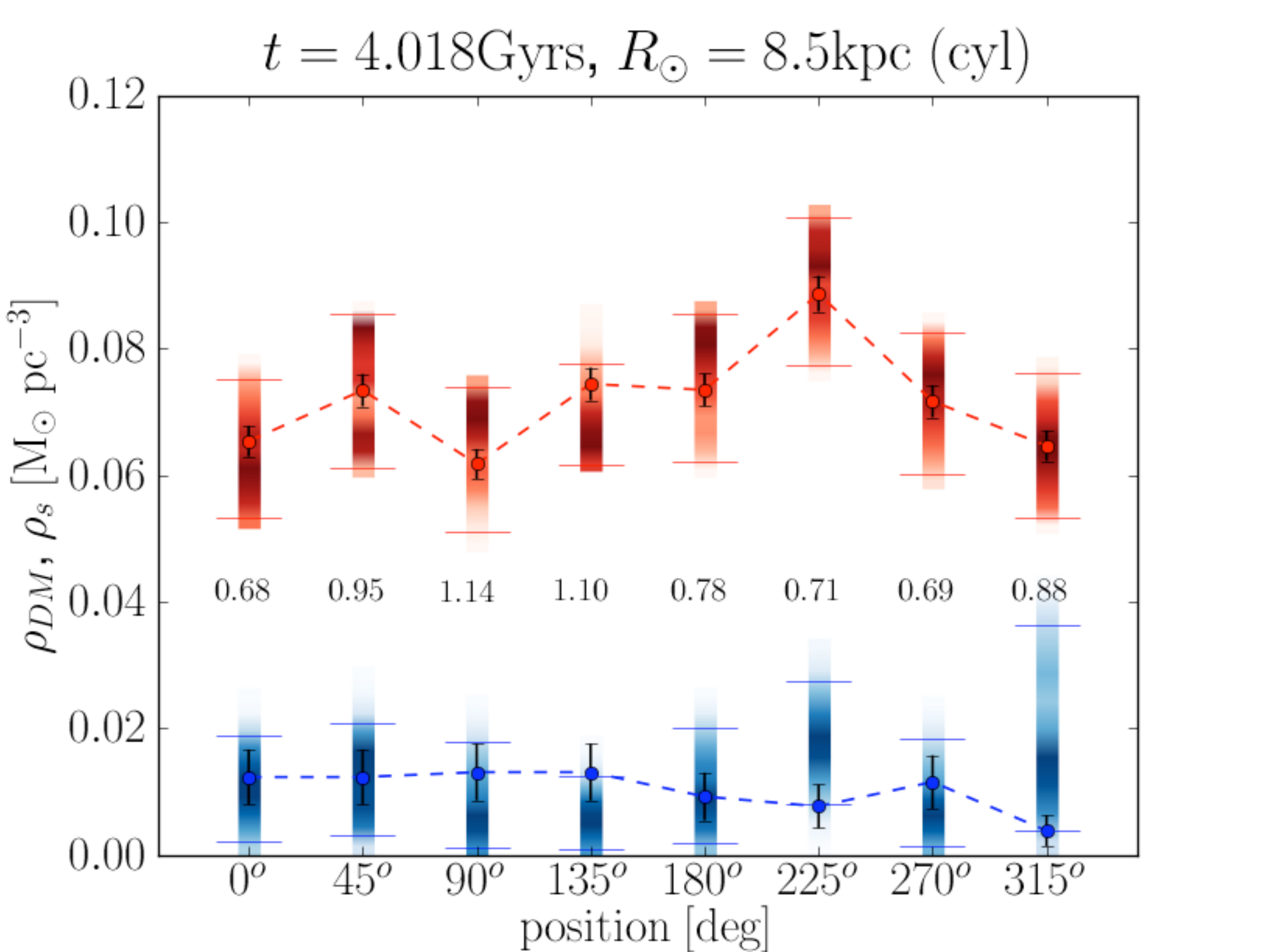}
\includegraphics[width=0.5\textwidth]{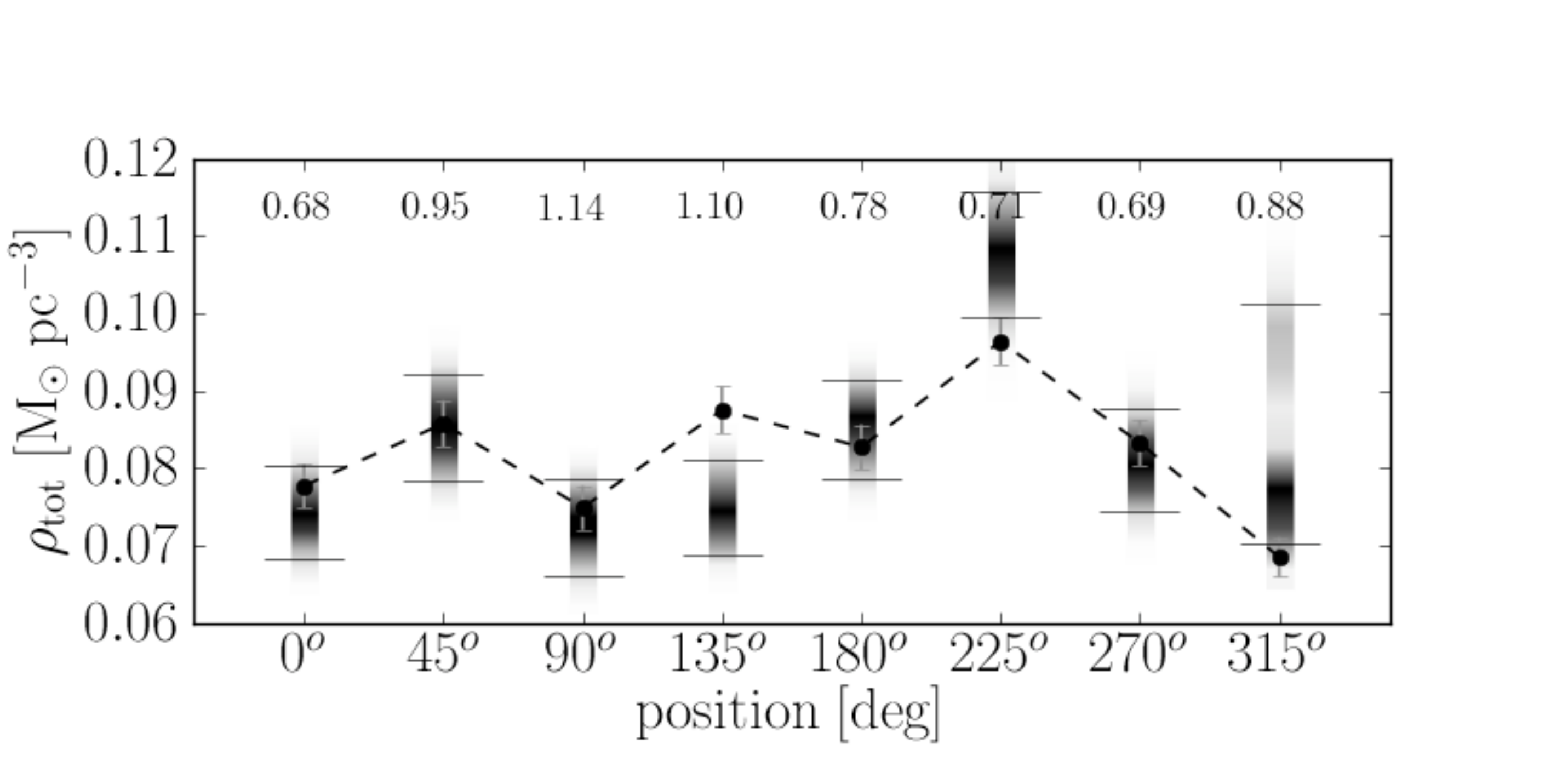}
\caption{Models explored by the MCMC for the MA method for 8 cylindrical `Solar neighbourhood' volumes applied to the evolved simulation at $R=8.5$\,kpc. Upper panel: recovered values of dark and visible matter density. Lower panel: recovered values of the total (dark+visible) matter density. The symbols and colours are as in Figure \ref{fig:820int}.}
\label{fig:MA+cyl}
\end{figure}

In this Appendix, we consider also the effects of sample error on the evolved simulation. We show the results for smaller cylindrical volumes at $R=8.5$\,kpc identical to those used to study the unevolved simulation in Section  \ref{results}. These volumes have a sampling and a shape similar to the {\it Hipparcos} data analysed by \cite{holmberg_local_2000} ($\sim 2000-3000$ within $|z|<200$\,pc). In Figure \ref{fig:MA+cyl}, we show the results for the MA method using cylindrical volumes. Now, due to the smaller volume sampled, the velocity dispersion $\vztwo(z)$ is quite noisy. To deal with this problem, we use the MCMC to marginalise over the velocity errors. At each iteration at the MCMC, we draw a value of $\vztwo(z)$ for each $z$-bin. We assume a Gaussian error distribution with a width corresponding to the uncertainty on $\vztwo(z)$. (Note that this approach is readily adapted to real data where $\vztwo(z)$ is also likely to be noisy and uncertain.) As can be seen in Figure \ref{fig:MA+cyl}, we can recover the correct value of the local visible, dark matter and total densities inside the errors for most of the volumes. Because of the poorer sampling, the uncertainties on the local density values are larger than the those obtained with the wedges (see Figure \ref{fig:moment-met}).

\section*{Acknowledgments}
We would like to thank Lawrence M. Widrow and John Dubinski for providing us with their latest version of the GalactICS code and the parameters for the initial condition of our Milky Way like galaxy; Chris Flynn for giving us the {\it Hipparcos} data from \cite{holmberg_local_2004} and for many useful comments; and Hanni Lux for providing the structure of the MCMC code used in this work. Justin I. Read would like to acknowledge support from SNF grant PP00P2\_128540/1.

\bibliographystyle{mn2e}
\bibliography{paperbibl}
\label{lastpage}
\end{document}